\documentclass[]{emulateapj}
\usepackage{lscape}
\submitted{Accepted to ApJ}

\newcommand{\Ha }{H$\alpha$}
\newcommand{\HI}{H{\sc i}}
\newcommand{\Dpak } {DensePak}
\newcommand{\kms} {km s$^{-1}$}
\newcommand{\siis}{[\mbox{S\,{\sc ii}}]$\lambda$6717}
\newcommand{\siio}{[\mbox{S\,{\sc ii}}]$\lambda$6731}
\newcommand{\nii}{[\mbox{N\,{\sc ii}}]$\lambda$6584}
\newcommand {\ml} {$\Upsilon_{*}$}

\shorttitle{Mass Models of LSB Galaxies}
\shortauthors{Kuzio de Naray et al.}

\begin{document}
\title{Mass Models for Low Surface Brightness Galaxies with High
  Resolution Optical Velocity Fields}
\author{Rachel Kuzio de Naray\altaffilmark{1,2}}
\affil{Center for Cosmology, Department of Physics and Astronomy, University of California, 
  Irvine, CA 92697-4575}
\altaffiltext{1}{NSF Astronomy and Astrophysics Postdoctoral Fellow}
\altaffiltext{2}{Visiting Astronomer, Kitt Peak
    National Observatory, National Optical Astronomy Observatory,
    which is operated by the Association of Universities for Research
    in Astronomy, Inc. (AURA) under cooperative agreement with the
    National Science Foundation.}
\email{kuzio@uci.edu}
\author{Stacy S. McGaugh}
\affil{Department of Astronomy, University of Maryland, College Park,
  MD 20742-2421}
\email{ssm@astro.umd.edu}
\and
\author{W.~J.~G.~ de Blok}
\affil{Department of Astronomy, University of Cape Town,
 Rondebosch 7700, South Africa}
\email{edeblok@circinus.ast.uct.ac.za}

\begin{abstract}
We present high-resolution optical velocity fields from \Dpak\ 
integral field spectroscopy, along with derived rotation curves, for 
a sample of low surface brightness galaxies.  In the limit of no 
baryons, we fit the NFW and pseudoisothermal halo models to the data 
and find the rotation curve shapes and halo central densities to be 
better described by the isothermal halo.  
For those galaxies with photometry, we present halo fits for three  
assumptions of the stellar mass-to-light ratio, \ml.  We find that the
velocity contribution from the baryons is significant enough in the 
maximum disk case that maximum disk and the NFW halo are mutually 
exclusive.    We find a substantial cusp mass 
excess at the centers of the galaxies, with at least two times more 
mass expected in the cuspy CDM halo than is allowed by the data. 
We also find that to reconcile the data with $\Lambda$CDM, $\sim$20 
\kms\ noncircular motions are needed and/or the power spectrum has a 
lower amplitude on the scales we probe.
\end{abstract}


\section{Introduction}
The behavior of cold dark matter (CDM) on galaxy scales has been 
discussed extensively in the literature, with particular attention 
given to constraints from rotation curves of low surface brightness 
(LSB) galaxies.  Numerical simulations of CDM predict cuspy halos with 
a density distribution showing a $\rho$ $\sim$ $r^{-\alpha}$ ($\alpha$ 
$\gtrsim$ 1) behavior, regardless of mass \citep*[e.g.,][]{ColeLacey, 
NFW96, NFW97, AvilaReese,  Klypin,Diemand}.  However, numerous studies of the 
rotation curves of LSB and dwarf galaxies have found the data to be 
inconsistent with a cuspy halo, and instead to be more consistent 
with a halo having a nearly constant density core: $\rho$ $\sim$ 
$r^{-\alpha}$ ($\alpha$ $\approx$ 0) \citep*[e.g.,][]{Flores, DMV, 
MRdB,Marchesini,dBBM,Gentile, Gentile05,Simon05, Kuzio, Spano07}.  This 
interpretation of the observations has been met with skepticism, 
though, as cored halos lack 
theoretical and cosmological motivation.  Systematic effects in the 
data (most notably beam smearing, slit misplacement, and noncircular 
motions) have also been used to argue against the presence of cored 
halos \citep*[e.g.,][]{vandenB00, Rob, Simon03, Rhee04}, although some 
authors find the magnitude of these effects to be insufficient for 
masking the presence of a cusp \citep{dBBM}.

To address these concerns, the most recent observations have been made 
using integral field spectrographs \citep[e.g.,][]{Chemin04, Gentile, 
Gentile05, Kuzio, Simon05, Swaters03}.  By their nature, these high-resolution 
two-dimensional velocity fields eliminate concerns about long-slit 
placement, and highlight the presence of noncircular motions.  They 
also typically probe in detail the innermost regions of the galaxies 
where the cusp-core conflict is most severe.  Even with these 
improved observations, the data remain more consistent with cored 
halos and tend to have concentrations too low for $\Lambda$CDM 
(Kuzio de Naray et al.~2006; Gentile et al.~2007; but see Swaters 
et al.~2003b).

In a previous paper 
\citep[hereafter K06]{Kuzio}\defcitealias{Kuzio}{K06}, we presented 
high-resolution optical velocity fields and derived rotation curves for 11 LSB 
galaxies, along with pseudoisothermal (core) and NFW (cusp) halo fits 
in the limit of no baryons. LSB galaxies are thought to be dark 
matter-dominated down to small radii (de Blok \& McGaugh 1996, 1997; 
Boriello \& Salucci 2001; but see Fuchs 2003) with the light simply 
providing a tracer for the dark matter, so neglecting the baryons is 
not entirely unreasonable.  The stellar mass contribution in these 
systems is low, which reduces errors involving the uncertainty in the stellar 
mass-to-light ratio, \ml, and in turn, the isolation of the dark 
matter component.  While halo fits are often made under the assumption 
that all observed rotation is due to the dark matter, it is not 
strictly true.  In this paper, we present halo fits for four 
assumptions about \ml\ for the galaxies in \citetalias{Kuzio} with 
available photometry.  We also present observed \Dpak\ velocity fields, 
rotation curves and halo fits for six new galaxies, along with 
additional \Dpak\ observations, updated rotation curves and halo 
fits for three galaxies in \citetalias{Kuzio}.

In \S\ 2 we describe our sample of galaxies and the \Dpak\ 
observations; the data reduction is discussed in \S\ 3.  The 
observed velocity fields and derived rotation curves are presented 
in \S\ 4 and the zero disk halo fits are presented in \S\ 5.  Mass 
models for the galaxies with photometry are presented in \S\ 6 and 
\S\ 7.  A discussion of the mass model results is given in \S\ 8.  Our 
conclusions are stated in \S\ 9.

\section{Sample and Observations}

During the nights of 2006 March 28-29, April 6-8, August 28-30 and 
September 25-27 we observed 17 LSB galaxies using the \Dpak\ 
Integrated Field Unit (IFU) on the 3.5 m 
WIYN\footnote{Based on observations obtained at the WIYN 
Observatory.  The WIYN Observatory is a joint facility of the
University of Wisconsin-Madison, Indiana University, Yale University,
and the National Optical Astronomy Observatory.} telescope at the Kitt 
Peak National Observatory (KPNO).  UGC 4325, DDO 64, and F583-1 were 
observed during March and April.  These observations augment previous 
\Dpak\ observations published in \citetalias{Kuzio}.  The 14 targets 
that were observed in August and September were selected primarily 
from the Nearby Galaxies Catalogue \citep{Tully}.  Selection criteria 
for these galaxies included positions satisfying 18$^{h}$ $\lesssim$ $\alpha$ 
$\lesssim$ 08$^{h}$ and 10\degr\ $\lesssim$
$\delta$ $\lesssim$ 50\degr, inclinations between 30\degr\ and
85\degr, heliocentric velocities $\lesssim$ 3000 
\kms, and an estimated $V_{flat}$ (approximated by 
$V_{flat}$ $\sim$ 0.5$W_{20}$(sin i)$^{-1}$) between roughly 50 \kms\ 
and 100 \kms.  Additionally, we selected those galaxies which appear
to have diffuse \Ha\ emission and lack indicators of significant
noncircular motions (e.g., strong bars or gross asymmetries).  We also
targeted galaxies with previous long-slit rotation curves or \Ha\ imaging.

Because of inclement weather and telescope scheduling, we did not have 
\Ha\ imaging or long-slit rotation curves for most of the galaxies in 
the sample prior to making the \Dpak\ observations.  
Without these data, there is no way of 
knowing how much \Ha\ emission will be detected by the 
IFU until the observations are actually made.  Thus,
our observed sample of galaxies is signal-limited;  the IFU fibers
detected sufficient \Ha\ emission to create a useable velocity field in just
under 50\% of our target galaxies. 

The observing setup and procedure were identical to that used in 
\citetalias{Kuzio}.
The IFU orientation on the sky and the 
total number of pointings per galaxy were tailored to each galaxy so 
that the critical central regions were covered by the \Dpak\ fibers.
Each exposure was 1800 s, and two exposures were taken at each 
pointing.  To provide wavelength calibration, a CuAr lamp was observed 
before and after each pointing.  We used the 860 line mm$^{-1}$ 
grating in second order, centered near \Ha, giving a 58 \kms\ velocity 
resolution.  The \Dpak\ fibers are 3\arcsec, however, we achieved 
$\sim$2\arcsec\ resolution by shifting the \Dpak\ array by small 
amounts so that the spaces between the fibers were observed.  The 
distances to the galaxies in the sample are such that a 3\arcsec\ 
fiber provides subkiloparsec resolution.

\section{Data Reduction}
The data were reduced following the procedure described in 
\citetalias{Kuzio}.  Briefly, the observations were reduced in 
IRAF\footnote{IRAF is distributed by 
the National Optical Astronomy Observatory, which is operated by the 
Association of Universities for Research in Astronomy (AURA), Inc., 
under agreement with the National Science Foundation.} using the 
\texttt{HYDRA} package.  Night-sky emission lines \citep{Osterbrock} 
were used for wavelength calibration, and velocities were measured by fitting 
Gaussians to both the sky lines and the four galactic emission lines 
of interest: \Ha, \nii, \siis\ and \siio.  The average error on 
individual emission-line velocities due to centroiding accuracies was 
roughly 1.5 \kms.  The velocity assigned to each fiber was the 
arithmetic mean of the measured emission-line velocities in the
fiber. The error on the fiber velocity was the maximum difference 
between the measured velocities and the mean.  Many of these errors were 
less than 5 \kms, although a few were as high as $\sim$20 \kms.  
If only \Ha\ was observed in a fiber, the observed \Ha\ velocity was 
taken as the fiber velocity.  Without other lines to determine the 
maximum difference, we adopt 10 \kms\ as a conservative error estimate 
based on experience with long-slit data \citep[e.g.][]{MRdB}.

Using the input shifts at the telescope, individual \Dpak\ pointings 
were combined to construct the observed velocity field.  An \Ha\ flux 
image of the galaxy constructed from the \Dpak\ observations was 
compared to an actual \Ha\ image of the galaxy obtained at the KPNO 2 m 
telescope to confirm the accuracy of the offsets.  The fiber maps were 
registered to the galaxy image by the correspondence of flux maxima 
through the fibers to observed features like individual H{\sc ii} regions.  
We find the accuracy  of the fiber positions 
($\sim$0\arcsec.6) to be consistent with our results in 
\citetalias{Kuzio}.  Both large (1\arcsec$-$2\arcsec) and small 
($\sim$0\arcsec.7) shifts can be made confidently.

Rotation curves were derived from the observed velocity fields using 
the tilted-ring fitting program \texttt{ROTCUR} \citep{Begeman} following the 
procedure outlined in \citetalias{Kuzio}.   To construct a rotation curve, 
\texttt{ROTCUR} requires the systemic velocity, inclination, kinematic center,
and position angle of the galaxy in addition to the observed fiber 
velocities.  For a more complete description of \texttt{ROTCUR} as 
applied here, the reader is referred to \citetalias{Kuzio}.

\begin{deluxetable*}{lccccccccccc}
\tabletypesize{\scriptsize}
\tablecaption{Properties of Observed Galaxies}
\tablewidth{0pt}
\tablehead{
 &\colhead{R.A.} &\colhead{Decl.} &\colhead{$\mu_{0}$(B)}
 &\colhead{Distance} &\colhead{$i$} &\colhead{$V_{hel}$}
 &\colhead{$R_{max}$} &\colhead{$V_{max}$} &\colhead{P.A.} &\colhead{$\sigma$} &\colhead{}\\
\colhead{Galaxy} &\colhead{(J2000.0)} &\colhead{(J2000.0)}
&\colhead{(mag arcsec$^{-2}$)} &\colhead{(Mpc)} &\colhead{(deg)}
&\colhead{(\kms)} &\colhead{(kpc)} &\colhead{(\kms)} &\colhead{(deg)} &\colhead{(\kms)} &\colhead{References}\\
\colhead{(1)} &\colhead{(2)} &\colhead{(3)} &\colhead{(4)}
&\colhead{(5)} &\colhead{(6)} &\colhead{(7)} &\colhead{(8)}
&\colhead{(9)} &\colhead{(10)} &\colhead{(11)}
 &\colhead{(12)}
}
\startdata
UGC 4325 &08 19 20.5 &+50 00 35 &22.5$^{a}$ &10.1 &41 &514 &2.9 &110
&52 &9.0 &2,2,2\\
DDO 64 &09 50 22.4 &+31 29 16 &\nodata &6.1 &60 &517 &1.9 &60 &97 &7.9
&2,2,2\\
F583-1 &15 57 27.5 &+20 39 58 &24.1 &32 &63 &2256 &5.8 &72 &355 &8.7 &4,4,4\\
NGC 7137$^{b}$   &21 48 13.0 &+22 09 34 &20.7$^{c}$ &22.5 &38 &1669
&3.2 &62 &44 &7.1 &1,10,12\\
UGC 11820 &21 49 28.4 &+14 13 52 &23.7 &13.3 &50 &1088 &2.9 &93 &309 &8.9 &14,7,9\\
UGC 128 &00 13 50.9 &+35 59 39 &24.2 &60 &57 &4509 &13.5 &133 &62 &15.5 &8,3,3\\
UGC 191  &00 20 05.2 &+10 52 48 &22.7 &17.6 &39 &1139 &2.4 &97 &156 &7.7 &14,13,12\\
UGC 1551   &02 03 37.5 &+24 04 32 &22.5 &20.2 &63 &2663 &3.8 &83 &114 &9.7 &5,11,12\\
NGC 959$^{d}$   &02 32 23.9 &+35 29 41 &21.9$^{e}$ &7.8 &51 &590 &1.6
&77 &64 &8.7 &6,7,7
\enddata
\tablecomments{Units of right ascension are hours, minutes, and
  seconds, units of declination are degrees, arcminutes, and
  arcseconds. Col.(1): Galaxy name. Col.(2): Right Ascension. Col.(3):
  Declination. Col.(4): Central surface brightness in
  $B$-band. Col.(5): Distance. Col.(6): Inclination. Col.(7):
  Heliocentric systemic velocity. Col.(8): Maximum radius of the
  DensePak rotation curve. Col.(9): Maximum velocity of the DensePak
  rotation curve. Col.(10): Position angle of major axis; see \S 3
  for details. Col.(11): Velocity dispersion of the \Dpak\ data.  Col.(12): References for surface brightness, distance, and inclination.\\
\indent $^{a}$ Converted from $R$ band assuming $B$ $-$ $R$ = 0.9.\\
\indent $^{b}$ NGC 7137 = UGC 11815.\\
\indent $^{c}$ Converted from $V$ band assuming $B$ $-$ $V$ = 0.57.\\
\indent $^{d}$ NGC 959 = UGC 2002.\\
\indent $^{e}$ Converted from $I$ band assuming $V$ $-$ $I$ = 0.90 and $B$ $-$ $V$ = 0.53.\\
\indent REFERENCES $-$ (1) Baggett et al.\ 1998.  (2) de Blok \& Bosma
2002.  (3) de Blok \& McGaugh 1996.  (4) de Blok, McGaugh, \& Rubin
2001.  (5) de Jong 1996.  (6) Heraudeau \& Simien 1996.  (7)
James et al.\ 2004.  (8) McGaugh 2005.  (9) McGaugh, Rubin, \& de Blok
2001.  (10) Rosenberg \&
Schneider 2003.  (11) Swaters \& Balcells 2002.  (12) Tully 1988.  (13)
van Zee \& Haynes 2006.  (14) van Zee et al.\ 1997.}
\end{deluxetable*}

Because the \Dpak\ data cover the centers of the galaxies and probe 
the regime of solid-body rotation, neither the galaxy center nor the 
inclination could be determined from the observations.  The velocity 
field centers were therefore fixed to the optical centers of the 
galaxies, as determined by the centroid of ellipses fit in the surface 
photometry, and the inclination was fixed to published values 
\citep{Tully, dBM96, MRdB, dBMR, dBB, James}.  The systemic velocities 
were determined by \texttt{ROTCUR}.  We used \texttt{ROTCUR} to 
determine the position angle of the major axis, using published 
long-slit values as the initial guess.  If the position angle could 
not be well-constrained by \texttt{ROTCUR}, then it was fixed to the 
long-slit value.  The position angles of UGC 4325, 
DDO 64, and F583-1 were not well-constrained by \texttt{ROTCUR}, and 
remained fixed at the values used in 
\citetalias{Kuzio}.  The position angle of the major axis was 
well-constrained by \texttt{ROTCUR} for all of the galaxies in the new 
sample except for UGC 128.  In the case of UGC 128, the position angle 
was set to the position angle of the \HI\ velocity field of 
\citet{vanderHulst}.    For reasons described in \citetalias{Kuzio}, 
we did not impose a minimum error on the rotation curve points.  
However, we added in quadrature to the error bars on the final 
\texttt{ROTCUR} rotation curve  the velocity error from Gaussian 
centroiding accuracy, typically $\sim$1.5 \kms, corrected for inclination.  
We have also not corrected the rotation velocities for asymmetric drift, 
as the corrections are typically only $\sim$2 \kms\ 
\citep[see also][]{dBB}.

\section{Observed Velocity Fields and Derived Rotation Curves }

In this section, we present the \Dpak\ fiber positions, observed
velocity fields, and rotation curves in Figures 1-3.  A
description of each galaxy is given, and the properties of the galaxies
for which a rotation curve was derived are listed in Table 1.

\subsection{Extended Observations of Previously Observed Galaxies}
Additional \Dpak\ observations were made of three galaxies published
in \citetalias{Kuzio}.  In this section, we describe the positions of the 
new \Dpak\ pointings, present the augmented velocity fields, and
discuss the resulting changes in the rotation curves.

$\textbf{\textit{UGC 4325}$-$}$  There were four new \Dpak\ pointings
roughly through the galaxy center from SE to NW.  All eight pointings
are shown on the \Ha\ image (Figure 1) of the galaxy.  The fiber velocities 
were the average of the \Ha\ and
\siis\ lines.  The position angle of the major axis remained fixed at
the average of the position angles of previous long-slit observations
\citep{dBB, Rob}.  With the addition of the new \Dpak\ pointings, the
\Dpak\ rotation curve extends to $\sim$ 60$\arcsec$ and remains
in excellent agreement with the long-slit \Ha\ rotation curve of
\citet{dBB}.

$\textbf{\textit{DDO 64}$-$}$  There were two new \Dpak\ pointings
along the SE end of DDO 64.  All five pointings are shown on the \Ha\
image (Figure 1) of the galaxy.  The fiber
velocities were the average of the \Ha\ and \siis\ lines.  The
position angle remained fixed at the value determined in \citet{dBB}.
With the addition of the new \Dpak\ pointings and an \Ha\ image, the
positions of the \Dpak\ fibers on the galaxy have been updated and the
center of the velocity field has moved 13\arcsec E and 
2\arcsec.5 S. The
new \Dpak\ rotation curve is now consistent with the long-slit \Ha\
rotation curve of \citet{dBB} in the inner 20\arcsec.  Beyond
20\arcsec, the \Dpak\ rotation curve shows a more linear rise with
less scatter than its previous version.

$\textbf{\textit{F583-1}$-$}$  There were two new \Dpak\ pointings, 
one to the north and one to the east, on this galaxy.  All three 
pointings are shown on the $R$-band image (Figure 1) of the galaxy from
\citet{DMV}. The fiber velocities were the  average of the \Ha, \siis,
and \siio\ lines.  The receding side of the galaxy and more of the
minor axis are now covered with the additional pointings.  The
position angle remained fixed at the value listed in \citet{MRdB}.
The new \Dpak\ rotation curve shows less scatter than the curve
derived from a single \Dpak\ pointing, and it displays a higher level
of consistency with the long-slit \Ha\ rotation curve of \citet*{dBMR}.

\begin{figure*}
\includegraphics[scale=0.20]{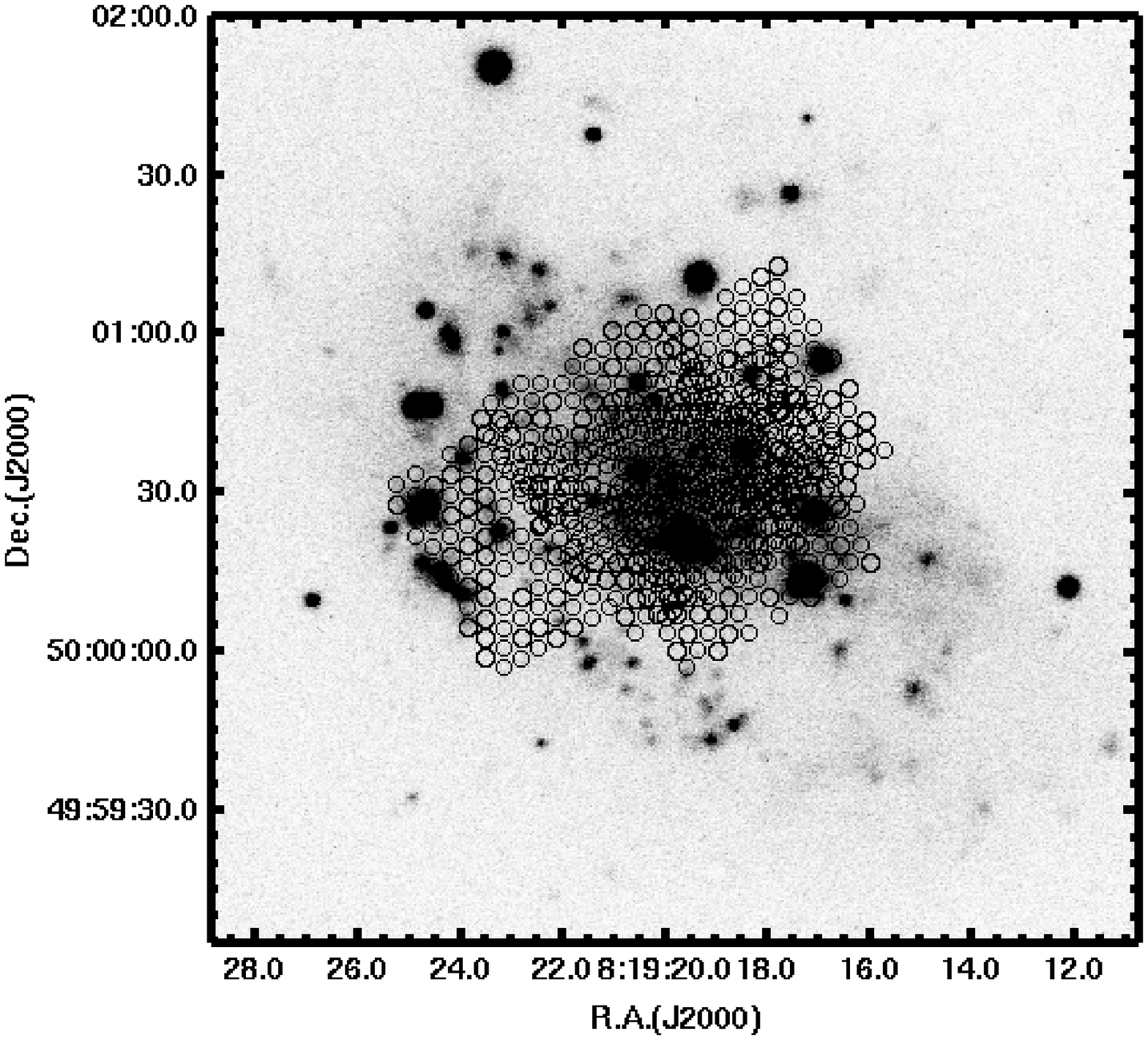}
\hfill
\includegraphics[scale=0.195]{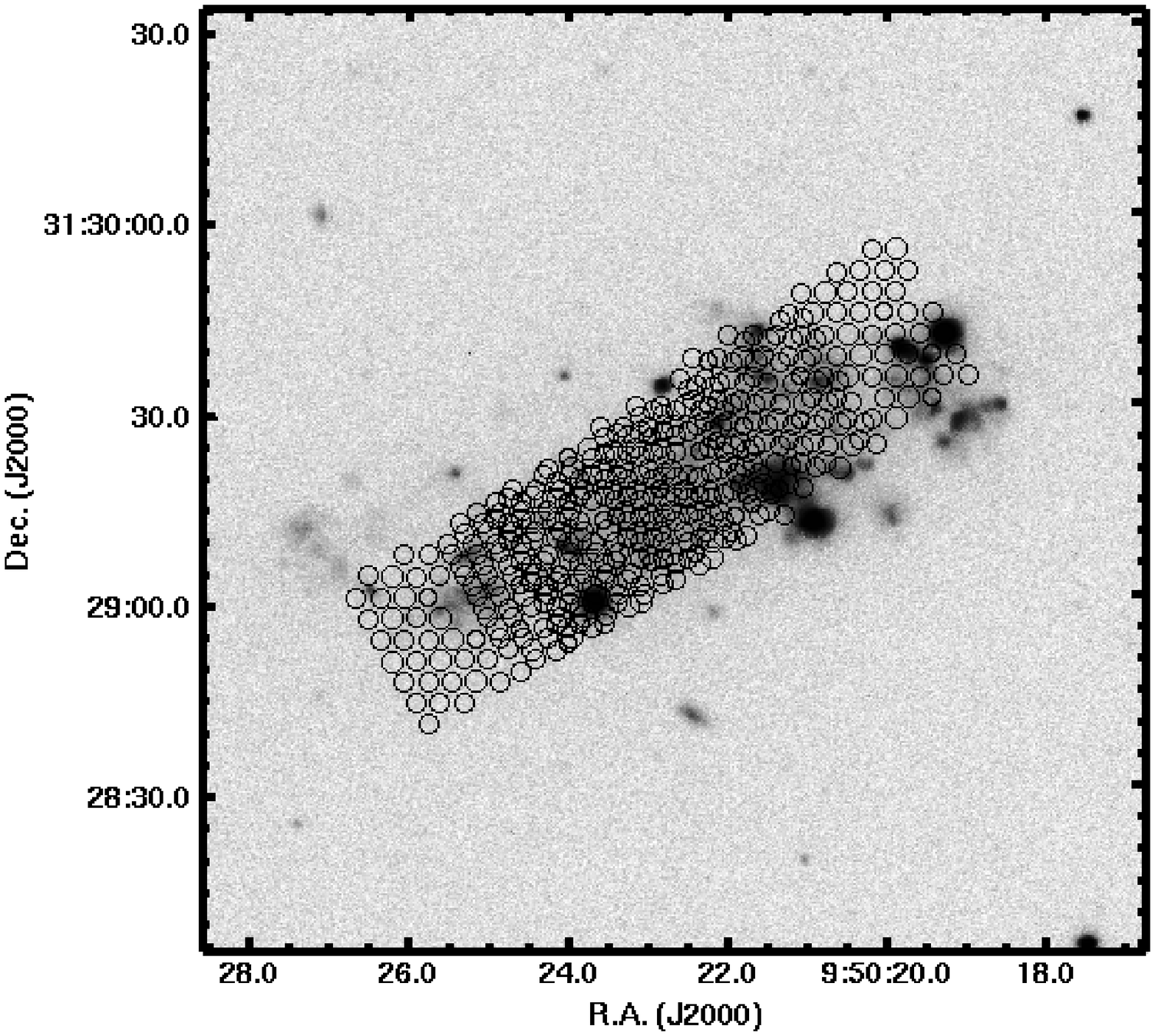}
\hfill
\includegraphics[scale=0.195]{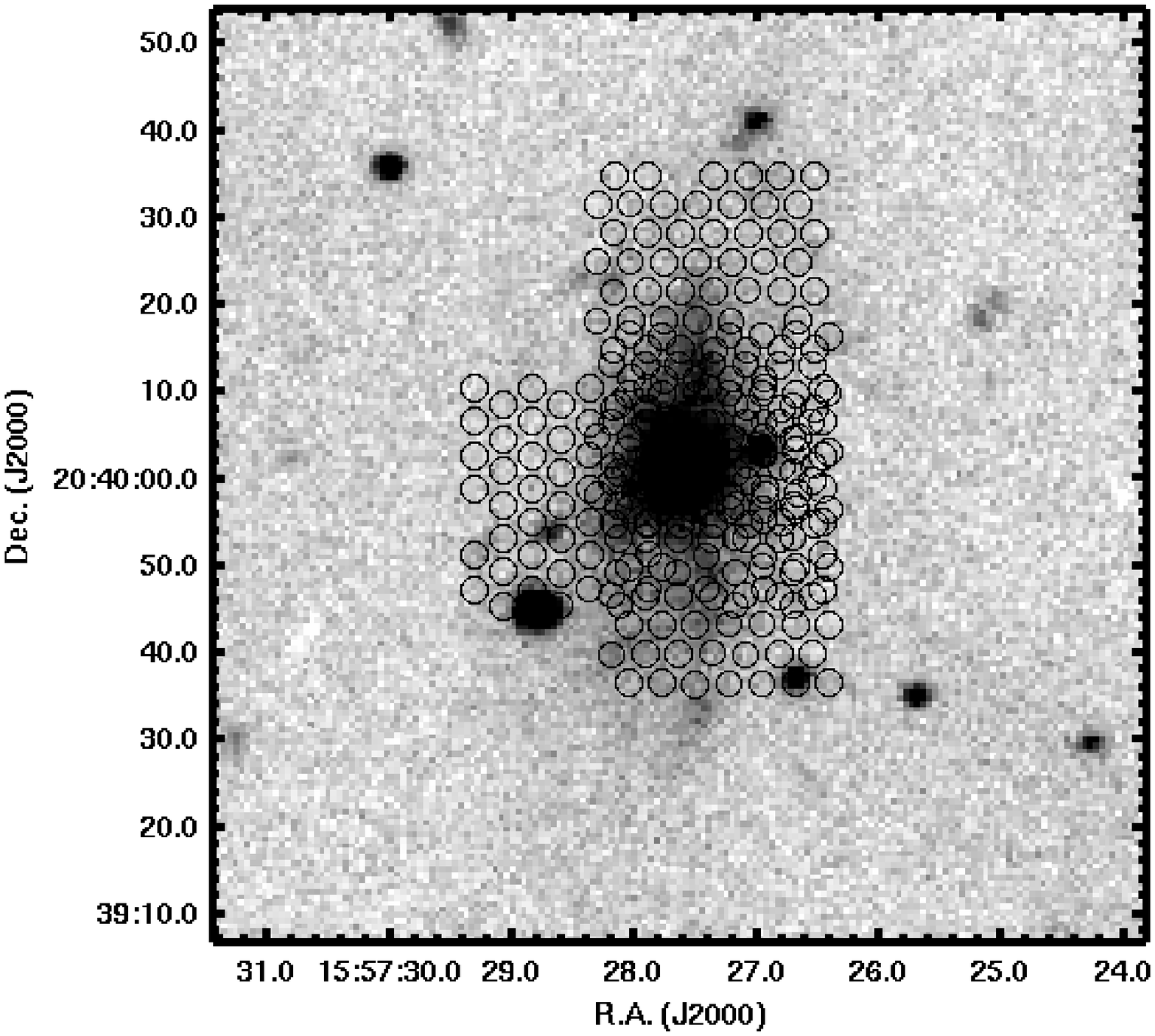}\\
\includegraphics[scale=0.34]{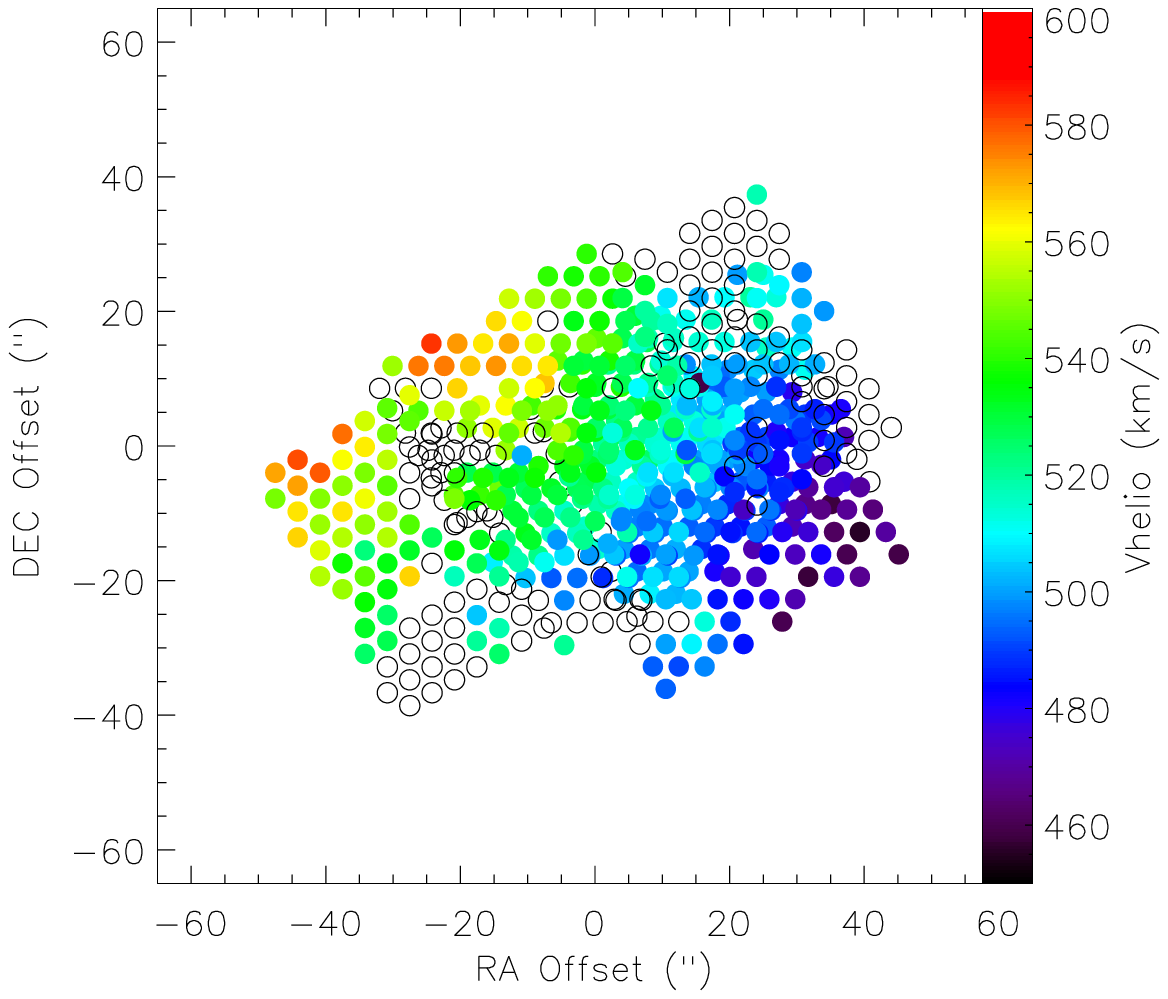}
\hfill
\hfill
\hfill
\includegraphics[scale=0.34]{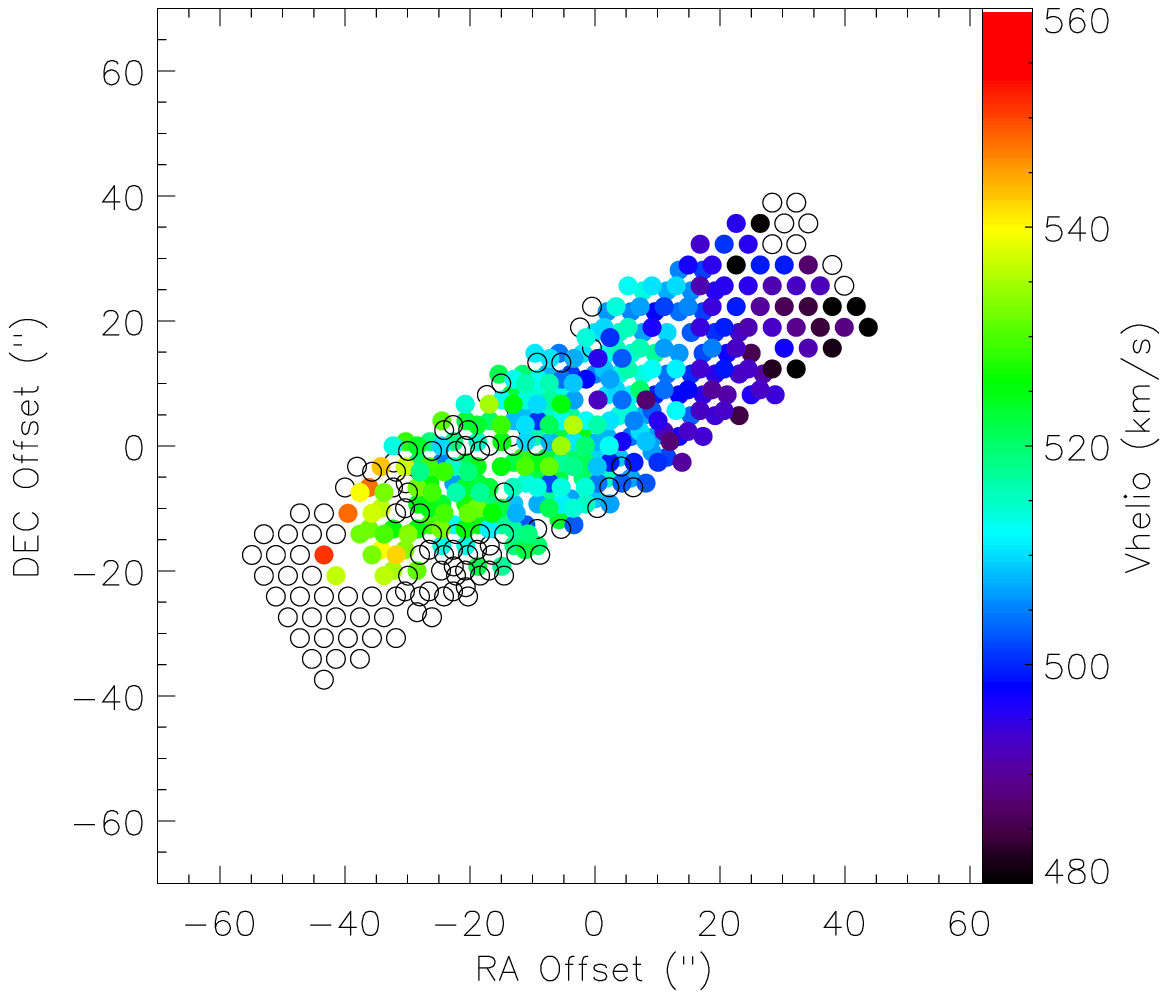}
\hfill
\hfill
\hfill
\includegraphics[scale=0.34]{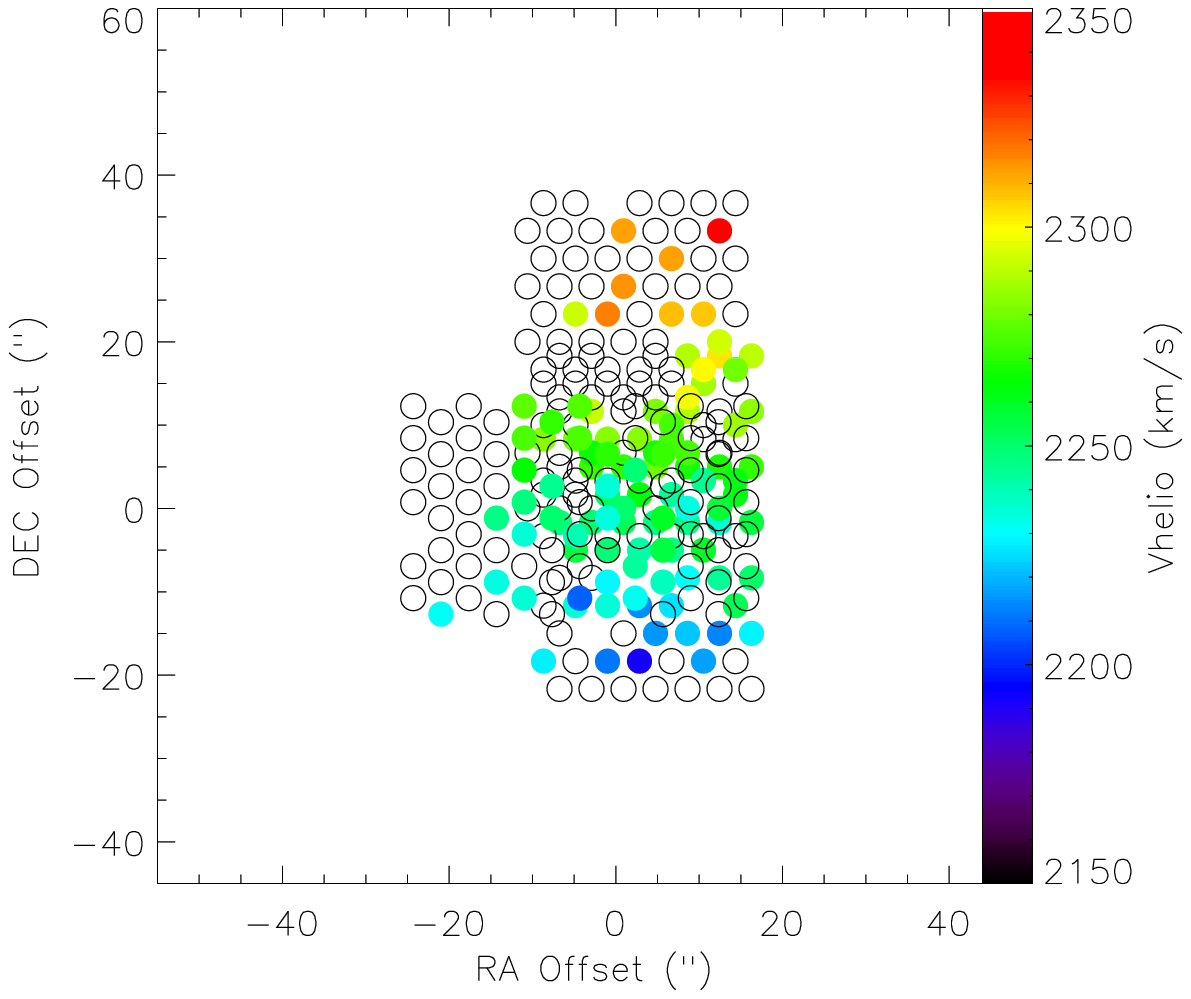}\\
\includegraphics[scale=0.21]{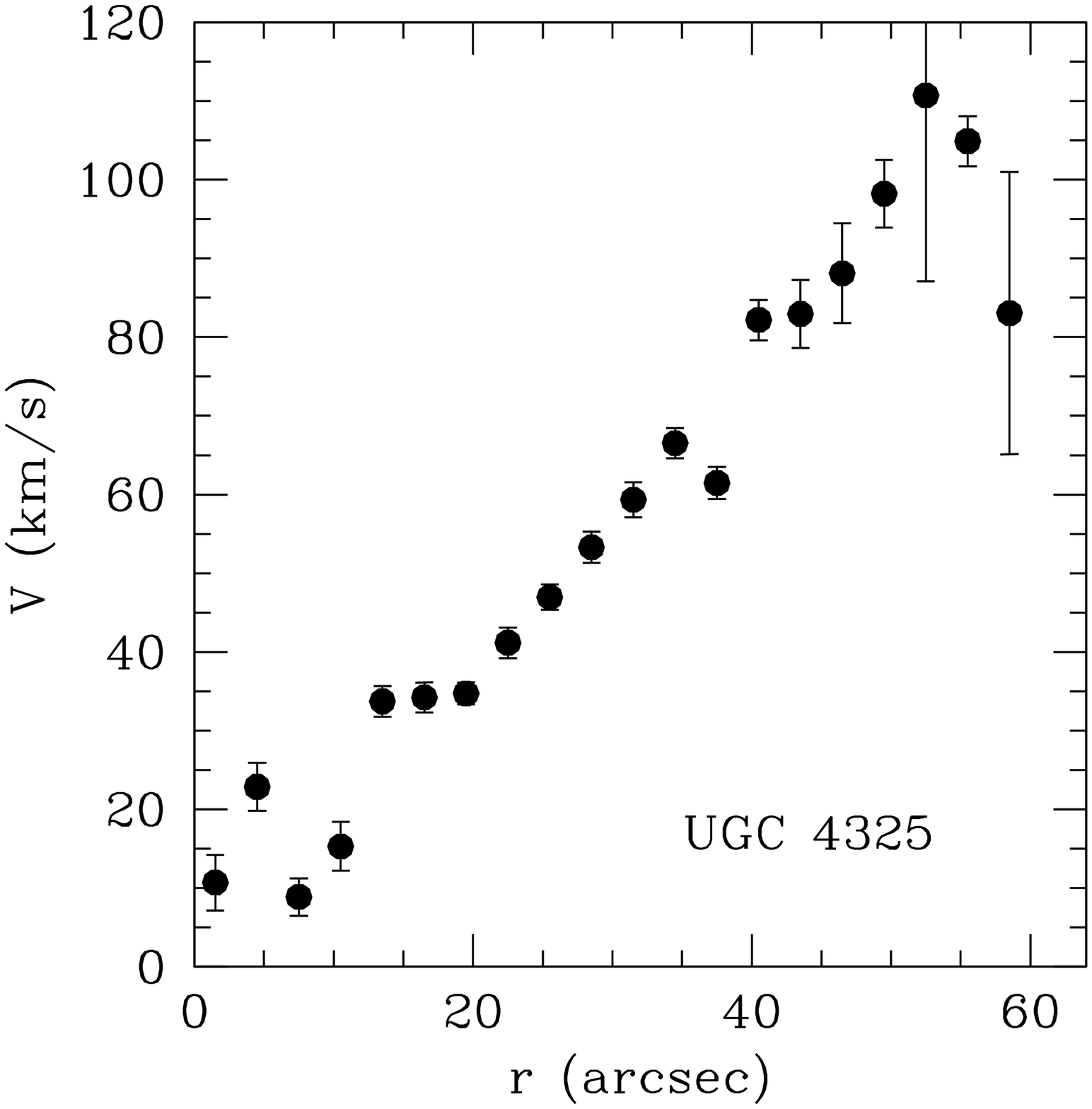}
\hfill
\includegraphics[scale=0.21]{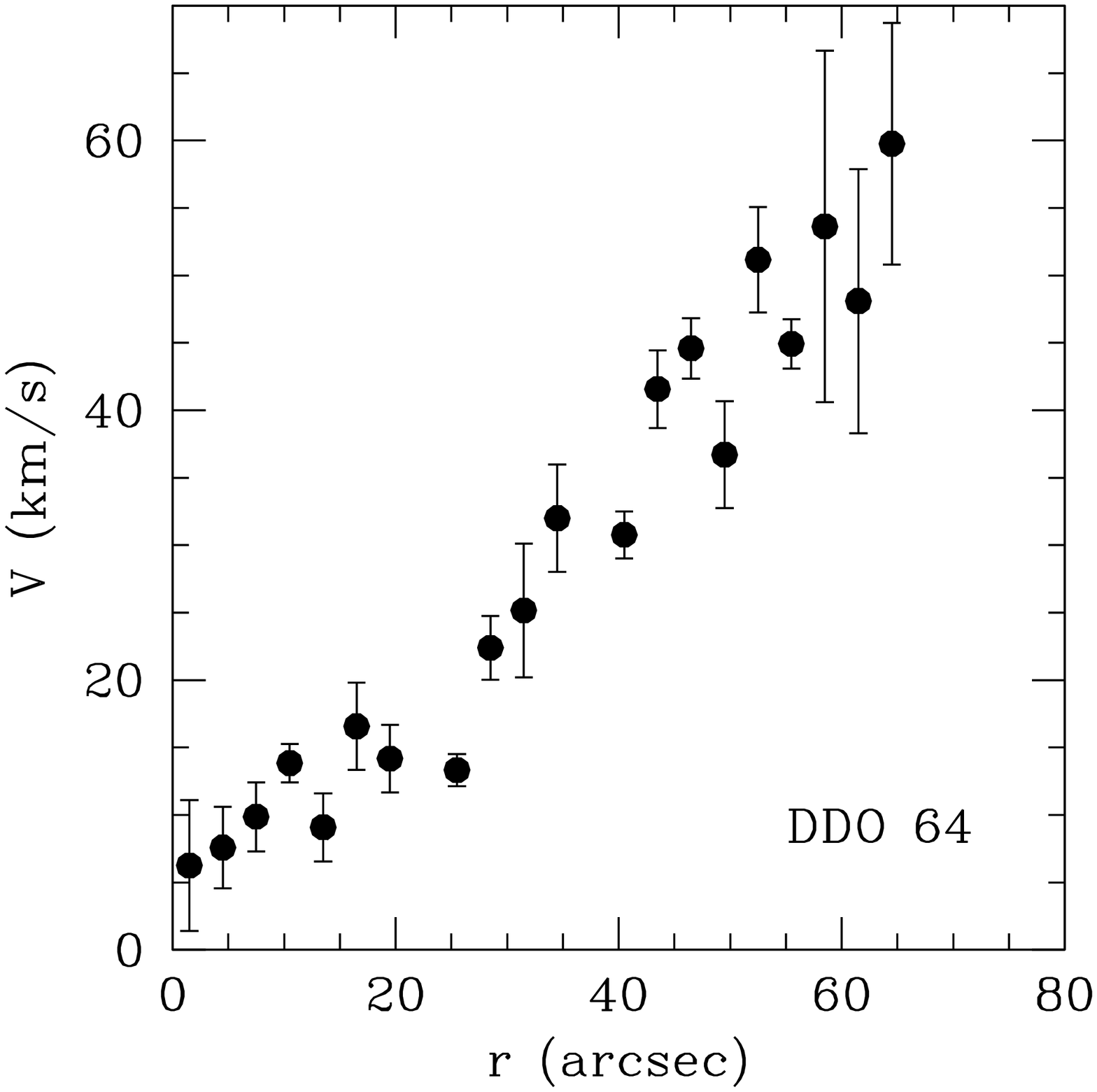}
\hfill
\includegraphics[scale=0.21]{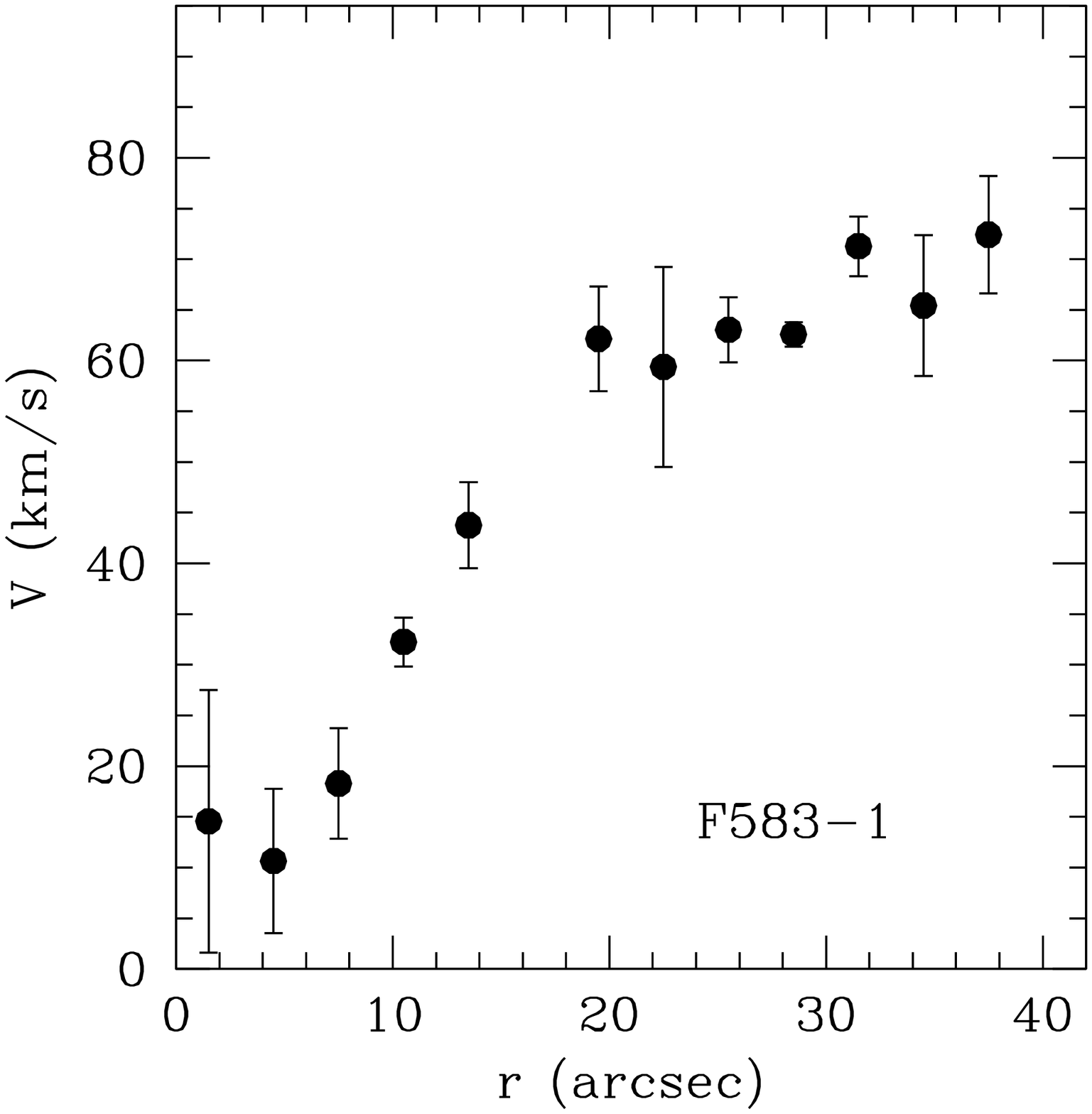}\\
\includegraphics[scale=0.21]{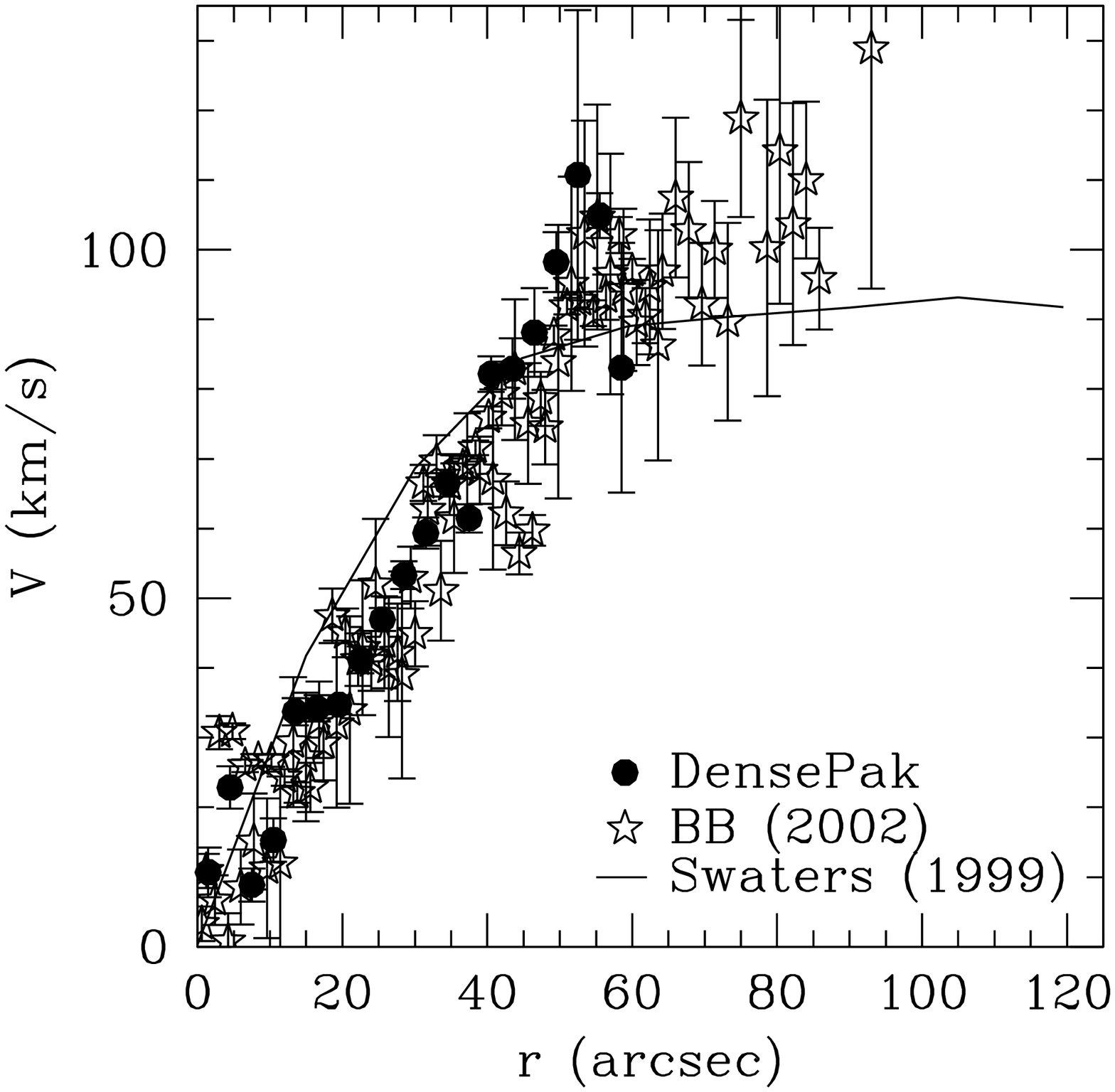}
\hfill
\includegraphics[scale=0.21]{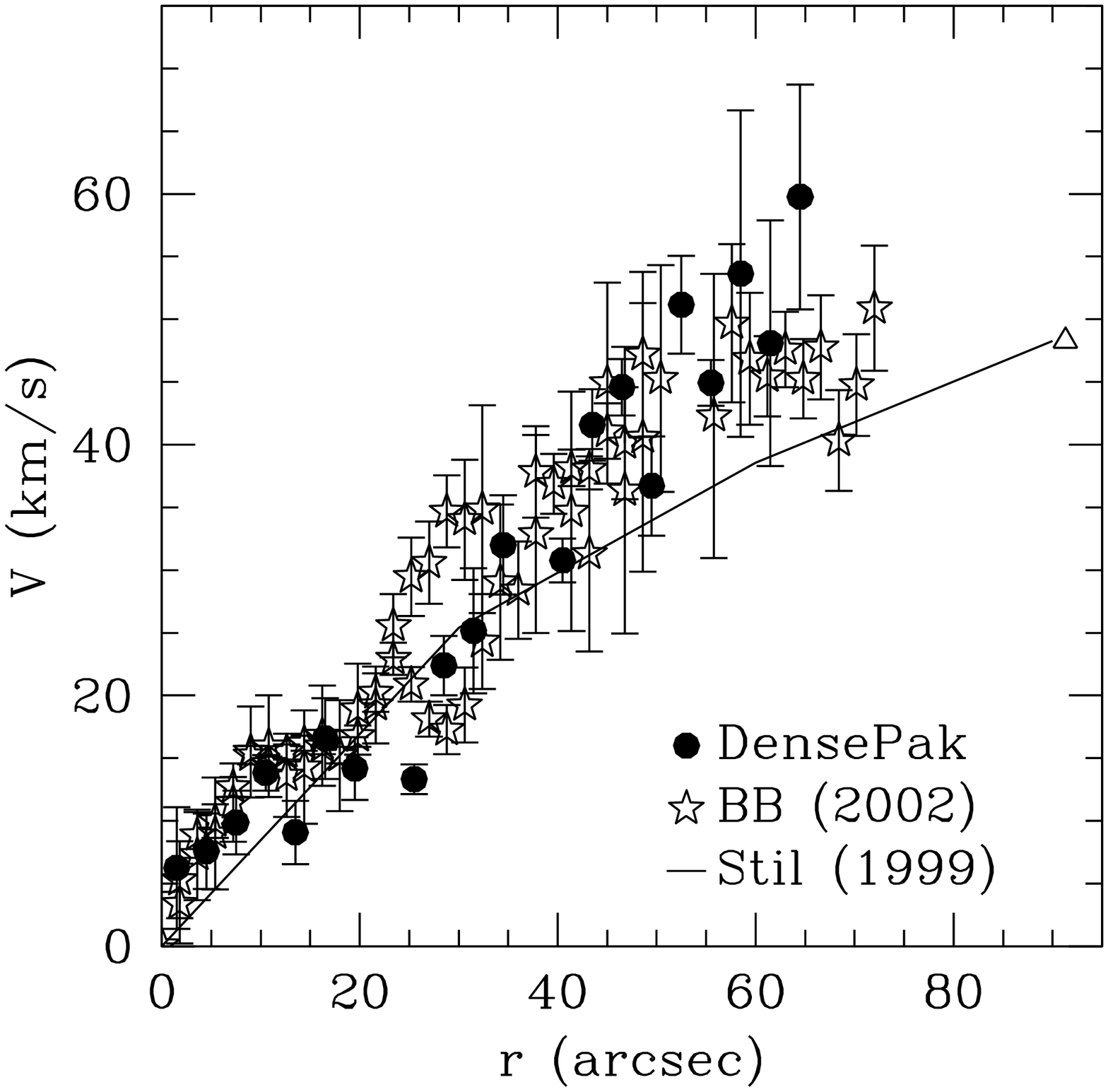}
\hfill
\includegraphics[scale=0.21]{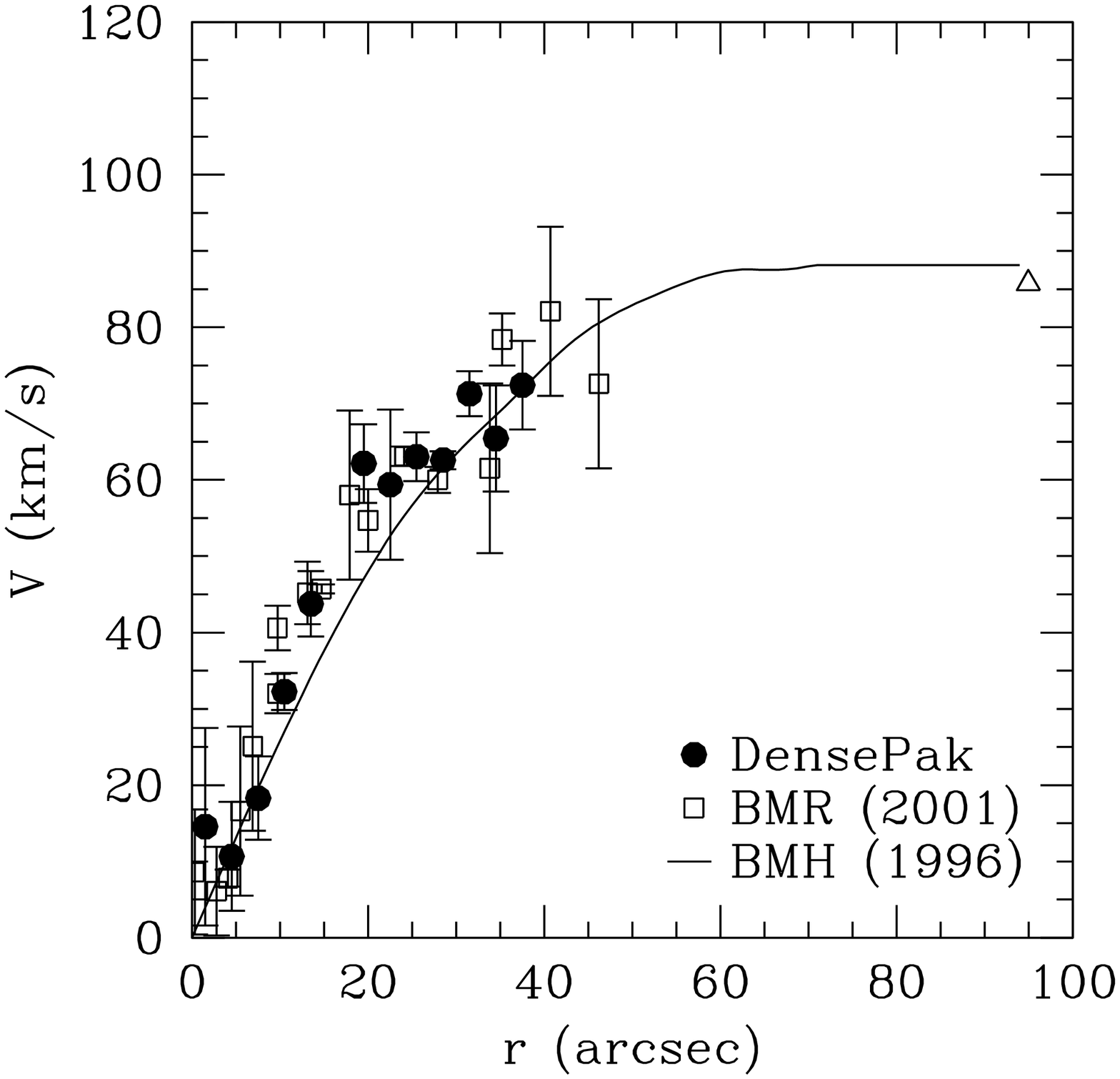}
\begin{quote}
\caption{Additional observations of UGC 4325, DDO 64, and F583-1.
  $\textit{Top row:}$ Position of \Dpak\ array on the \Ha\ images of
  UGC 4325 and DDO 64 and the $R$-band image of F583-1.
  $\textit{Second row:}$ Observed \Dpak\ velocity field with new pointings.
 Empty fibers are those without detections.  $\textit{Third row:}$
 Updated \Dpak\ rotation curves.  $\textit{Bottom row:}$ Updated
 \Dpak\ rotation curves plotted with long-slit \Ha\ and \HI\ rotation
 curves.  [{\it See the
  electronic edition of the Journal for a color version of this figure.}] }
\end{quote}
\end{figure*}

\begin{figure*}
\includegraphics[scale=0.30]{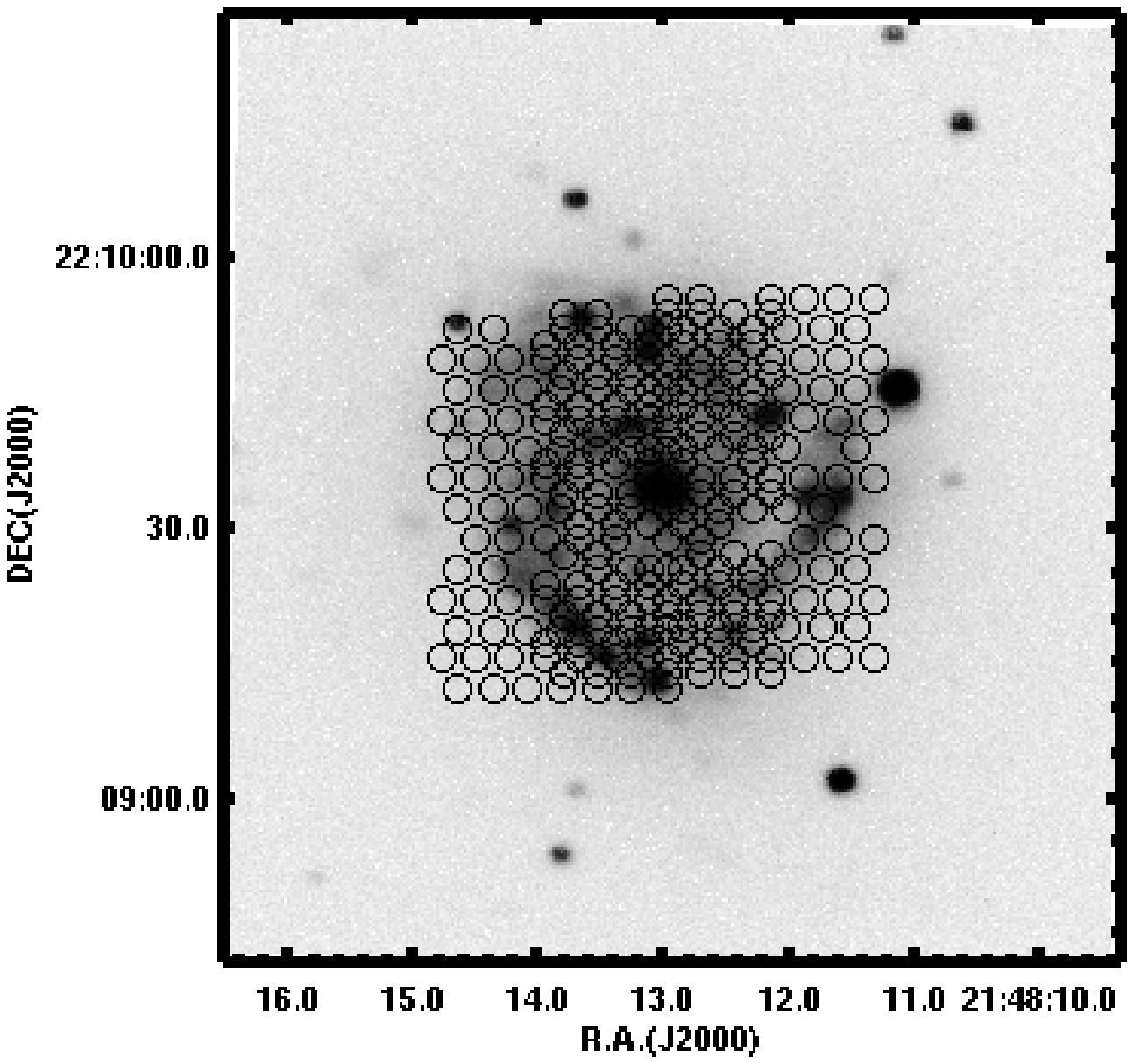}
\hfill
\includegraphics[scale=0.30]{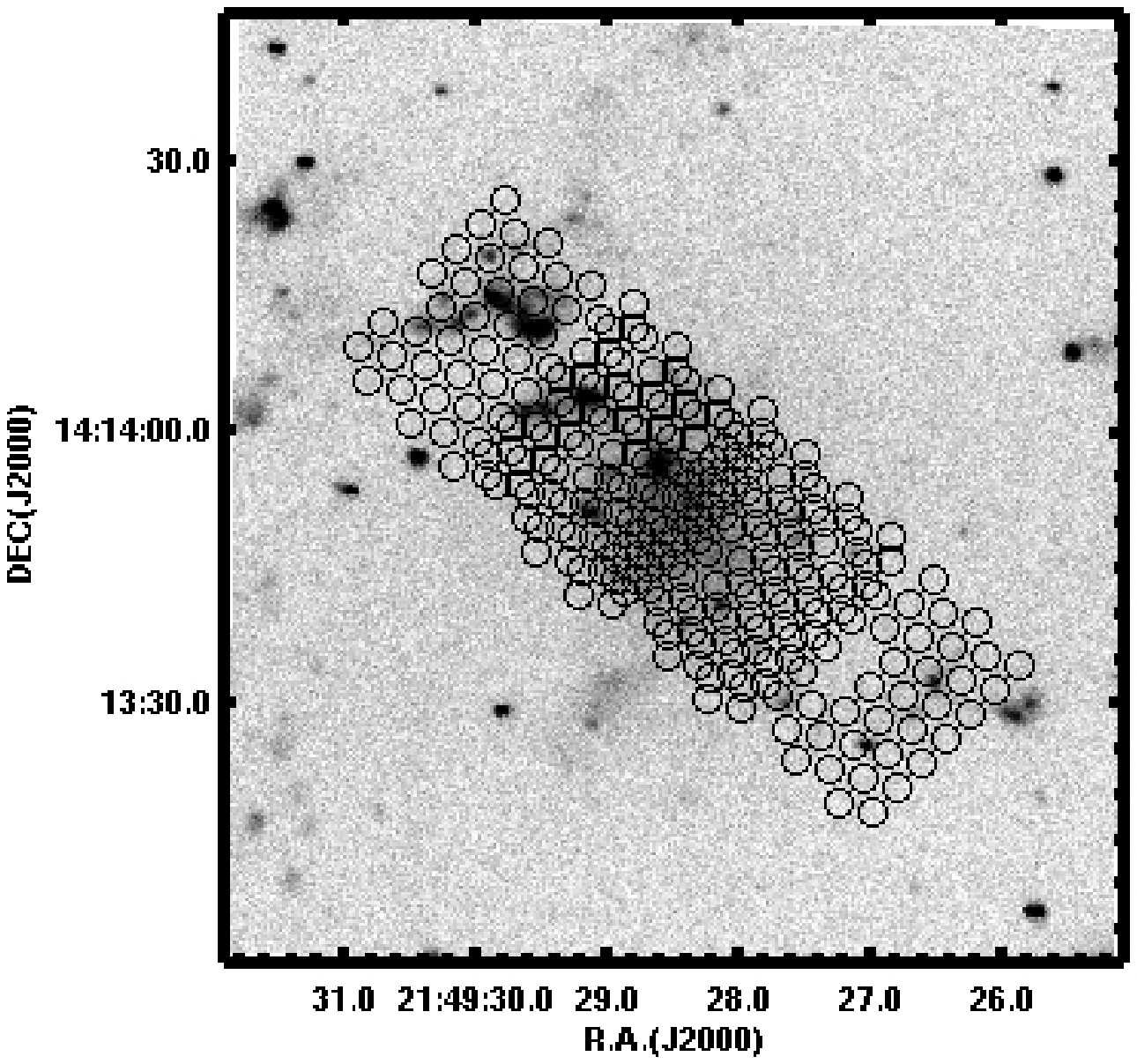}
\hfill
\includegraphics[scale=0.30]{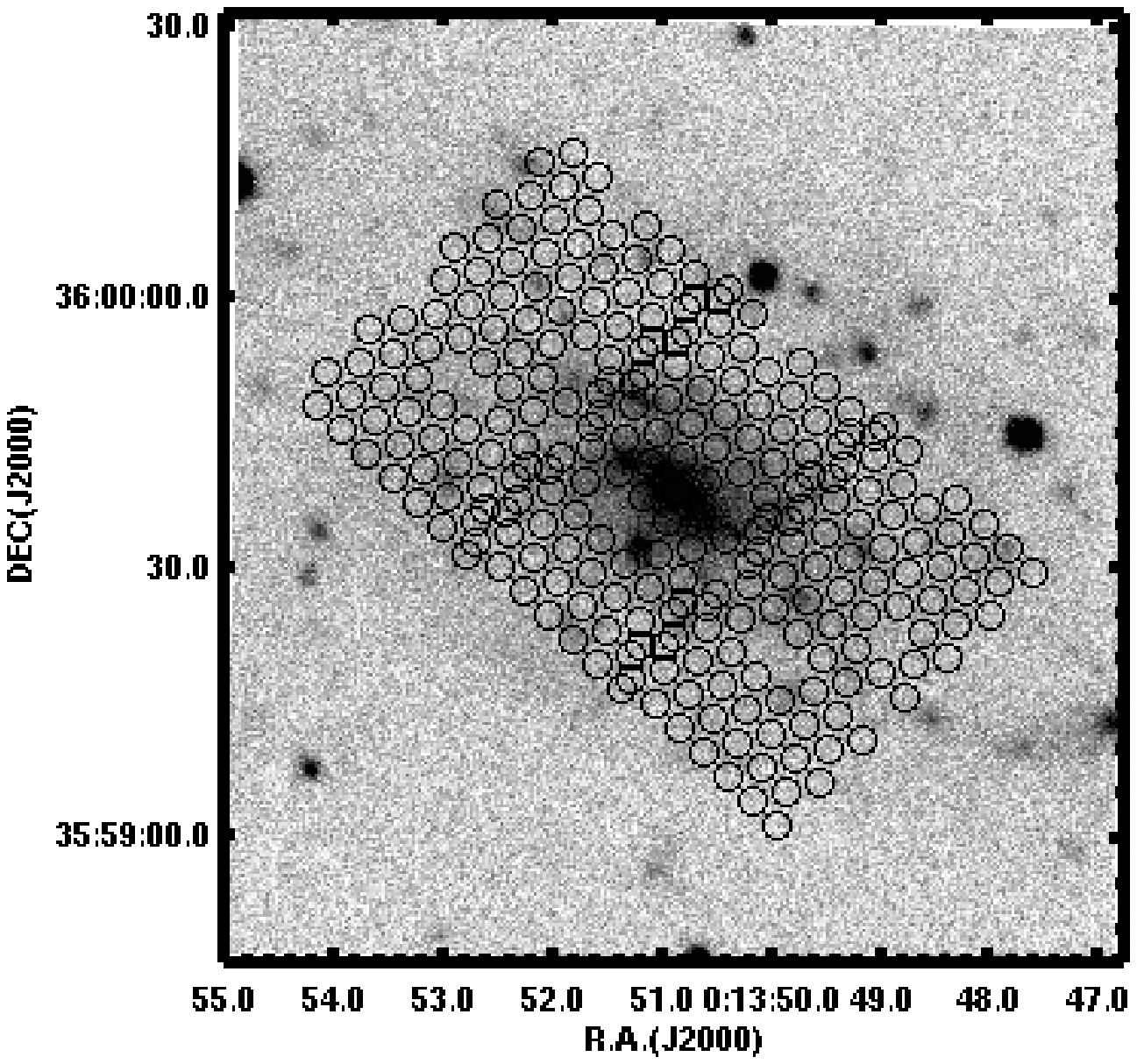}\\
\includegraphics[scale=0.34]{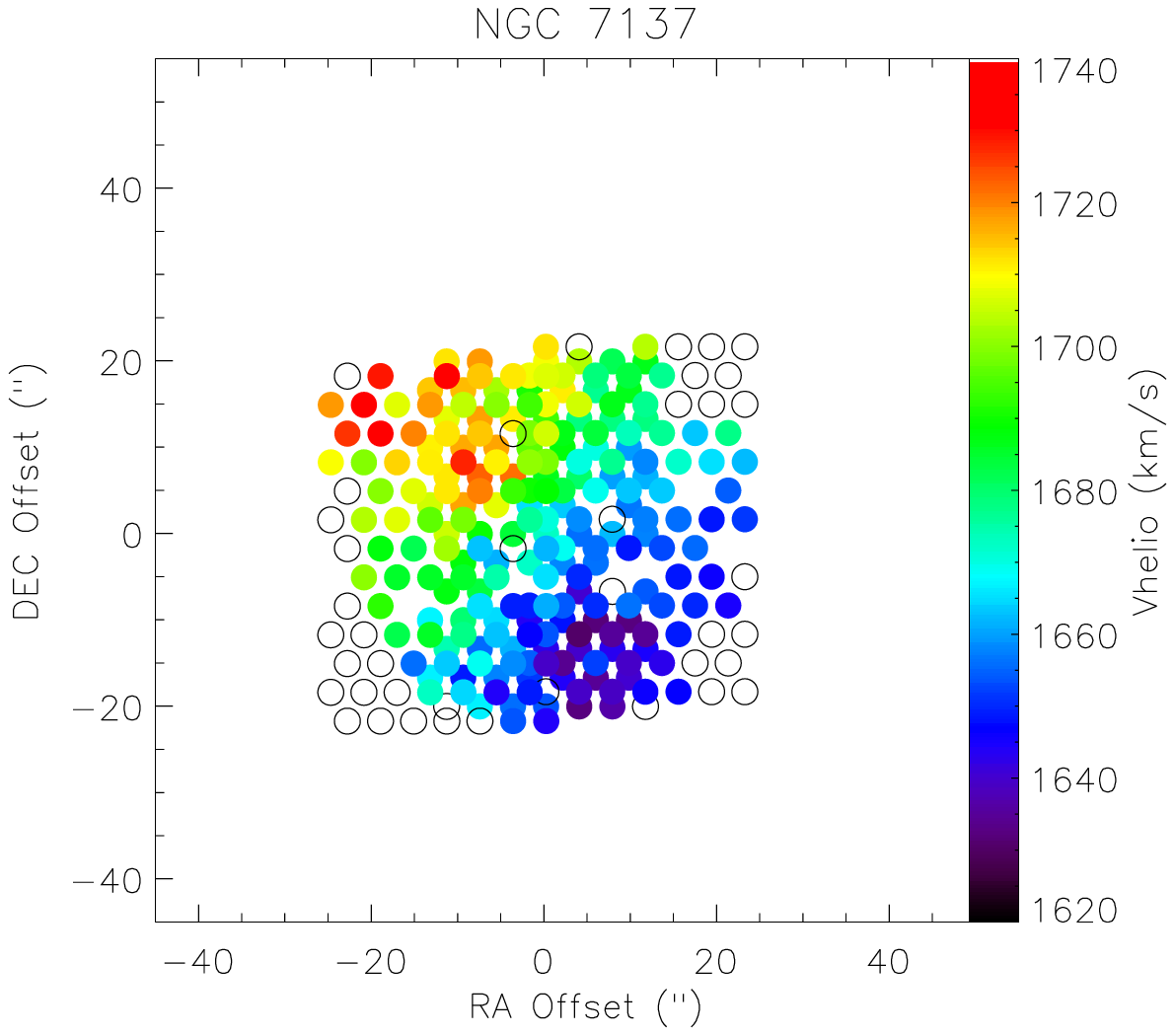}
\hfill
\hfill
\hfill
\includegraphics[scale=0.34]{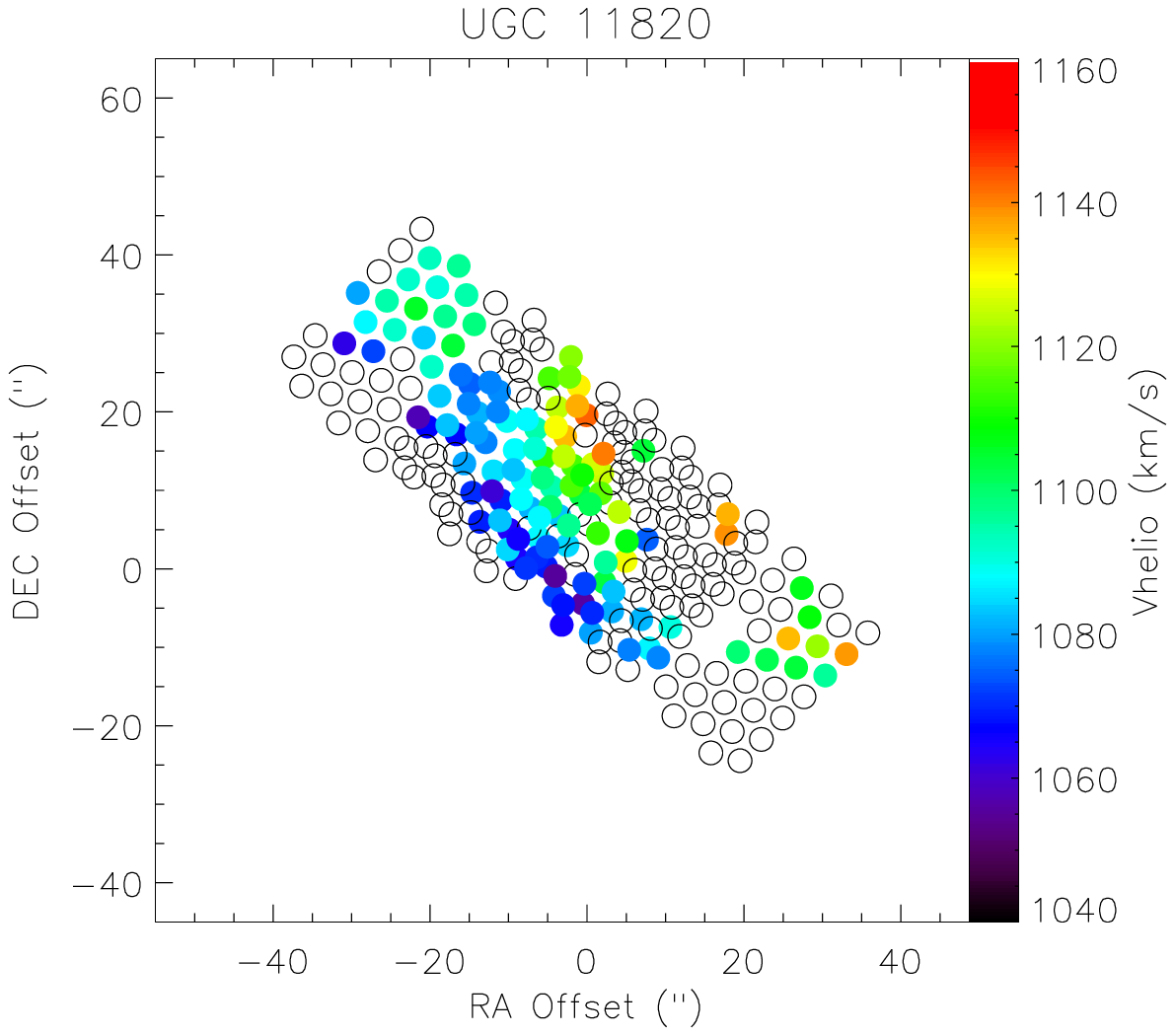}
\hfill
\hfill
\hfill
\includegraphics[scale=0.34]{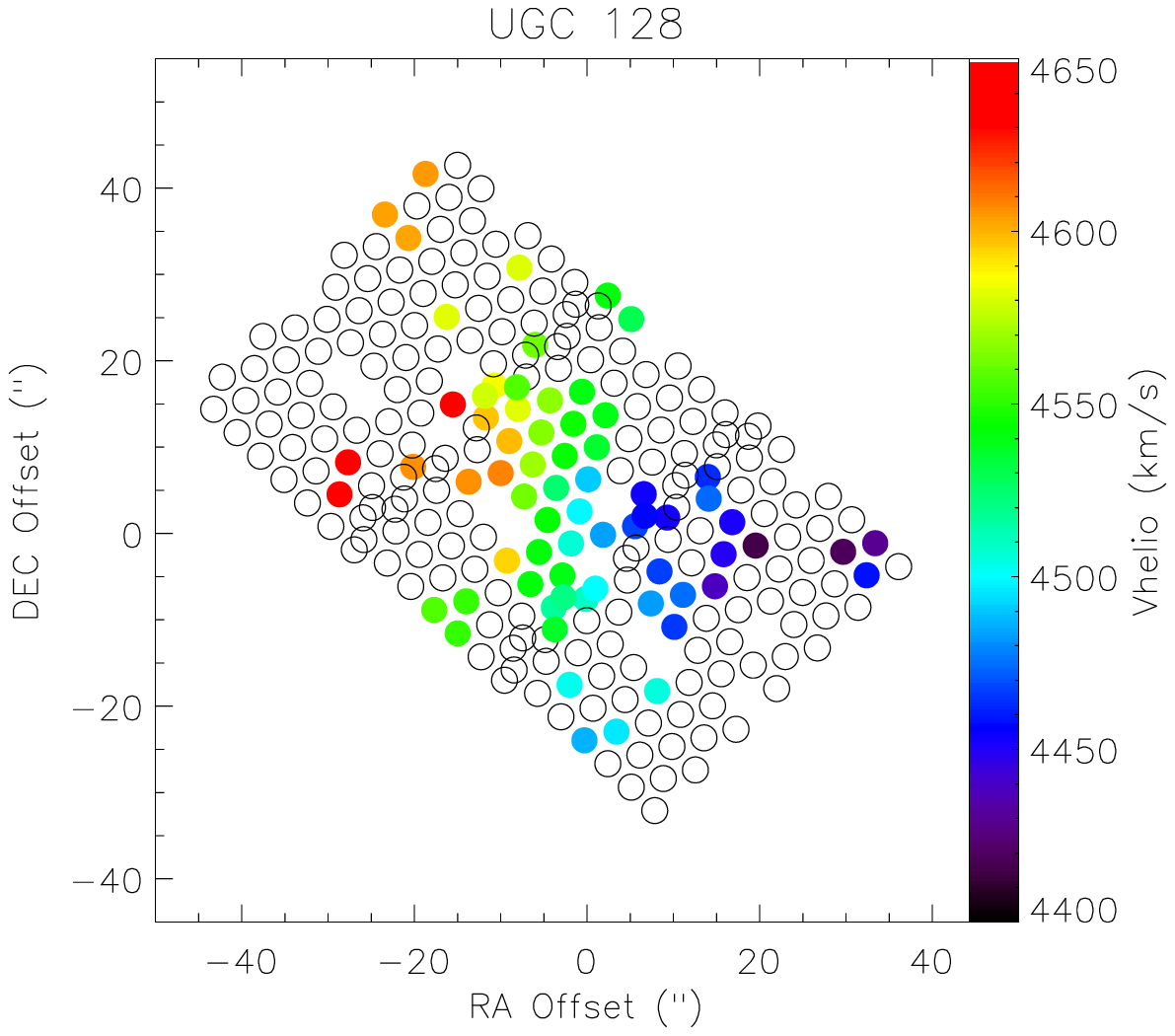}\\
\includegraphics[scale=0.21]{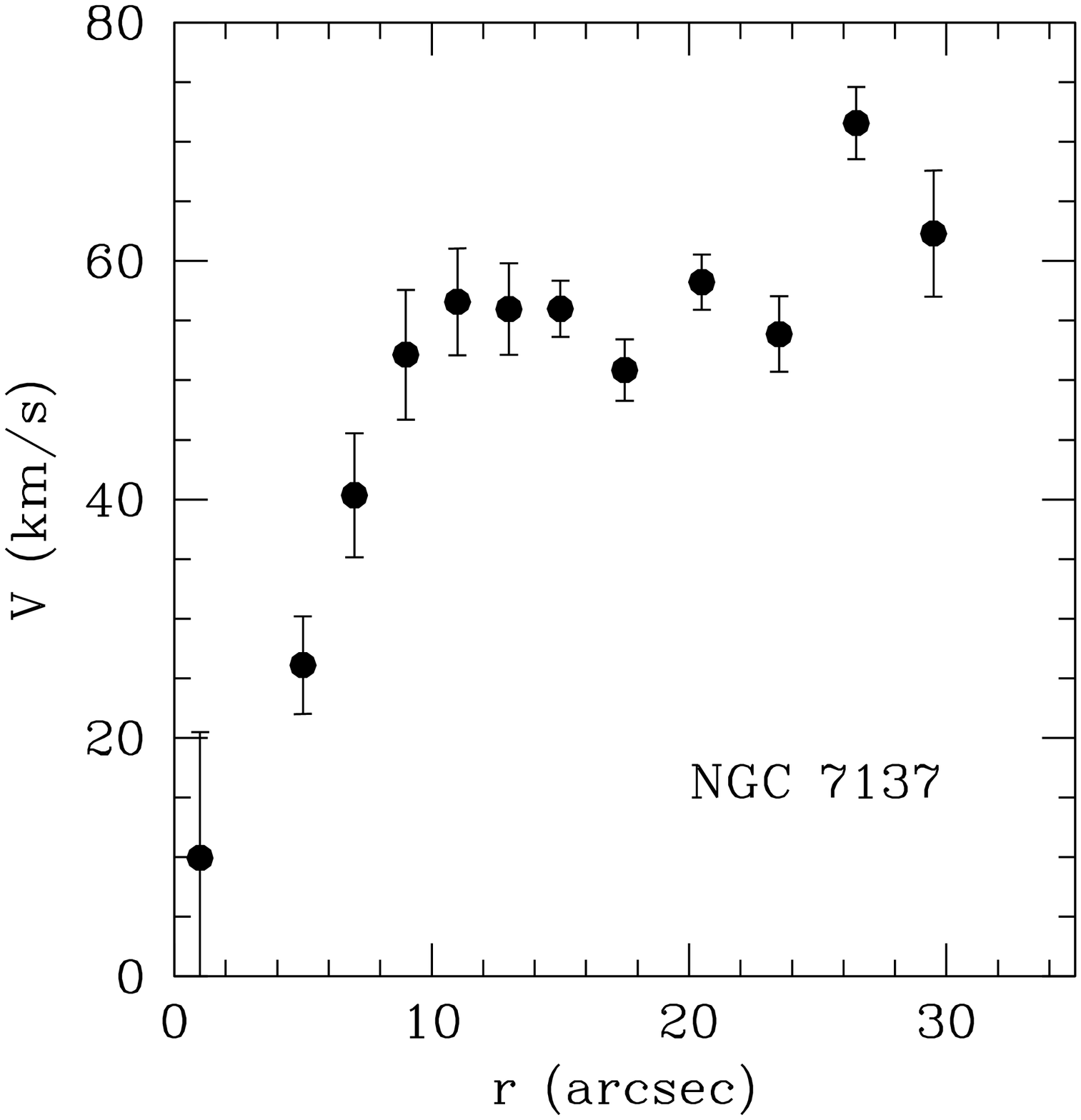}
\hfill
\includegraphics[scale=0.21]{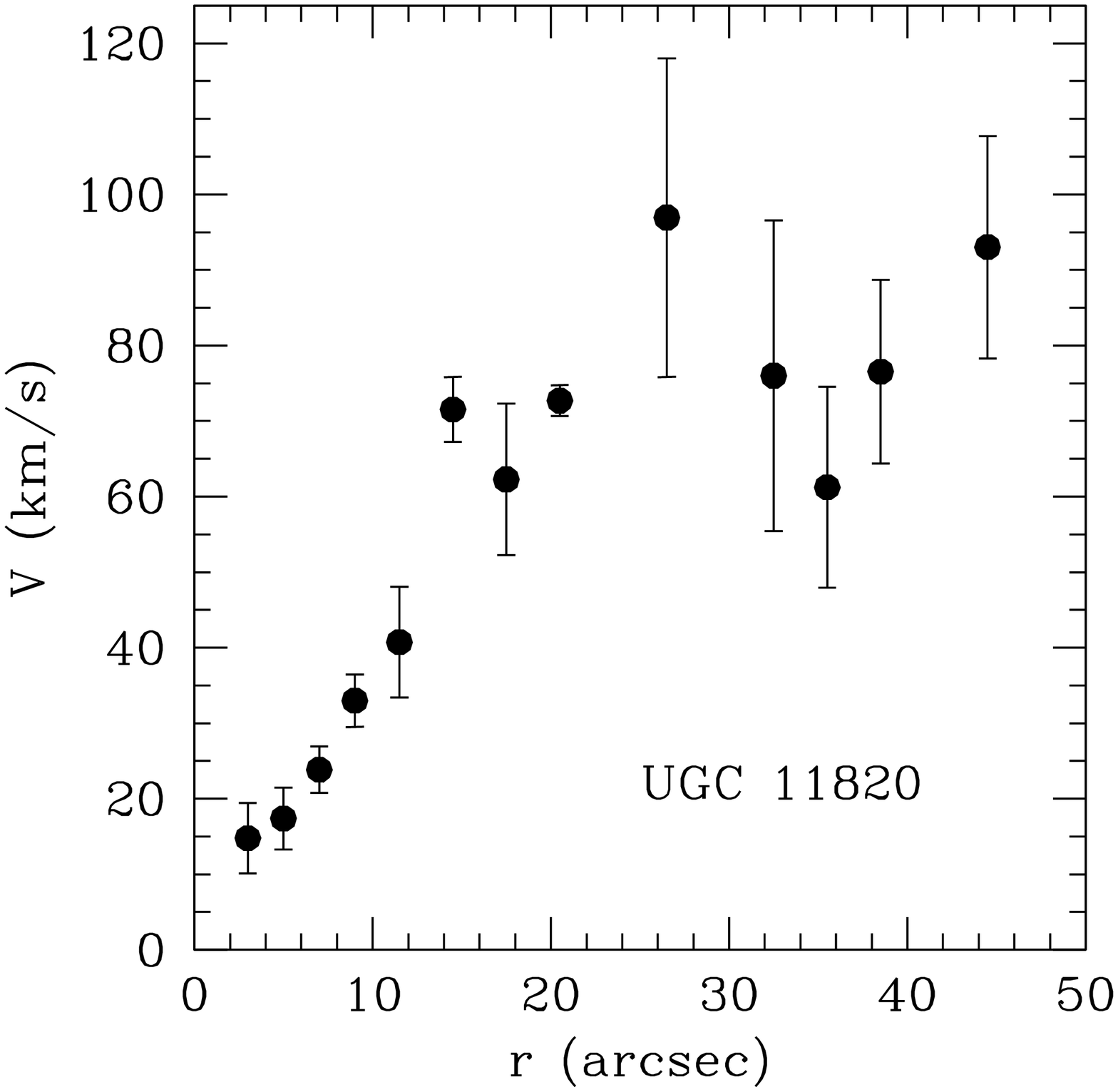}
\hfill
\includegraphics[scale=0.21]{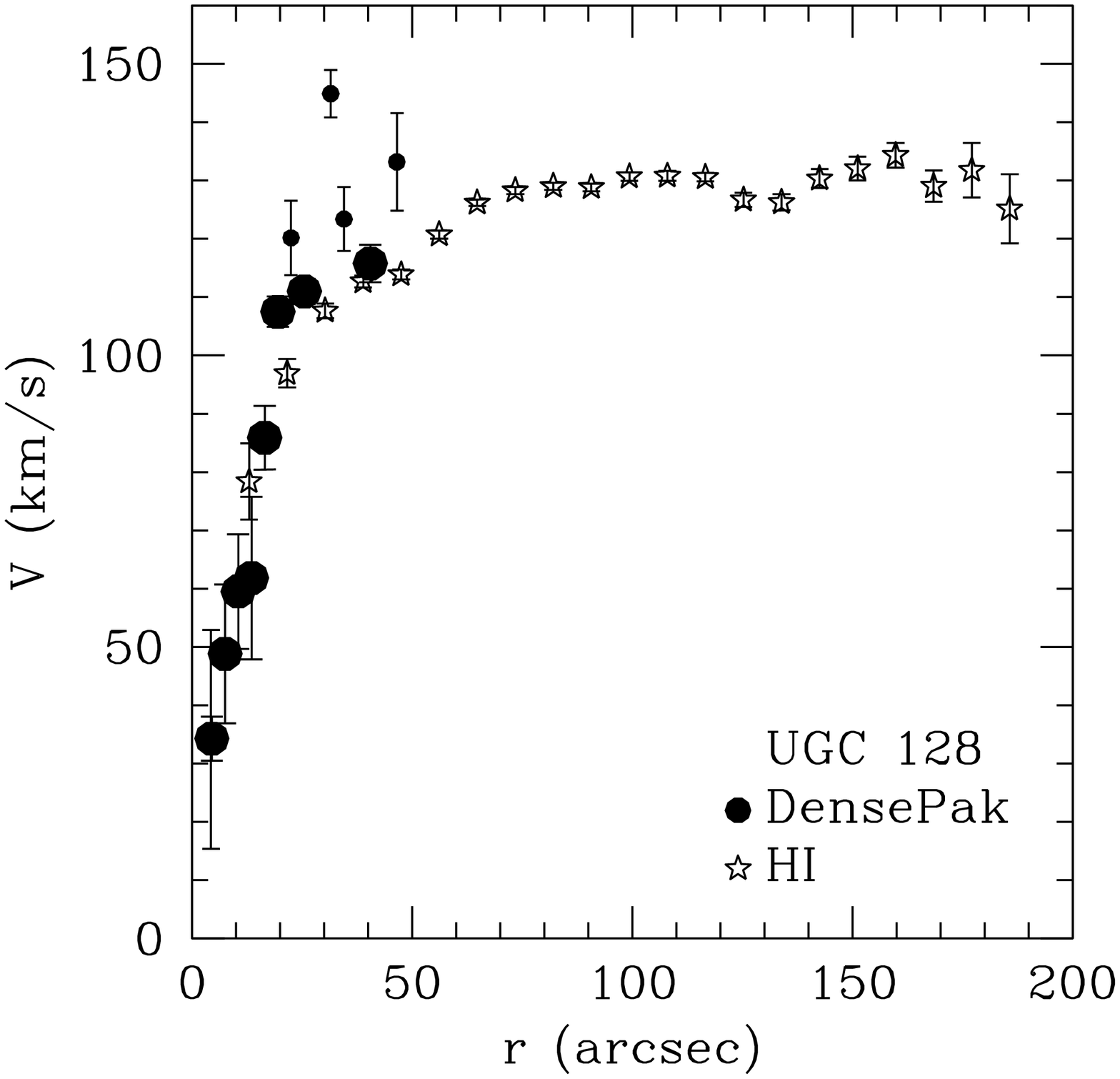}
\begin{quote}
\caption{Results for NGC 7137, UGC 11820, \& UGC 128.  
 $\textit{Top row:}$ Position of \Dpak\ array on the \Ha\ images of the
  galaxies.  $\textit{Middle row:}$ Observed \Dpak\ velocity field.
 Empty fibers are those without detections.  $\textit{Bottom row:}$
 \Dpak\ rotation curves.  The UGC 128 \Dpak\ points at large radii with 
$\sigma$ $>$ 4 \kms\ are plotted as smaller points.  The \Dpak\ rotation 
 curve of UGC 128 is
 plotted with the \HI\ rotation curve of \citet{VerheijendB} (stars).
[{\it See the
  electronic edition of the Journal for a color version of this figure.}]}
\end{quote}
\end{figure*}

\begin{figure*}
\includegraphics[scale=0.30]{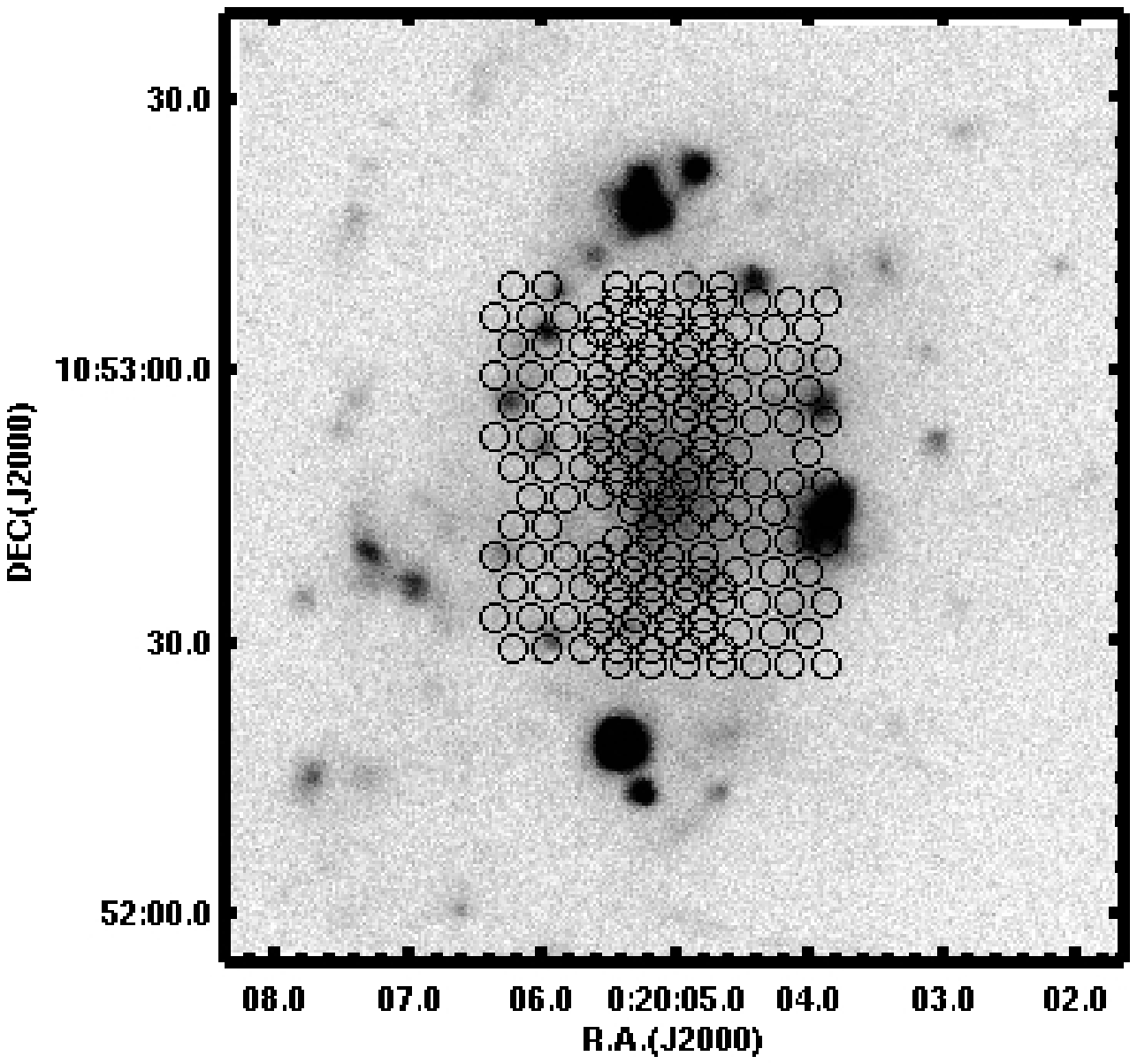}
\hfill
\includegraphics[scale=0.30]{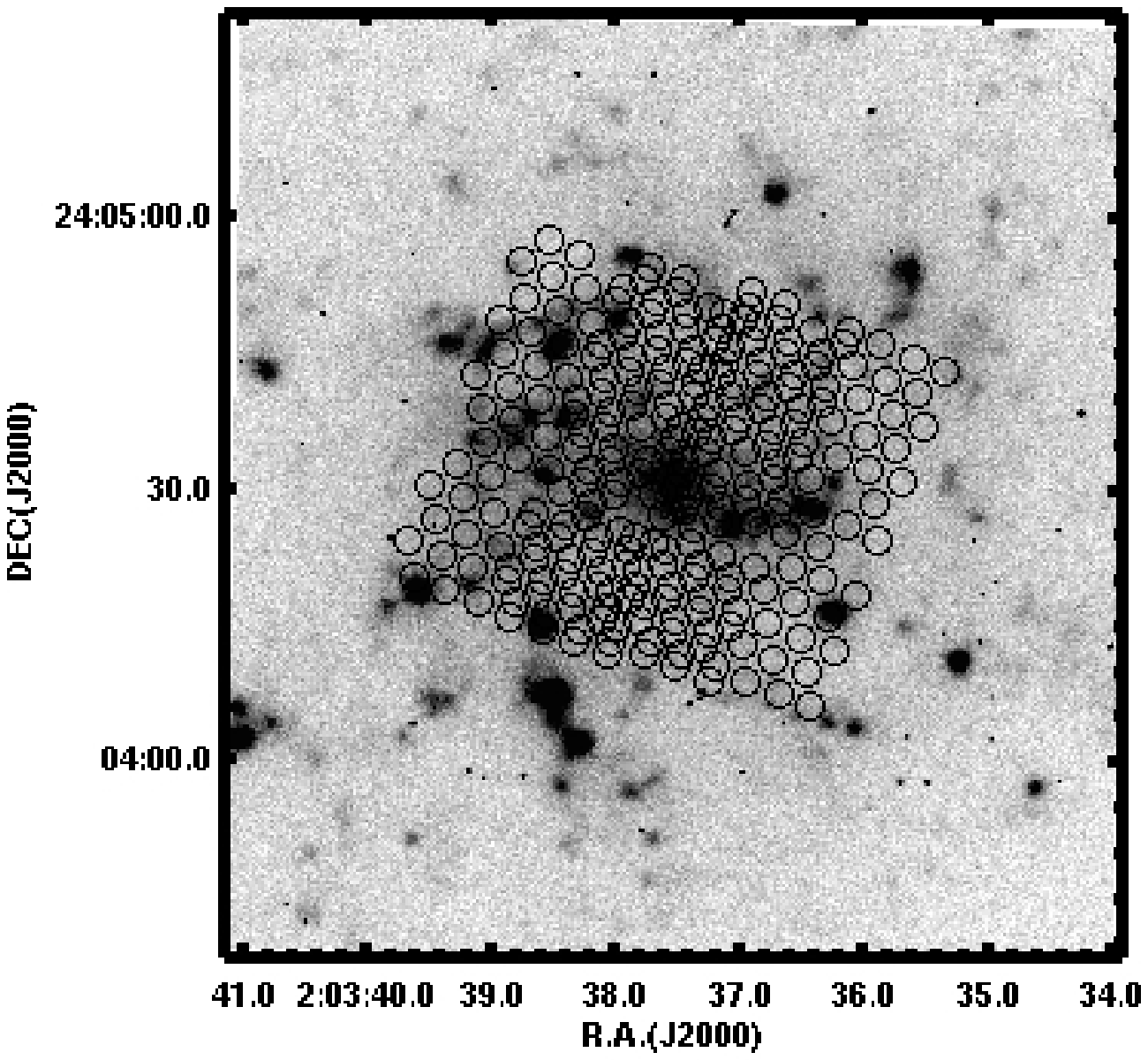}
\hfill
\includegraphics[scale=0.30]{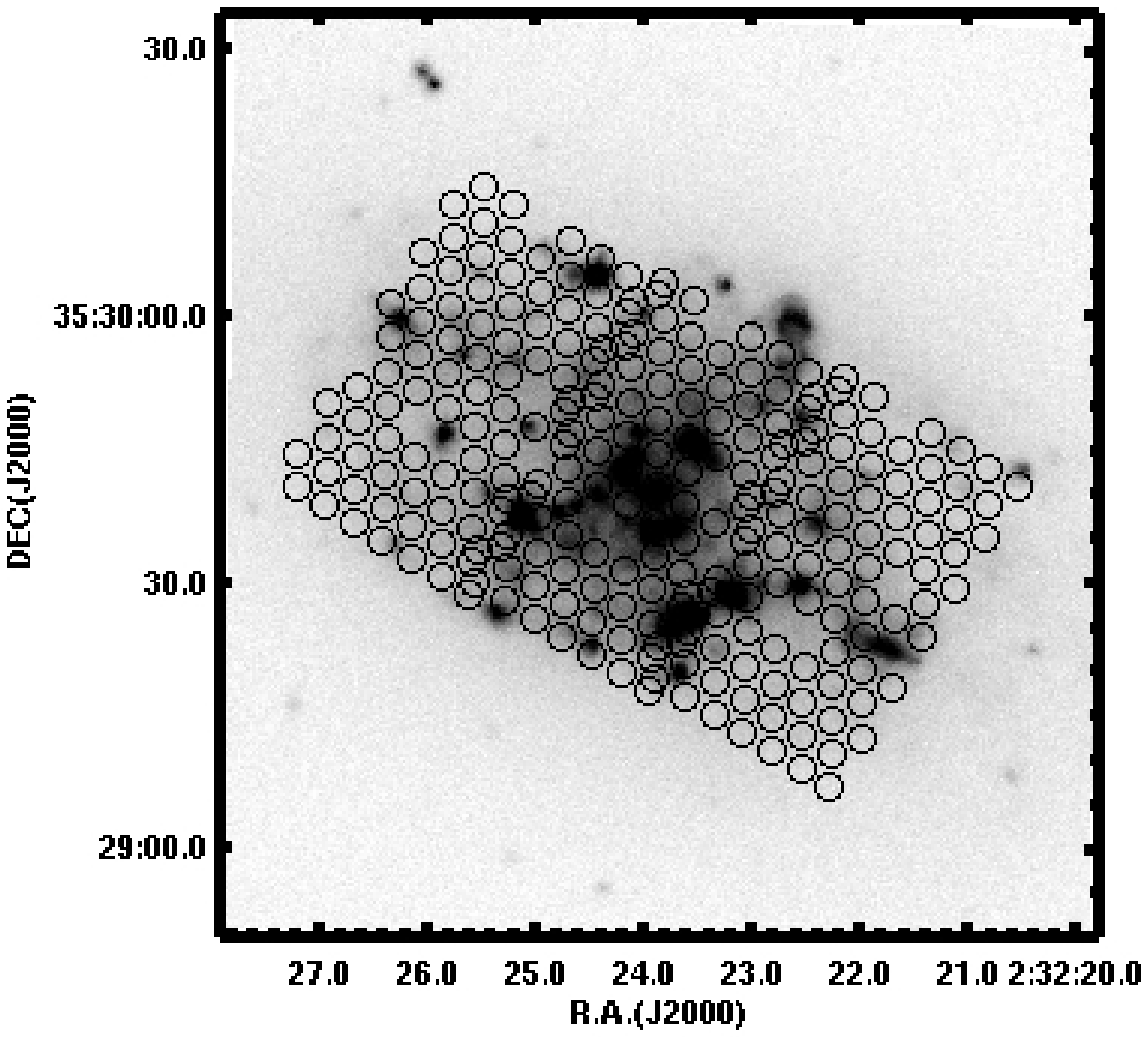}\\
\includegraphics[scale=0.34]{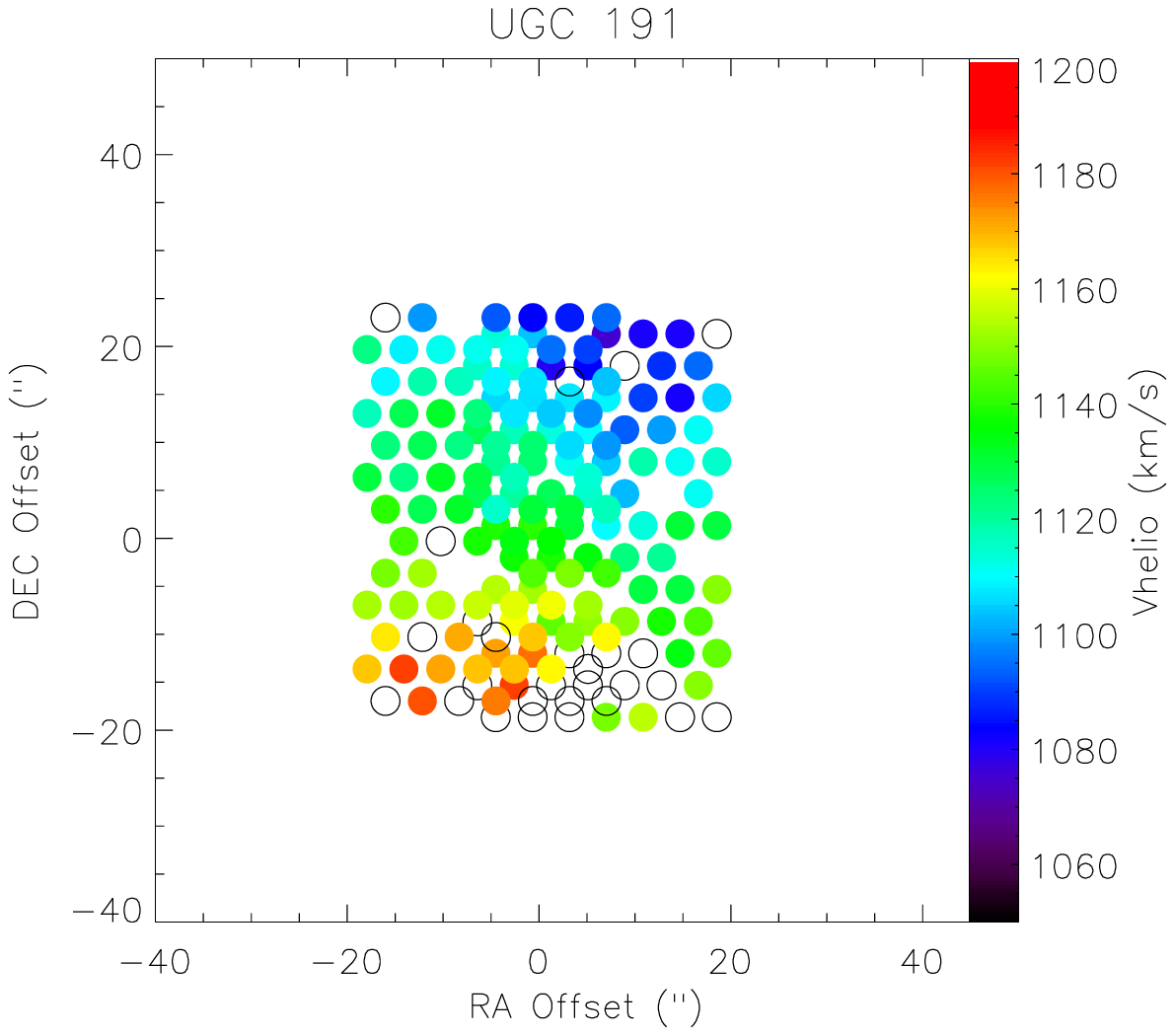}
\hfill
\hfill
\hfill
\includegraphics[scale=0.34]{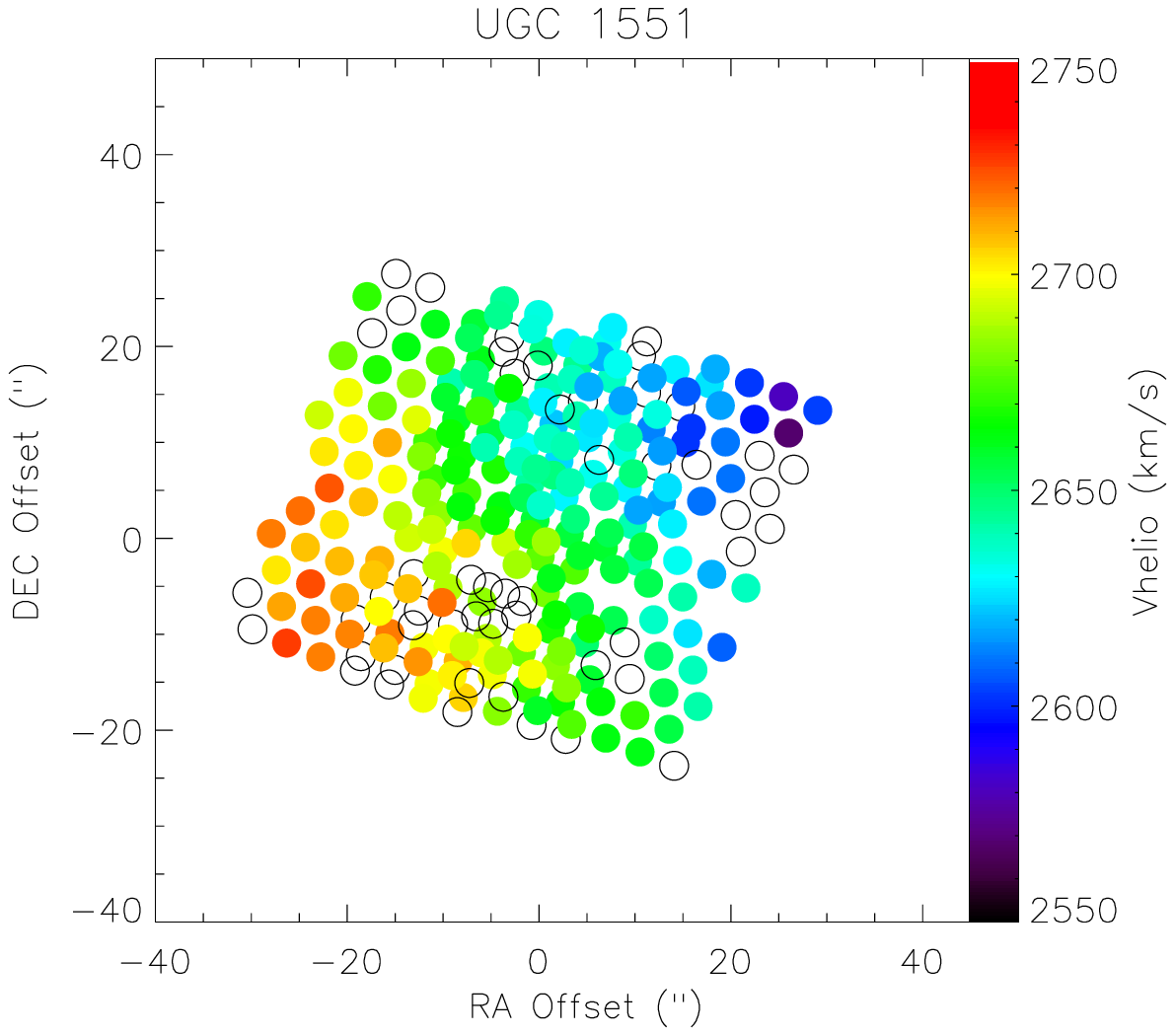}
\hfill
\hfill
\hfill
\includegraphics[scale=0.34]{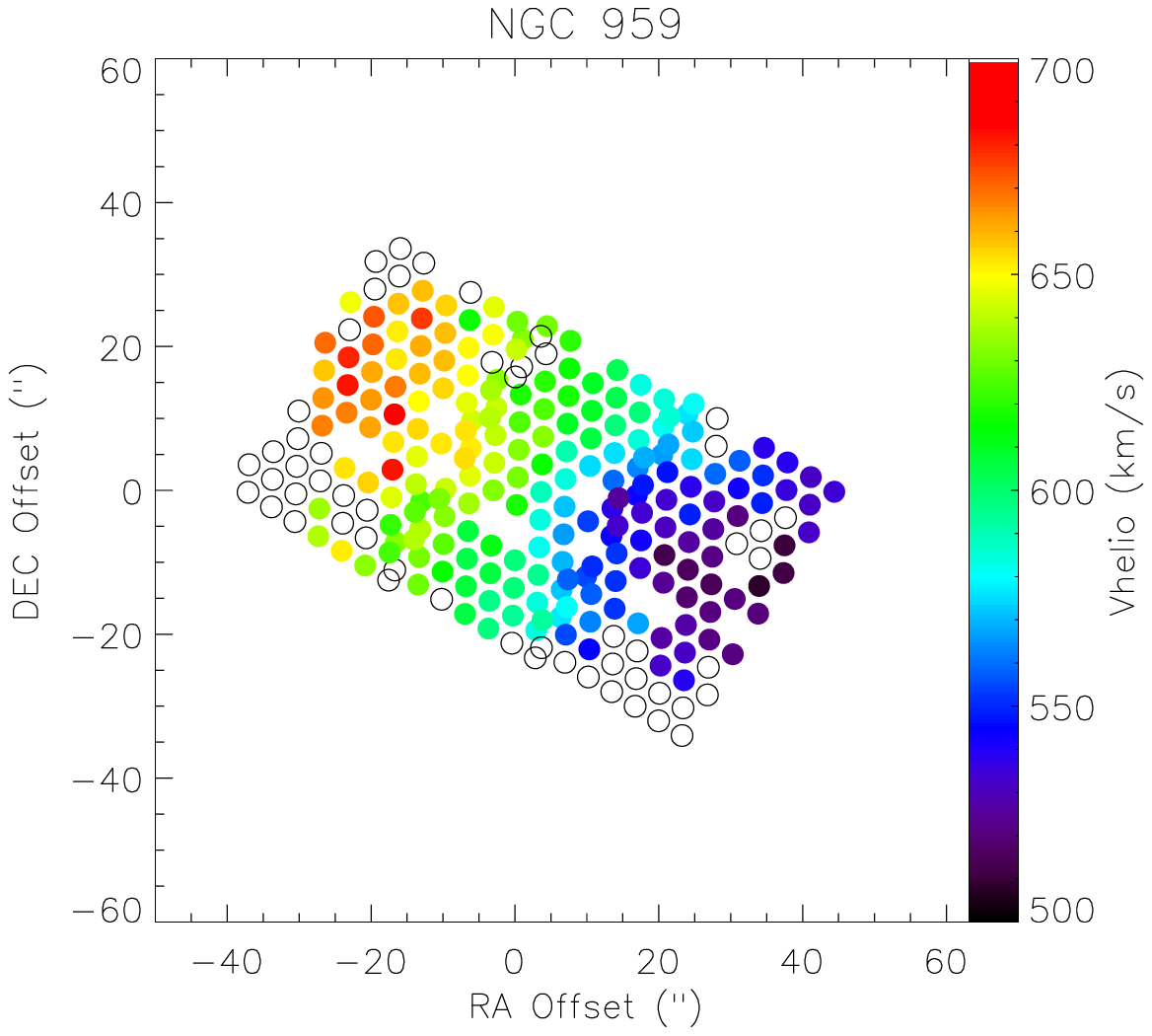}\\
\includegraphics[scale=0.21]{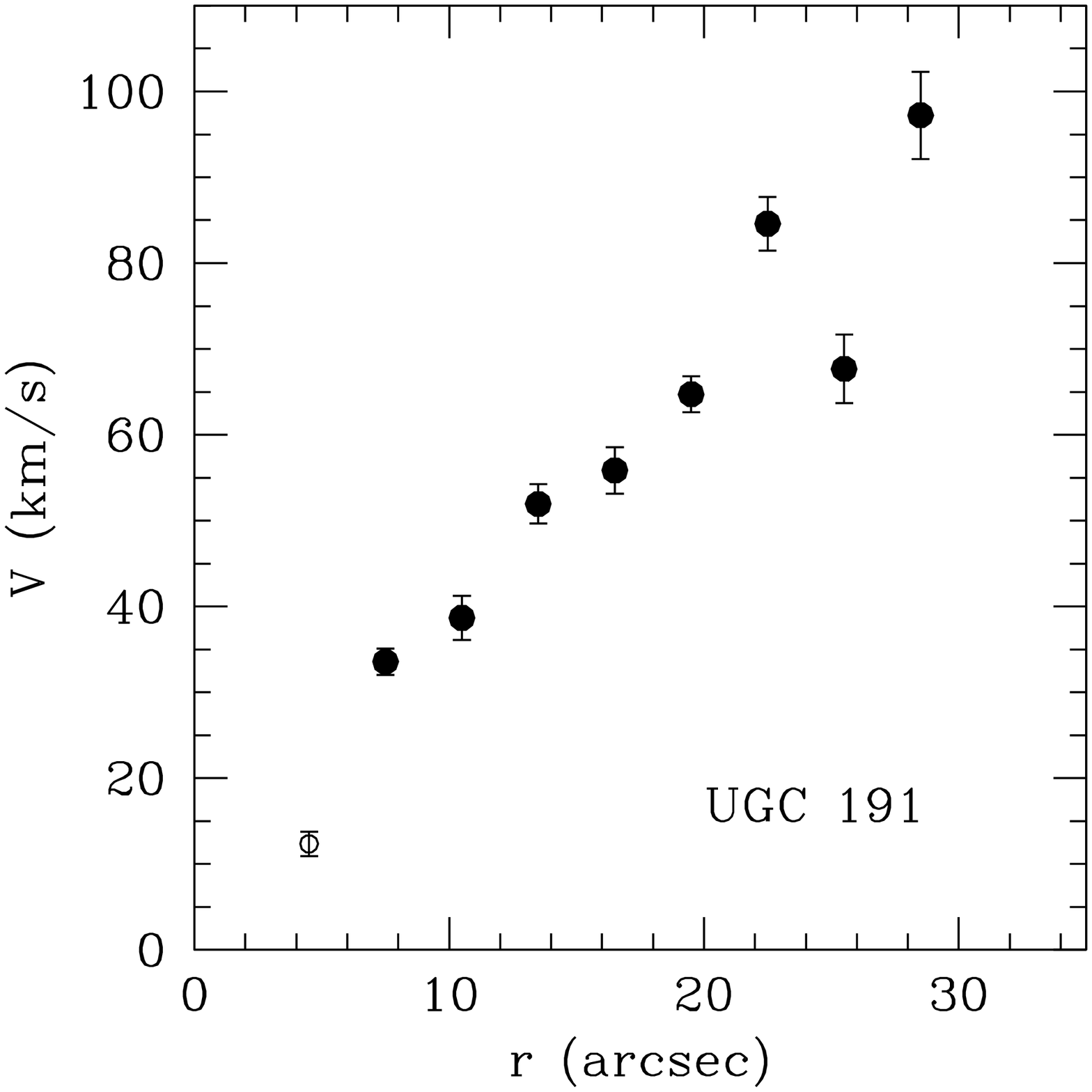}
\hfill
\includegraphics[scale=0.21]{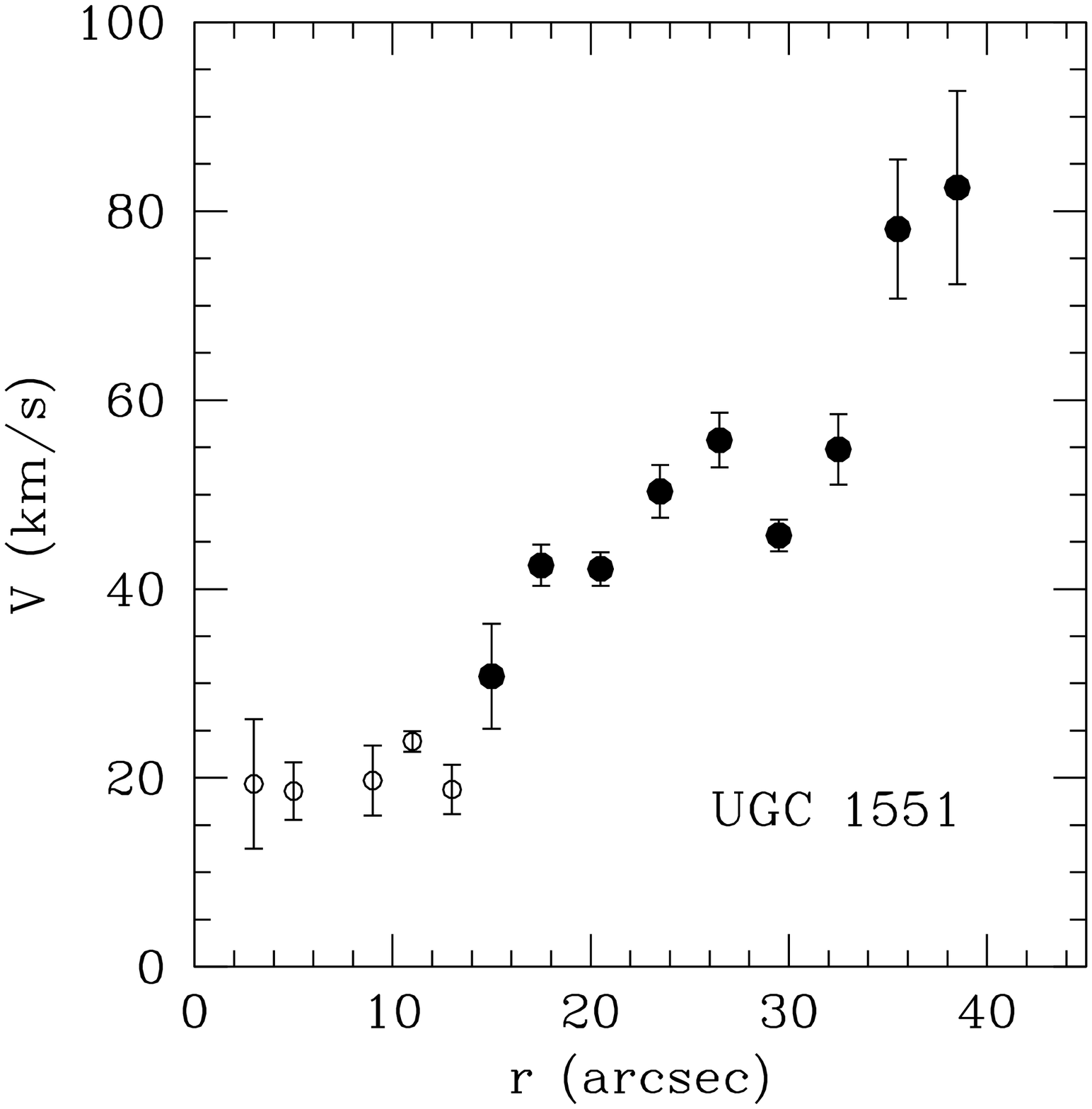}
\hfill
\includegraphics[scale=0.21]{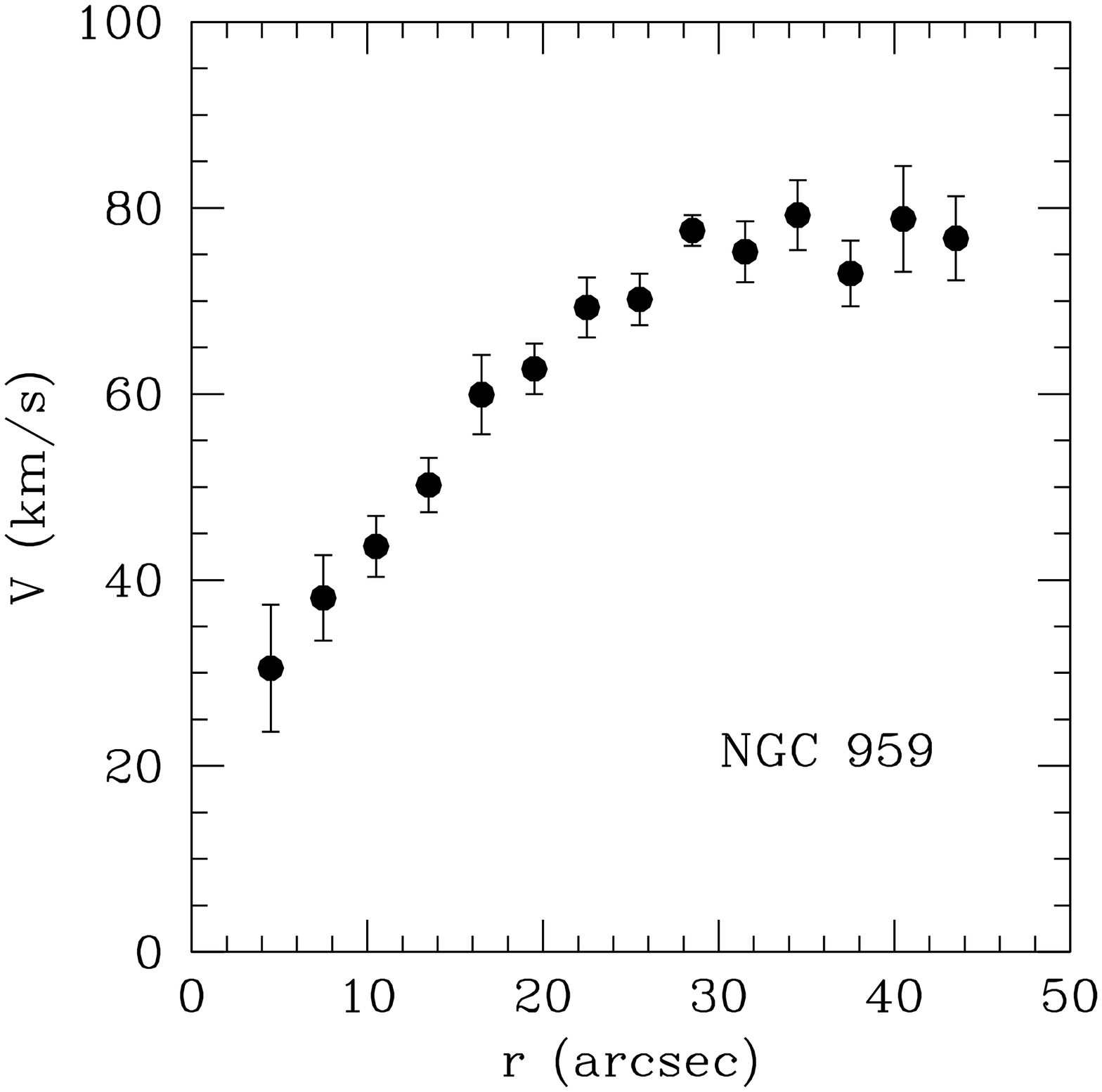}
\begin{quote}
\caption[\Dpak\ observations of UGC 191, UGC 1551, \& NGC
959]{Results for UGC 191, UGC 1551, \& NGC 959.  $\textit{Top
    row:}$ Position of \Dpak\ array on the \Ha\ images of the
  galaxies.  $\textit{Middle row:}$ Observed \Dpak\ velocity field.
 Empty fibers are those without detections.  $\textit{Bottom row:}$
 \Dpak\ rotation curves.  The open points in the UGC 191 and UGC 1551
 rotation curves were excluded from the halo fits. [{\it See the
  electronic edition of the Journal for a color version of this figure.}]  }
\end{quote}
\end{figure*}

\begin{figure*}
\includegraphics[scale=0.23]{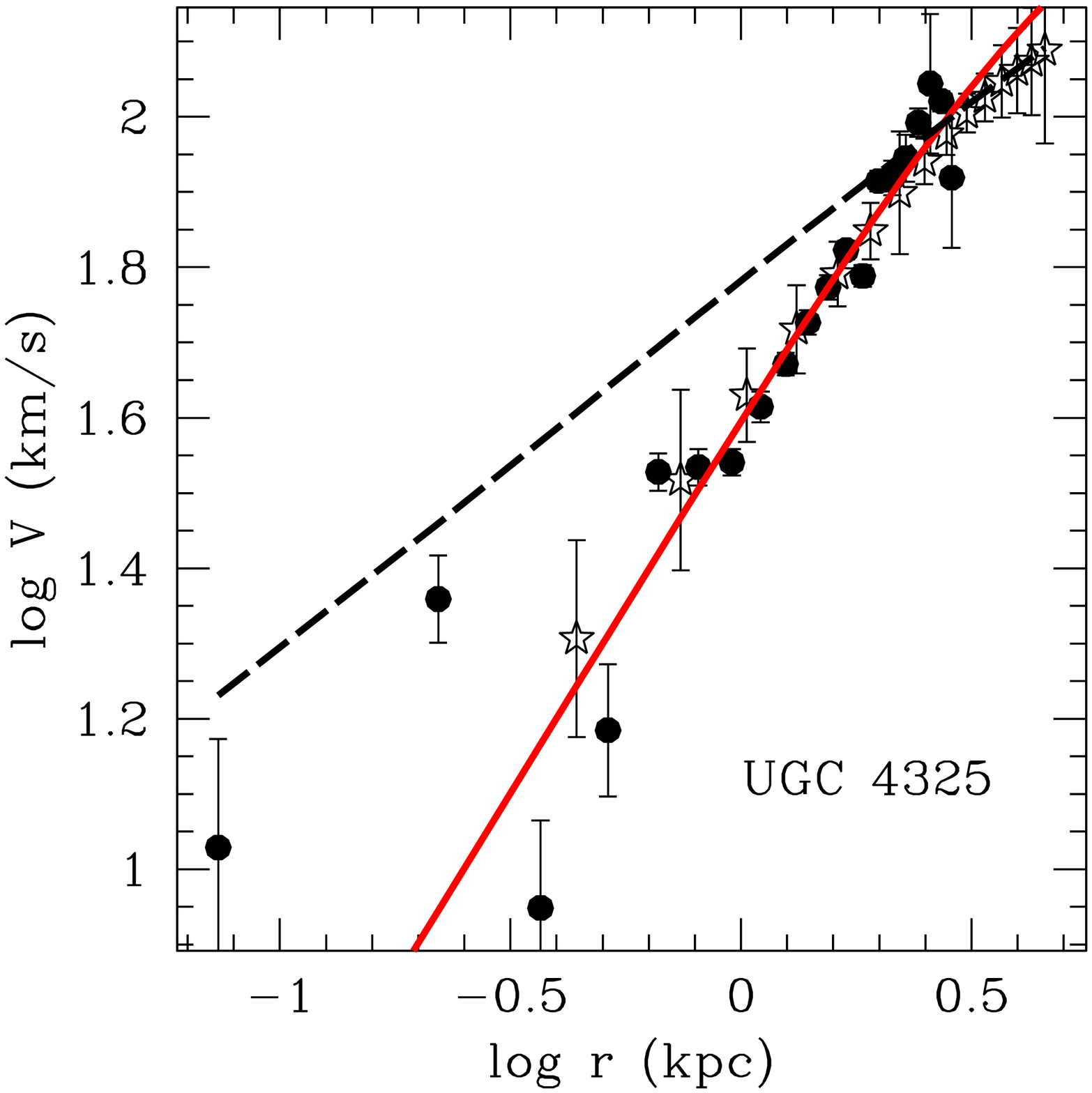}
\hfill
\includegraphics[scale=0.23]{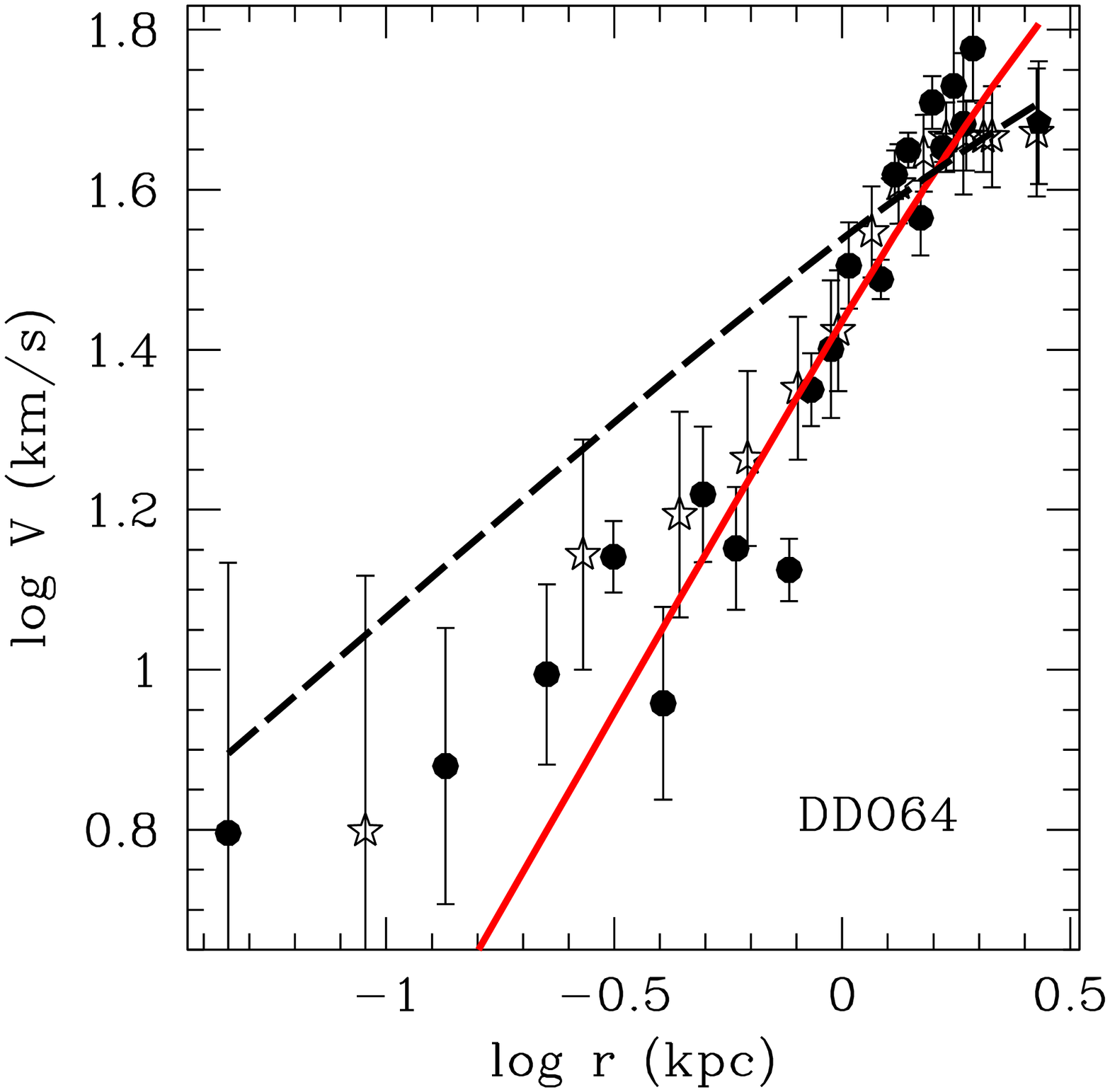}
\hfill
\includegraphics[scale=0.23]{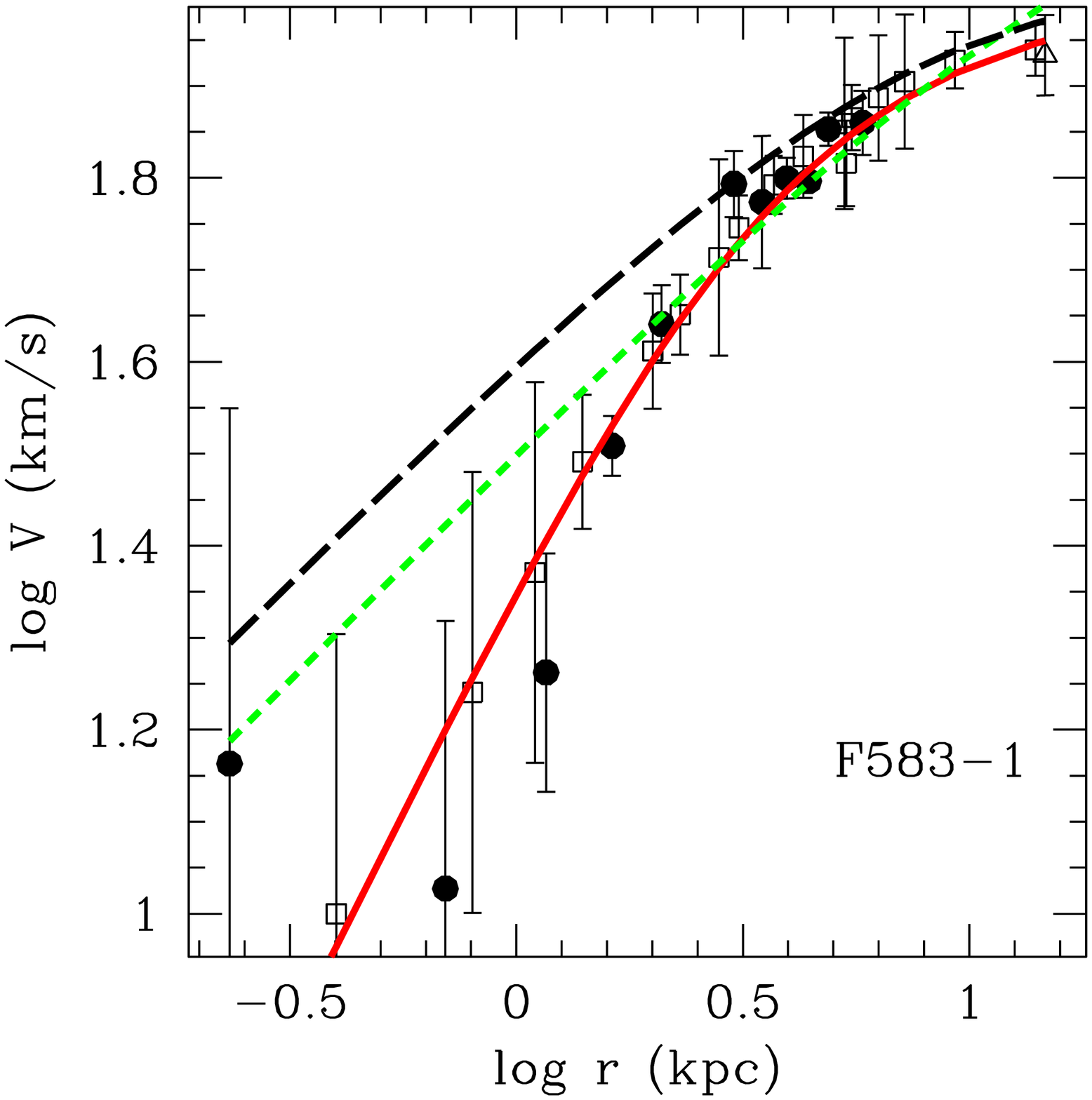}\\
\includegraphics[scale=0.23]{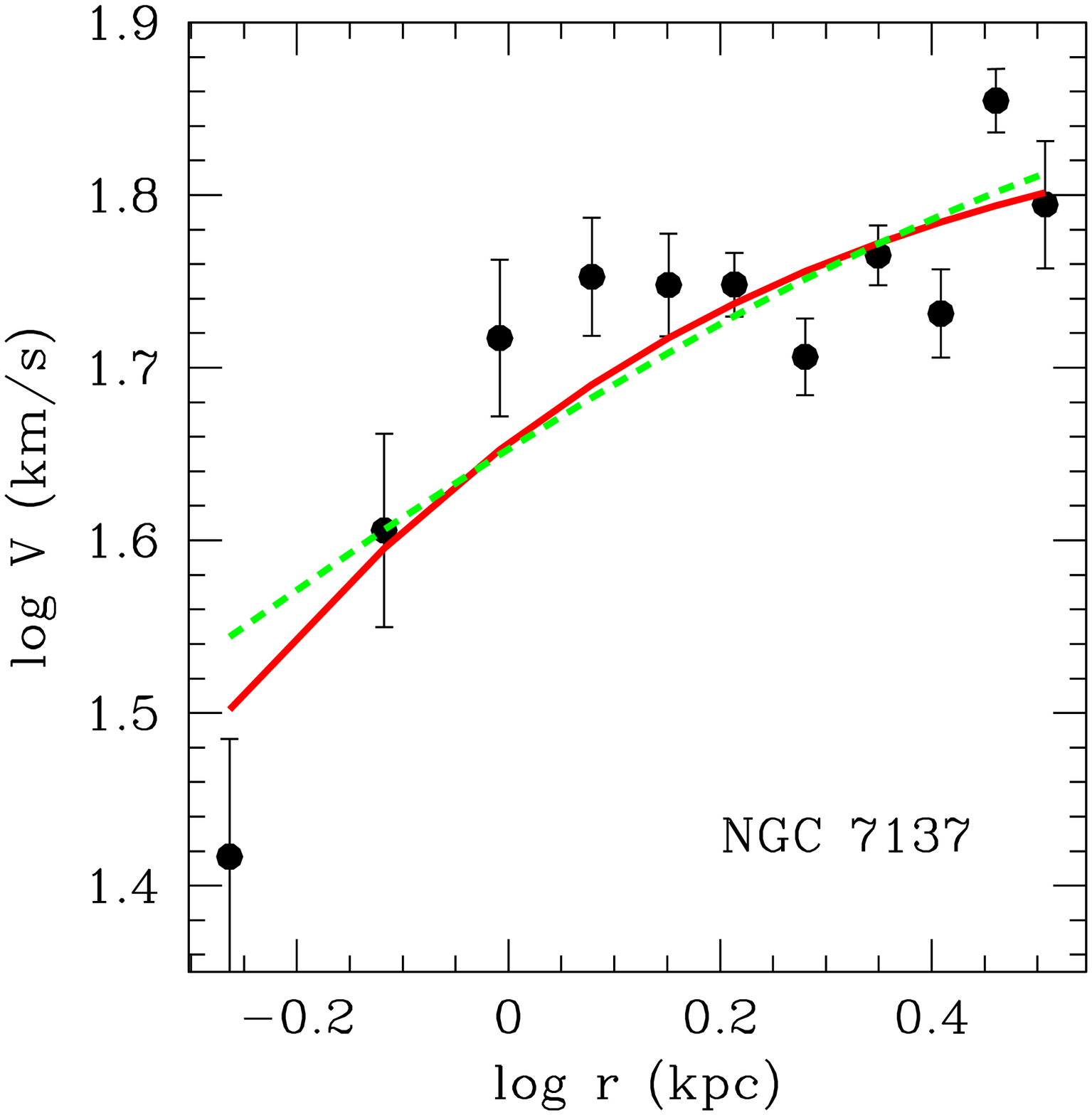}
\hfill
\includegraphics[scale=0.23]{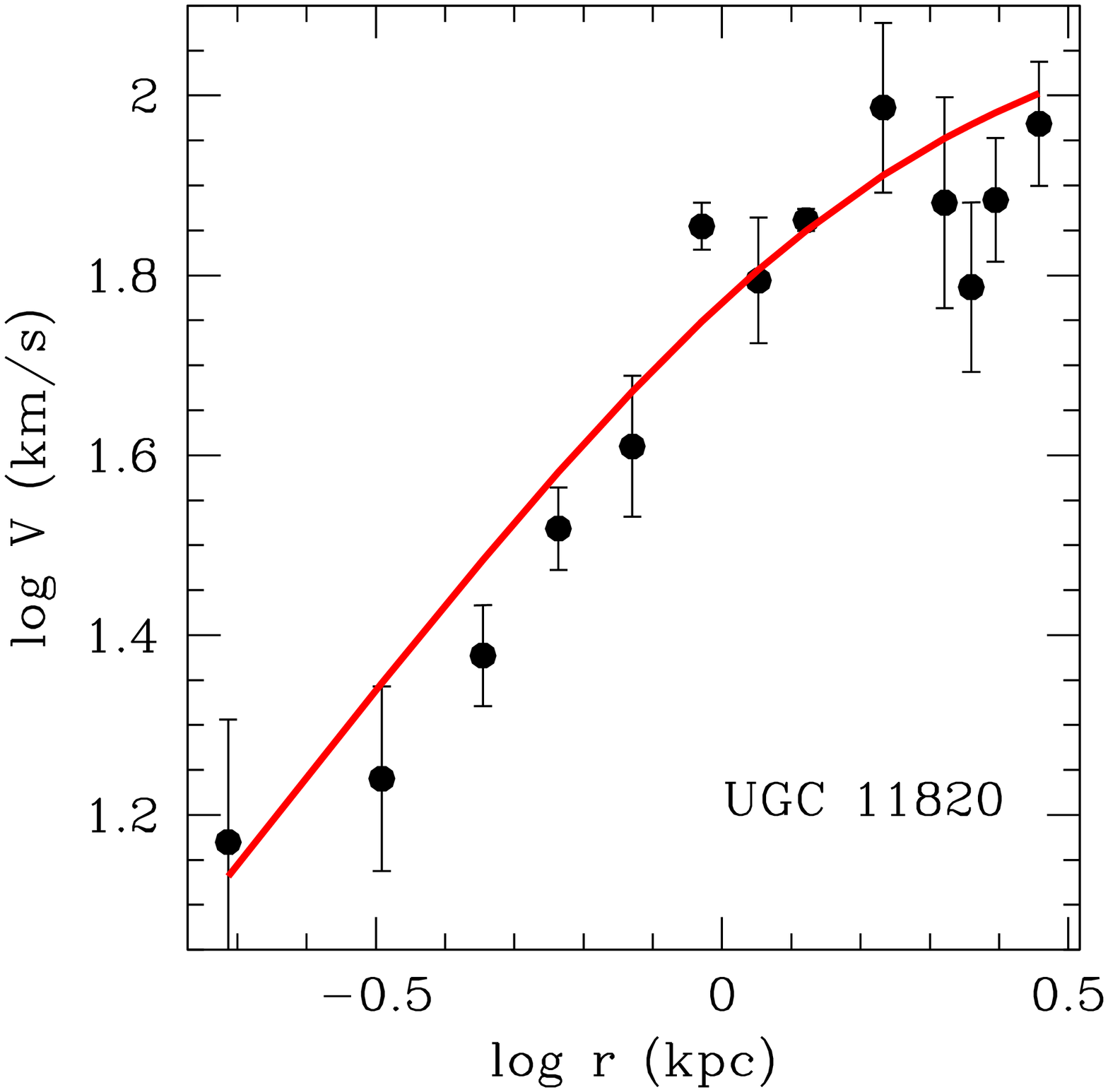}
\hfill
\includegraphics[scale=0.23]{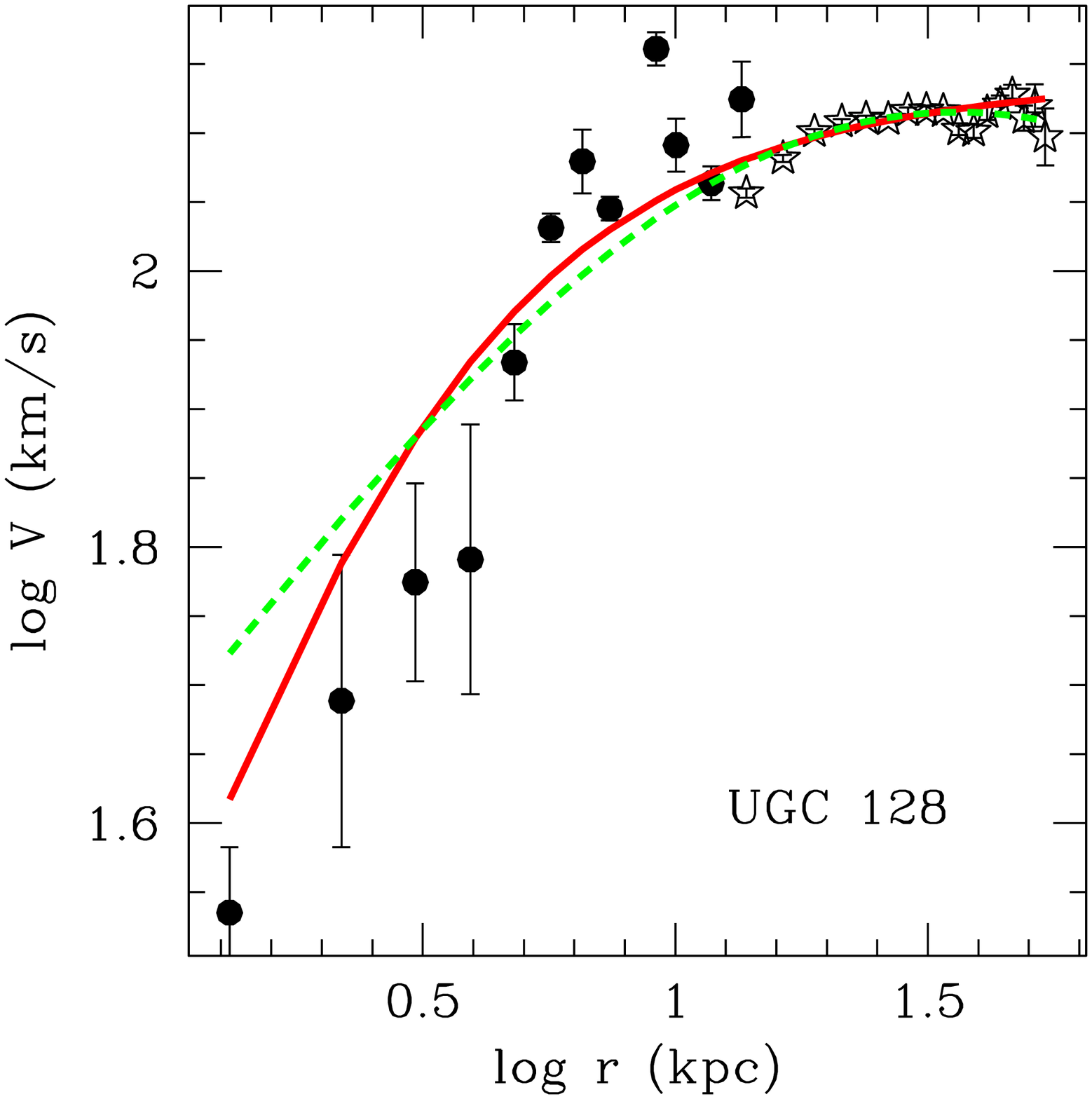}\\
\includegraphics[scale=0.23]{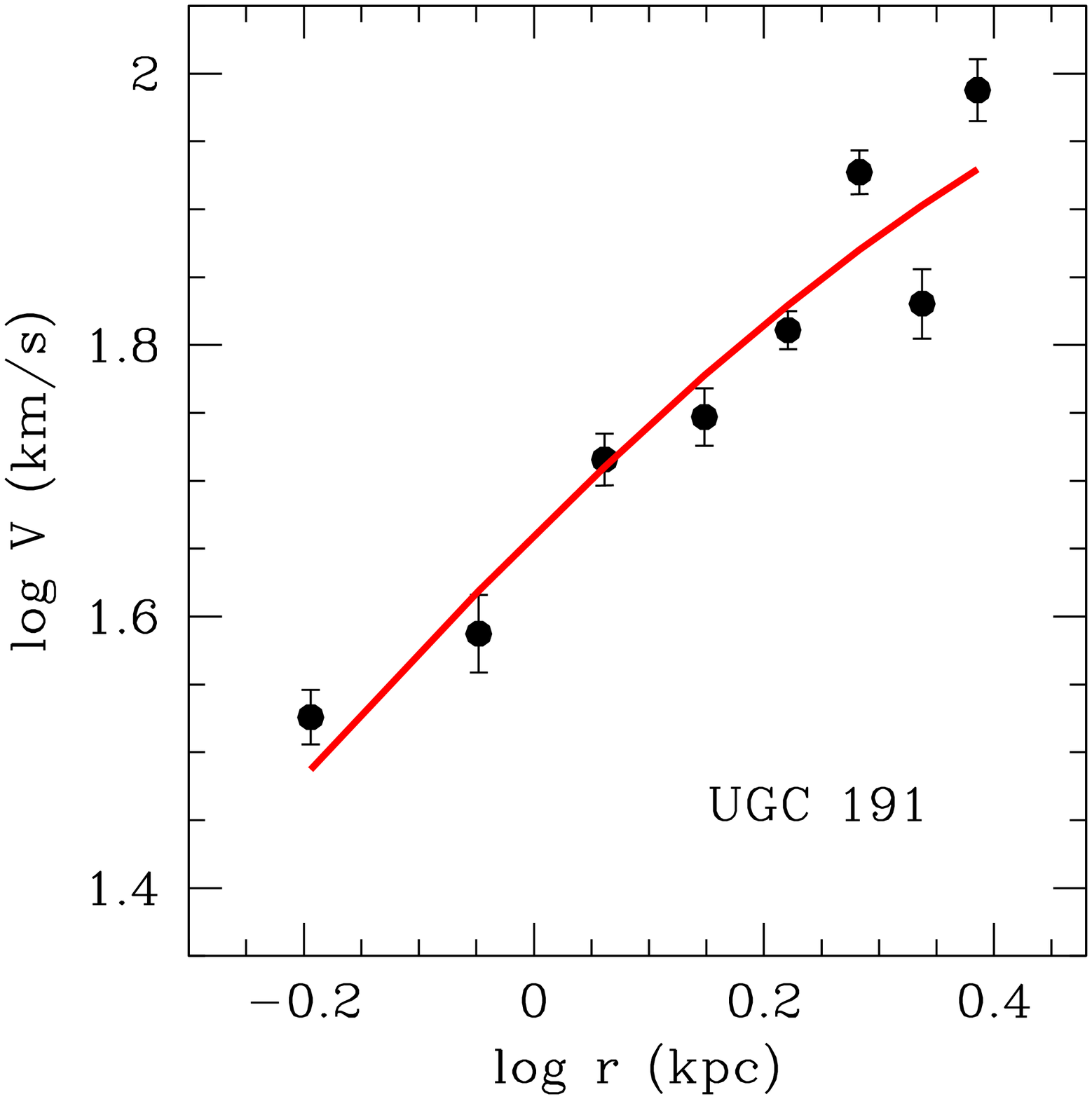}
\hfill
\includegraphics[scale=0.23]{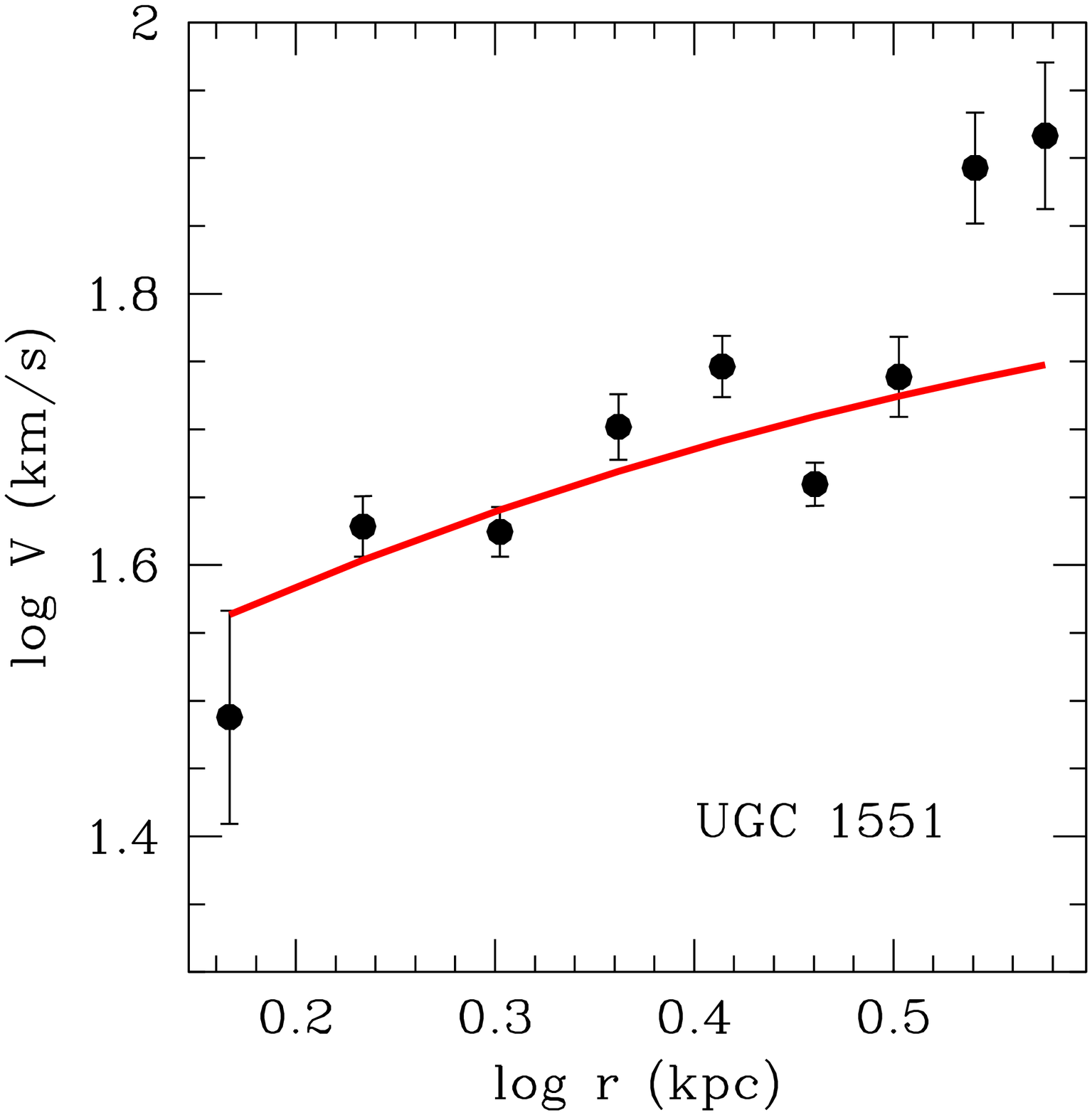}
\hfill
\includegraphics[scale=0.23]{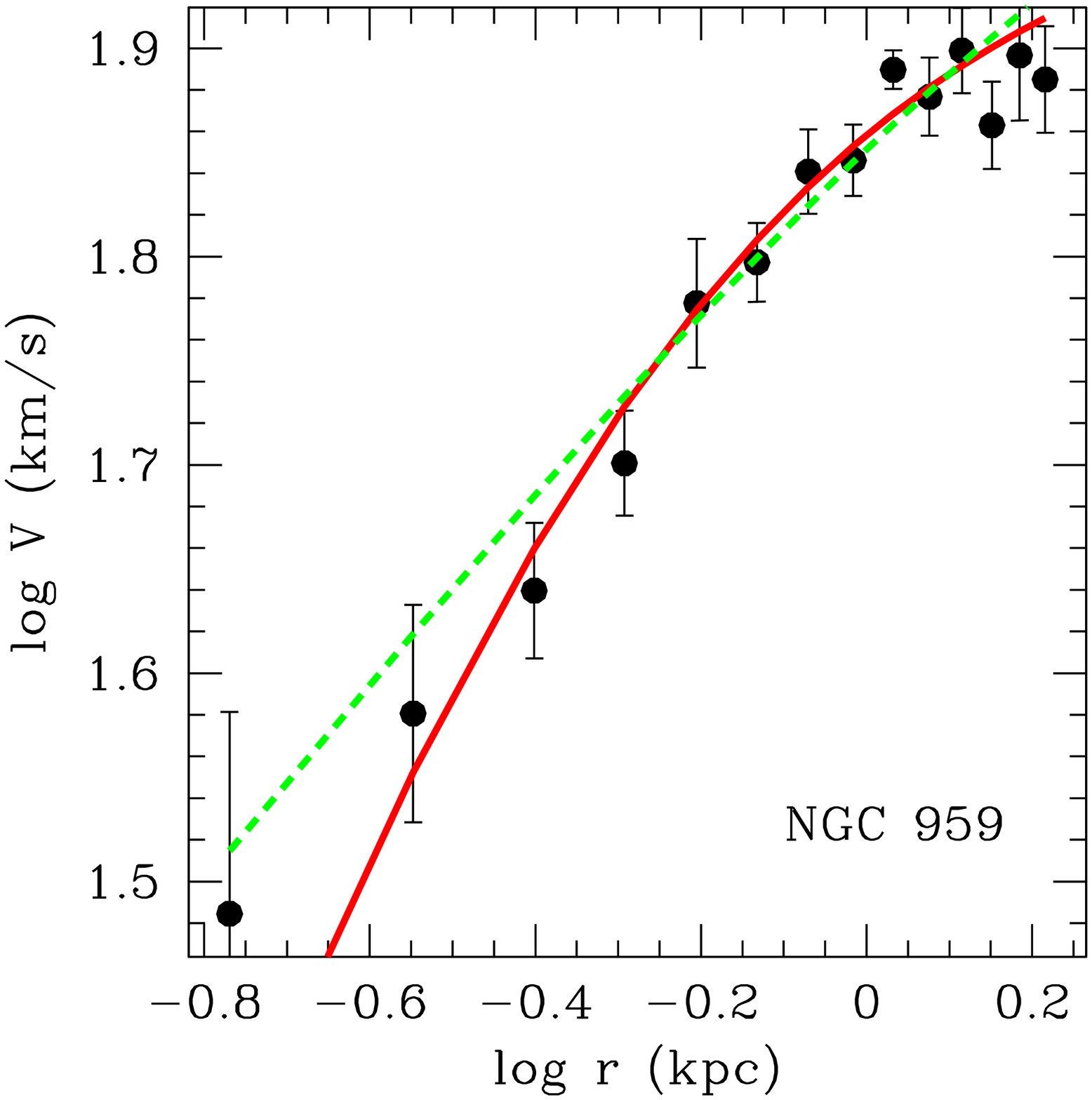}
\begin{quote}
\caption{Halo fits to the
\Dpak\ rotation curves.  The red solid line is the best-fit
isothermal halo, the green short-dashed line is the best-fit unconstrained
NFW halo, and the black long-dashed line is the best-fit
NFW$_{constr}$ halo.  NFW$_{constr}$ fits were only made to UGC 4325,
DDO 64, and F583-1. [{\it See the
  electronic edition of the Journal for a color version of this figure.}]  }
\end{quote}
\end{figure*}

\begin{deluxetable*}{lcccccccccccccccccc}
\tabletypesize{\scriptsize}
\tablecaption{Isothermal Halo Parameters}
\tablecolumns{16}
\tablewidth{0pt}
\tablehead{
 &\multicolumn{7}{c}{ZERO DISK} &\colhead{} &\multicolumn{7}{c}{MINIMAL DISK}\\
\cline{2-8} \cline{10-16}
\colhead{Galaxy} &\colhead{$R_{c}$} &\colhead{} &\colhead{$\rho_{0}$} &\colhead{} &\colhead{$\chi^{2}_{r}$} &\colhead{} &\colhead{$\Upsilon_{*}$} &\colhead{} &\colhead{$R_{c}$} &\colhead{} &\colhead{$\rho_{0}$} &\colhead{} &\colhead{$\chi^{2}_{r}$} &\colhead{} &\colhead{$\Upsilon_{*}$}
}
\startdata
UGC 4325 &4.1$\pm$0.3 & &88$\pm$3 & &3.2 & &0.0 & &4.6$\pm$1.5 & &77$\pm$5 & &3.1 & &0.57\\
F563-V2 &1.5$\pm$0.1 & &119$\pm$6 & &0.71 & &0.0 & &1.4$\pm$0.2 & &114$\pm$20 & &0.68 & &0.44\\
F563-1 &2.1$\pm$0.1 & &67$\pm$2 & &0.43 & &0.0 & &2.0$\pm$0.2 & &61$\pm$10 & &0.48 & &0.68\\
DDO 64 &3.3$\pm$0.5 & &43$\pm$3 & &3.2 & &0.0 & &4.1$\pm$3.8 & &37$\pm$6 & &3.3 & &0.62\\
F568-3 &3.8$\pm$0.2 & &27$\pm$1 & &1.2 & &0.0 & &4.0$\pm$0.5 & &22$\pm$3 & &1.5 & &0.66\\
UGC 5750 &5.7$\pm$0.4 & &7.1$\pm$0.3 & &0.83 & &0.0 & &6.5$\pm$1.1 & &5.2$\pm$0.6 & &0.84 & &0.62\\
NGC 4395 &0.7$\pm$0.1 & &258$\pm$9 & &2.9 & &0.0 & &0.57$\pm$0.05 & &318$\pm$42 & &2.7 & &0.70\\
F583-4 &1.3$\pm$0.1 & &67$\pm$2 & &0.67 & &0.0 & &1.2$\pm$0.2 & &66$\pm$16 & &0.62 & &0.53\\
F583-1 &2.5$\pm$0.1 & &30$\pm$2 & &0.50 & &0.0 & &2.4$\pm$0.2 & &30$\pm$3 & &0.59 & &0.62\\
NGC 7137 &0.6$\pm$0.1 & &274$\pm$17 & &3.1 & &0.0 & &\nodata & &\nodata & &\nodata & &\nodata\\
UGC 11820 &1.1$\pm$0.1 & &274$\pm$21 & &2.9 & &0.0 & &\nodata & &\nodata & &\nodata & &\nodata\\
UGC 128 &2.3$\pm$0.1 & &66$\pm$0.6 & &8.5 & &0.0 & &\nodata & &\nodata & &\nodata & &\nodata\\
UGC 191 &1.7$\pm$0.1 & &138$\pm$10 & &5.8 & &0.0 & &\nodata & &\nodata & &\nodata & &\nodata\\
UGC 1551 &1.3$\pm$0.2 & &57$\pm$5 & &5.5 & &0.0 & &\nodata & &\nodata & &\nodata & &\nodata\\
NGC 959 &0.4$\pm$0.1 & &1117$\pm$29 & &1.2 & &0.0 & &\nodata & &\nodata & &\nodata & &\nodata\\
\hline\\
\hline
\hline

&\multicolumn{7}{c}{POPSYNTH} &\colhead{} &\multicolumn{7}{c}{MAXIMUM DISK}\\
\cline{2-8} \cline{10-16}
\colhead{Galaxy} &\colhead{$R_{c}$} &\colhead{} &\colhead{$\rho_{0}$} &\colhead{} &\colhead{$\chi^{2}_{r}$} &\colhead{} &\colhead{$\Upsilon_{*}$} &\colhead{} &\colhead{$R_{c}$} &\colhead{} &\colhead{$\rho_{0}$} &\colhead{} &\colhead{$\chi^{2}_{r}$} &\colhead{} &\colhead{$\Upsilon_{*}$}\\
\hline\\
UGC 4325$^{a}$ &6.2$\pm$3.3 & &66$\pm$5 & &3.2 & &1.14  & &$\textit{2.5}$ & &$\textit{45}$ & &$\textit{11}$ & &4.5\\
F563-V2 &1.4$\pm$0.2 & &102$\pm$20 & &0.75 & &0.88 & &4.7$\pm$5.9 & &7.1$\pm$8.8 & &1.4 & &4.0\\
F563-1$^{b}$ &2.1$\pm$0.2 & &54$\pm$10 & &0.50 & &1.36 & &4.2$\pm$1.2 & &13$\pm$6 & &0.83 & &6.5\\
 &\nodata & &\nodata & &\nodata & &\nodata &   &12.5$\pm$9.0 & &2.7$\pm$1.6 & &1.3 & &10.0\\
DDO 64$^{a}$ &5.9$\pm$11.1 & &32$\pm$5 & &3.5 & &1.24 & &$\textit{1.5}$ & &$\textit{24}$ & &$\textit{2.2}$ & &5.0\\
F568-3 &4.7$\pm$0.8 & &16$\pm$3 & &1.8 & &1.32 & &6.1$\pm$1.8 & &10$\pm$3 & &2.5 & &2.3\\
UGC 5750 &7.5$\pm$1.6 & &4.1$\pm$0.5 & &0.95 & &1.24 & &9.8$\pm$3.2 & &2.6$\pm$0.5 & &1.3 & &2.2\\
NGC 4395$^{a}$ &0.50$\pm$0.04 & &355$\pm$50 & &2.8 & &1.40 & &$\textit{19}$ & &$\textit{0.32}$ & &$\textit{4.1}$ & &9.0\\
F583-4$^{a}$ &1.2$\pm$0.2 & &63$\pm$16 & &0.62 & &1.06 & &$\textit{8}$ & &$\textit{1.4}$ & &$\textit{1.2}$ & &10.0\\
F583-1$^{b}$ &2.5$\pm$0.2 & &27$\pm$3 & &0.60 & &1.24 & &3.6$\pm$0.5 & &14$\pm$2 & &0.83 & &5.0\\
   &\nodata & &\nodata & &\nodata & &\nodata &     &6.9$\pm$1.7 & &5.2$\pm$0.9 & &2.0 & &10.0\\

\enddata
\tablecomments{$R_{c}$ is in kpc; $\rho_{0}$ is in 10$^{-3}$
  M$_{\sun}$ pc$^{-3}$.  Photometry is unavailable for NGC 7137, UGC
  11820, UGC 128, UGC 191, UGC 1551, \& NGC 959; halo fits beyond zero
  disk are not presented.\\
  $^{a}$ The baryons can account for most of the velocity in the maximum disk fit.  See text for details of fit.\\
  $^{b}$ As discussed in the text, F563-1 and F583-1 have two possible values of \ml$_{Max}$.}
\end{deluxetable*}

\begin{deluxetable*}{lcccccccccccccccccc}
\tabletypesize{\scriptsize}
\tablecaption{NFW Halo Parameters}
\tablecolumns{16}
\tablewidth{0pt}
\tablehead{
 &\multicolumn{7}{c}{ZERO DISK} &\colhead{} &\multicolumn{7}{c}{MINIMAL DISK}\\
\cline{2-8} \cline{10-16}
\colhead{Galaxy} &\colhead{c} &\colhead{} &\colhead{$V_{200}$} &\colhead{} &\colhead{$\chi^{2}_{r}$} &\colhead{} &\colhead{$\Upsilon_{*}$} &\colhead{} &\colhead{c} &\colhead{} &\colhead{$V_{200}$} &\colhead{} &\colhead{$\chi^{2}_{r}$} &\colhead{} &\colhead{$\Upsilon_{*}$}
}
\startdata
UGC 4325 &$\textit{6.9}$ & &$\textit{249}$ & &$\textit{39}$ & &0.0 & &$\textit{1.0}$ & &$\textit{1002}$ & &$\textit{12}$ & &0.57\\
F563-V2 &7.7$\pm$2.0 & &128$\pm$32 & &0.40& &0.0 & &8.4$\pm$1.5 & &105$\pm$17 & &0.46 & &0.44\\
F563-1 &7.8$\pm$1.3 & &106$\pm$10 & &0.88 & &0.0 & &7.6$\pm$1.3 & &100$\pm$9 & &0.89 & &0.68\\
DDO 64 &$\textit{9.2}$ & &$\textit{62}$ & &$\textit{12}$  & &0.0 & &$\textit{1.0}$ & &$\textit{376}$ & &$\textit{6.5}$ & &0.62\\
F568-3 &$\textit{8.2}$ & &$\textit{110}$ & &$\textit{12}$ & &0.0 & &$\textit{1.0}$ & &$\textit{465}$ & &$\textit{3.9}$ & &0.66\\
UGC 5750 &0.5$\pm$0.1 & &320$\pm$43 & &1.7 & &0.0 & &$\textit{1.0}$ & &$\textit{167}$ & &$\textit{1.7}$ & &0.62\\
NGC 4395 &10.1$\pm$0.6 & &77$\pm$4 & &2.1 & &0.0 & &11.5$\pm$1.0 & &63$\pm$4 & &2.1 & &0.70\\
F583-4 &5.5$\pm$2.2 & &92$\pm$32 & &0.41 & &0.0 & &5.7$\pm$1.4 & &83$\pm$18 & &0.41 & &0.53\\
F583-1$^{a}$ &4.5$\pm$0.8 & &120$\pm$20 & &1.7 & &0.0 & &5.1$\pm$1.2 & &102$\pm$21 & &1.8 & &0.62\\
NGC 7137 &15$\pm$3 & &56$\pm$10 & &3.4 & &0.0 & &\nodata & &\nodata &
&\nodata & &\nodata\\
UGC 11820 &\nodata & &\nodata & &\nodata & &\nodata & &\nodata &
&\nodata & &\nodata & &\nodata\\
UGC 128 &8.9$\pm$0.2 & &111$\pm$0.7 & &9.3 & &0.0 & &\nodata &
&\nodata & &\nodata & &\nodata\\
UGC 191 &\nodata & &\nodata & &\nodata & &\nodata & &\nodata &
&\nodata & &\nodata & &\nodata\\
UGC 1551 &\nodata & &\nodata & &\nodata & &\nodata & &\nodata &
&\nodata & &\nodata & &\nodata\\
NGC 959 &23$\pm$4 & &76$\pm$15 & &1.7 & &0.0 & &\nodata & &\nodata & &\nodata & &\nodata\\
\hline\\
\hline
\hline

&\multicolumn{7}{c}{POPSYNTH} &\colhead{} &\multicolumn{7}{c}{MAXIMUM DISK}\\
\cline{2-8} \cline{10-16}
\colhead{Galaxy} &\colhead{c} &\colhead{} &\colhead{$V_{200}$} &\colhead{} &\colhead{$\chi^{2}_{r}$} &\colhead{} &\colhead{$\Upsilon_{*}$} &\colhead{} &\colhead{c} &\colhead{} &\colhead{$V_{200}$} &\colhead{} &\colhead{$\chi^{2}_{r}$} &\colhead{} &\colhead{$\Upsilon_{*}$}\\
\hline\\
UGC 4325 &$\textit{1.0}$ & &$\textit{897}$ & &$\textit{12}$ & &1.14  & &$\textit{1.0}$ & &$\textit{283}$ & &$\textit{12}$ & &4.5\\
F563-V2 &7.8$\pm$1.7 & &104$\pm$20 & &0.52 & &0.88 & &$\textit{1.0}$ & &$\textit{203}$ & &$\textit{1.1}$ & &4.0\\
F563-1$^{b}$ &7.0$\pm$1.3 & &102$\pm$11 & &0.90 & &1.36 & &$\textit{1.0}$ & &$\textit{283}$ & &$\textit{1.0}$ & &6.5\\
       &\nodata & &\nodata & &\nodata & &\nodata & &$\textit{1.0}$ & &$\textit{210}$ & &$\textit{1.5}$ & &10.0\\
DDO 64 &$\textit{1.0}$ & &$\textit{332}$ & &$\textit{6.4}$ & &1.24 & &$\textit{1.0}$ & &$\textit{146}$ & &$\textit{2.6}$ & &5.0\\
F568-3 &$\textit{1.0}$ & &$\textit{404}$ & &$\textit{4.3}$ & &1.32 & &$\textit{1.0}$ & &$\textit{317}$ & &$\textit{5.2}$ & &2.3\\
UGC 5750 &$\textit{1.0}$ & &$\textit{141}$ & &$\textit{1.9}$ & &1.24 & &$\textit{1.0}$ & &$\textit{101}$ & &$\textit{2.4}$ & &2.2\\
NGC 4395 &12.5$\pm$1.1 & &61$\pm$4 & &2.2 & &1.40 & &\nodata & &\nodata & &\nodata & &9.0\\
F583-4 &5.8$\pm$1.5 & &86$\pm$19 & &0.40 & &1.06 & &\nodata & &\nodata & &\nodata & &10.0\\
F583-1$^{b}$ &4.9$\pm$1.2 & &110$\pm$26 & &1.9 & &1.24 & &$\textit{1.0}$ & &$\textit{298}$ & &$\textit{2.1}$ & &5.0\\
&\nodata & &\nodata & &\nodata & &\nodata & &$\textit{1.0}$ & &$\textit{173}$ & &$\textit{3.3}$ & &10.0\\

\enddata
\tablecomments{$V_{200}$ is in km s$^{-1}$.  Italicized halo
  parameters are forced fits.  See text for details.  Photometry is unavailable for NGC 7137, UGC
  11820, UGC 128, UGC 191, UGC 1551, \& NGC 959; halo fits beyond zero
  disk are not presented.\\
$^{a}$ The parameters of the updated $NFW_{constr}$ fit for the zero
disk case of F583-1 are $c$ = 8.7,
$V_{200}$ = 83, $\chi^{2}_{r}$ = 6.5.\\
$^{b}$ As discussed in the text, F563-1 and F583-1 have two possible values of \ml$_{Max}$.}
\end{deluxetable*}
\subsection{New Observations}

Of the 14 galaxies observed, there were eight galaxies for which 
meaningful velocity fields
could not be constructed.\footnote{F469-2, UGC 2034, UGC 2053, UGC
  11944, UGC 12048/NGC 7292,  UGC 12082, UGC 12212, and UGC 12632.} 
The \Ha\ emission in these galaxies was too faint to be detected
and/or not spread out enough across the fiber array.  Velocity fields
and rotation curves were derived for the remaining six galaxies in the
sample, and each is described below.

$\textbf{\textit{NGC 7137}$-$}$  There were three \Dpak\ pointings for
this galaxy.  The  pointings are shown on the \Ha\ image (Figure 2) 
of the galaxy.  Spiral arms are 
clearly visible in this galaxy.  The fiber velocities were the
average of the \Ha, \nii, \siis, and \siio\ lines.  \Ha\ emission was 
abundant and the majority of fibers detected emission.  The
inclination was fixed to the value in \citet{Tully}.  The position 
angle of the major axis was well-constrained by \texttt{ROTCUR}.  The
rotation curve rises steeply out to roughly 10$\arcsec$ then dips
slightly before rising again.

$\textbf{\textit{UGC 11820}$-$}$  This galaxy is both large and diffuse on
the sky.  There is a central concentration of \Ha\ emission, which is 
perhaps a bar, that runs roughly NE-SW and also two large, diffuse 
arms that extend from the end of the bar.  There were three \Dpak\ pointings
along the central feature.  The pointings are shown on the \Ha\ image  
(Figure 2) of the galaxy.  The 
fiber velocities were the average of the \Ha, \siis, and \siio\ lines.  
The \Ha\ emission was sparse and only roughly half of the fibers 
detected emission.  The inclination was fixed to the value listed in 
\citet{MRdB}.  The position angle of the major axis was well-constrained 
by \texttt{ROTCUR}.

$\textbf{\textit{UGC 128}$-$}$  There were three \Dpak\ pointings across this
galaxy.  The pointings are shown on the \Ha\ image (Figure 2) of the galaxy.  
The fiber velocities were the
average of the \Ha, \nii, and \siis\ lines.  The \Ha\ emission in this
galaxy was sparse, but the emission that was present was scattered
across the three \Dpak\ pointings such that both the approaching and
receding sides of the velocity field were mapped.  The position angle
was fixed to the position angle of the \HI\ velocity field of
\citet{vanderHulst} and the inclination was fixed to the value listed
in \citet{dBM96}.  The \Dpak\ rotation curve is plotted with the \HI\
rotation curve of \citet{VerheijendB}.  The \Dpak\ rotation curve is
largely consistent with if not slightly steeper than the \HI\ curve 
 and does not go out far enough to show a clear turn-over.

$\textbf{\textit{UGC 191}$-$}$  There were two \Dpak\ pointings for
this galaxy and they are shown on the \Ha\ image (Figure 3).  The fiber 
velocities were the
average of the \Ha, \siis, and \siio\ lines.  There is ample \Ha\
emission in the galaxy and almost every fiber had a detection.  The
inclination was fixed to the value in \citet{Tully}.  The position angle 
of the major axis was well-constrained by \texttt{ROTCUR}.  The
rotation curve rises linearly and has no clear turn-over.  

$\textbf{\textit{UGC 1551}$-$}$ There were three \Dpak\ pointings for
this galaxy.  The pointings are shown on the \Ha\ image (Figure 3) 
of the galaxy.  The fiber 
velocities were the average of the \Ha, \nii, \siis, and \siio\ lines.  
\Ha\ emission was detected in nearly all of the fibers.   The 
inclination was fixed to the value in \citet{Tully}.  The position angle 
of the major axis was well-constrained by \texttt{ROTCUR}.  There is a twist 
in the velocity field that is suggestive of the presence of a bar.  Noncircular
motions are probably important inside of 15$\arcsec$ where the rotation
curve is mostly flat.  Beyond 15$\arcsec$, there is a linear rise in the
rotation curve.

$\textbf{\textit{NGC 959}$-$}$  There were three \Dpak\ pointings
across the length of this galaxy.  The  pointings are shown on the
\Ha\ image (Figure 3) of the galaxy.  The
fiber velocities were the average of the \Ha, \nii, \siis, and \siio\
lines.  \Ha\ emission was abundant and the majority of fibers detected
emission.  The inclination was fixed to the value listed in
\citet{James}.  The position angle of the major axis was well-constrained 
by \texttt{ROTCUR}.  The rotation curve is well-behaved with a steady rise
and a turnover to  $V_{flat}$ $\sim$ 80 \kms.

\section{Zero Disk Halo Fits}
In this section, we present the pseudoisothermal and NFW halo fits 
to the \Dpak\ rotation curves in the zero disk case.  By ignoring the 
velocity contribution from the baryons and attributing all rotation to 
dark matter, we are able to put an upper limit on the slope and/or 
concentration of the halo density profile.  For those galaxies with 
photometry, halo fits for three assumptions about the stellar 
mass-to-light ratio are presented in \S\ 6.

\subsection{Halo Models}
The cuspy NFW halo and the cored pseudoisothermal halo are two of the 
most well-known competing descriptions of dark matter halos.  We 
provide a brief description of each below.

\subsubsection{NFW Profile}
Numerical simulations show that the density of CDM halos rises steeply 
toward the halo center.  The exact value of the inner slope of the CDM 
halo varies slightly depending on the simulation \citep[e.g.][]
{NFW96,NFW97,Moore, Reed, Navarro2004, Diemand}.  From an observational
perspective, there is very little to distinguish the various flavors 
of cuspy CDM halos, and we choose to fit the data with the NFW halo.

The NFW mass-density distribution is described as
\begin{equation}
\rho_{NFW}(R) = \frac{\rho_{i}}{(R/R_{s})(1 + R/R_{s})^{2}} ,
\end{equation}
in which $\rho_{i}$ is related to the density of the universe at the 
time of halo collapse, and $R_{s}$ is the characteristic radius of 
the halo.  The NFW rotation curve is given by
\begin{equation}
V(R) = V_{200}\sqrt{ \frac{\ln(1+cx) - cx/(1 + cx)}{x[\ln(1 + c) - c/(1+c)]}},
\end{equation}
with $x$ = $R$/$R_{200}$.  The rotation curve is parameterized by 
a radius $R_{200}$ and a concentration parameter $c$ = $R_{200}$/$R_{s}
$, both of which are directly related to $R_{s}$ and $\rho_{i}$. Here  
$R_{200}$ is the radius at which the density contrast exceeds 200,  
roughly the virial radius; $V_{200}$ is the circular velocity
at $R_{200}$ \citep{NFW96,NFW97}.  Because the NFW profile has a shallower 
slope than other cuspy halo models, it provides a lower limit on the 
slope of cuspy density profiles and, as such, gives the cuspy halo the 
best possible chance to fit the data.

\subsubsection{Pseudoisothermal Halo}
The pseudoisothermal halo describes a dark
matter halo that has a core of roughly constant density.  By
construction, it produces flat rotation curves at large radii. 
The density profile of the pseudoisothermal halo is
\begin{equation}
\rho_{iso}(R) = \rho_{0}[1 + (R/R_{C})^{2}]^{-1} ,
\end{equation}
with $\rho_{0}$ being the central density of the halo and $R_{C}$ 
representing the core radius of the halo.  The rotation curve 
corresponding to this density profile is
\begin{equation}
V(R) = \sqrt{4\pi G\rho_{0} R_{C}^{2}\left[1 - \frac{R_{C}}{R}\arctan\left(\frac{R}{R_{C}}\right)\right]} .
\end{equation}
The pseudoisothermal halo is empirically motivated and predates halo 
profiles stemming from numerical simulations.

\subsection{Halo Fits to Previously Observed Galaxies}

We find the best-fit zero disk 
case isothermal and NFW halos to the new \Dpak\
rotation curves of UGC 4325, DDO 64, and F583-1.  When available, 
the \Dpak\ rotation curves have been supplemented with previous 
smoothed long-slit \Ha\ and \HI\ rotation curves.  We use the entire 
long-slit rotation curve, and include only those \HI\ points that 
extend beyond the radial range of both the \Dpak\ and long-slit data.  
Uncertainties from possible resolution effects are avoided by using 
only the outer \HI\ points. 

For these three galaxies, we also fit an NFW halo called 
NFW$_{constrained}$ (hereafter NFW$_{constr}$), as
described in detail in \citetalias{Kuzio}.  This halo fit is motivated by 
NFW fits that have parameters which are unrealistic or inconsistent 
with $\Lambda$CDM.  Briefly, the constrained halo was 
required to match the velocities at the outer radii of each galaxy 
while constraining the concentration to agree with cosmology.  The 
concentrations were calculated using Equation (7) of \citet{dBBM}, 
which gives the concentration as a function of of $V_{200}$ 
\citep{NFW97}, and then adjusted to the cosmology of \citet{Tegmark} 
by subtracting 0.011 dex \citep*{McGaugh03}.

The halo fits are plotted over the data in Figure 4, and the halo 
parameters are listed in Tables 2 and 3.  For comparison,
the numbers mentioned in the text below are the fits from 
\citetalias{Kuzio}, unless specifically noted otherwise.

$\textbf{\textit{UGC 4325}$-$}$  The isothermal fit to UGC 4325
improves slightly with the addition of the four new \Dpak\ pointings 
($R_{c}$ = 3.3 $\pm$ 0.2; $\rho_{0}$ = 91 $\pm$ 4; $\chi^{2}$ = 3.8).
An unconstrained NFW halo still could not be fitted to the data, and the
poor quality of the NFW$_{constr}$ fit remains virtually
unimproved ($c$ = 6.9; $V_{200}$ = 249; $\chi^{2}$ = 40).  This galaxy 
remains best-described by the isothermal halo.

$\textbf{\textit{DDO 64}$-$}$  The two new \Dpak\ pointings on DDO 64
help to improve the quality of both the isothermal and NFW$_{constr}$
fits.  The values of the isothermal halo parameters remain within the
errors of the original values, but the value of $\chi^{2}_{r}$
decreases ($R_{c}$ = 4.4 $\pm$ 0.9; $\rho_{0}$ = 38 $\pm$ 3; 
$\chi^{2}$ = 5.5).  While no unconstrained NFW halo fit could be made, the
quality of the NFW$_{constr}$ halo significantly improves ($c$ = 9.2; 
$V_{200}$ = 62; $\chi^{2}$ = 20).  This does
not mean, however, that the NFW halo is a good fit to the data.  The
NFW$_{constr}$ halo continues to overshoot the data at radii interior
to where it was forced to match the data.  While it passes through the
large errorbars on the inner rotation curve points, significant
noncircular motions would need to be important all the way out to
$\sim$1 kpc in order to boost the observed velocities up to the
expected NFW velocities.

$\textbf{\textit{F583-1}$-$}$  The isothermal and NFW fits are
significantly better constrained by the addition of the two new \Dpak\
pointings.  F583-1 remains best-described by the isothermal halo.  The
values of the halo parameters remain within the errors of the previous
values, but the value of $\chi^{2}_{r}$ falls from 5.4 to 0.5 
($R_{c}$ = 2.7 $\pm$ 0.1; $\rho_{0}$ = 35 $\pm$ 2).  The
new values of the isothermal halo parameters are almost
indistinguishable from the values determined by \citet{dBMR}: $R_{c}$
= 2.44 $\pm$ 0.06, $\rho_{0}$ = 33.0 $\pm$ 1.1.  The values of the
parameters of the unconstrained NFW fit are also mostly unchanged, and
$\chi^{2}_{r}$ decreases to 1.7 from 8.7 ($c$ = 4.7 $\pm$ 0.7; 
$V_{200}$ = 133 $\pm$ 21).  The value of the
concentration, $c$ = 4.5, is still too low to be consistent with
$\Lambda$CDM.  The value of $\chi^{2}_{r}$ also drops for the
NFW$_{constr}$ fit ($\chi^{2}$ = 11 $\rightarrow$ 6.5), but the NFW 
velocities continue to over-predict
the observed velocities interior to where they were forced to match
the data.

\subsection{Halo Fits to New Observations}
 We find the best-fit zero disk case isothermal and NFW halos to the \Dpak\
rotation curves of NGC 7137, UGC 11820, UGC 128, UGC 191, UGC 1551, 
and NGC 959.  In the case of UGC 128, we combine the \Dpak\
rotation curve with the \HI\ rotation curve of \citet{VerheijendB},
using only those \HI\ points beyond the radial range of the \Dpak\
data.   We do not make an NFW$_{constr}$ fit to these six galaxies, as 
the radial range of the data does not extend beyond the rising part of 
the rotation curve.  In Figure 4 we plot the halo fits over the data, 
and list the halo parameters in Tables 2 and 3.

$\textbf{\textit{NGC 7137}$-$}$  The isothermal halo is a
slightly better fit to NGC 7137 than the NFW halo, but the difference 
is not particularly significant.   There are 
``bumps and wiggles'' in the rotation curve that simple, smooth halo 
models cannot fit.  The value of the concentration, $c$ = 15, for 
the NFW fit is on the high side of values expected in a $\Lambda$CDM 
cosmology.  

$\textbf{\textit{UGC 11820}$-$}$  UGC 11820 is fit relatively
well by the isothermal halo; no NFW halo could be fit to the rotation
curve. 

$\textbf{\textit{UGC 128}$-$}$  UGC 128 is more consistent with the 
isothermal halo than the NFW halo, though the value of $\chi^{2}_{r}$ 
is high for both fits and both halo models overestimate the rotation 
velocities at small radii.  The best-fitting concentration, $c$ = 8.9, is 
reasonable for a galaxy of this size in a $\Lambda$CDM cosmology.  
The best-fitting isothermal halo parameters for the combined \Dpak+\HI\ 
rotation curve are $R_{c}$ = 2.3$\pm$0.1 and $\rho_{0}$ = 66$\pm$0.6.  
\citet{dBM96} also fit an isothermal halo to the \citet{vanderHulst} 
\HI\ rotation curve and find $R_{c}$ = 4.0 and $\rho_{0}$ = 21.7.  
The differences in the halo parameters reflect the inclusion of the 
slightly steeper \Dpak\ rotation curve.

$\textbf{\textit{UGC 191}$-$}$   We exclude the innermost point of 
the UGC 191 rotation curve due to its unrealistically small formal 
error bar and the steep jump in velocity between it and the next 
rotation curve point.  Excluding this point from the halo
fits does not significantly alter the values of the halo parameters, but does
improve the values of the reduced $\chi^{2}$.   No NFW fit could be made to
the \Dpak\ rotation curve.  The data were fit with an
isothermal halo, albeit with a large $\chi^{2}_{r}$.  The UGC 191
\Dpak\ rotation curve is linearly rising and shows no turn-over over
in the radial range covered by the \Dpak\ data.  To be useful for 
distinguishing between halo models, data points at larger
radii are needed to better constrain the halo fits.  

$\textbf{\textit{UGC 1551}$-$}$  We exclude the first five points of
the UGC 1551 rotation curve because the twist in the velocity field 
suggests that a bar may be present in the galaxy.    An NFW fit could
not be made to the data;  an isothermal fit was made, but was not 
well-constrained.  Like UGC 191, data at larger radii are necessary 
for obtaining useful constraints on the halo models.  

$\textbf{\textit{NGC 959}$-$}$  NGC 959 is well
described by the isothermal halo and is a slightly better fit than the
NFW halo.  The value of the concentration, $c$ = 23, of the best fit NFW
halo is on the high side of expected values for galaxies in a
$\Lambda$CDM cosmology.

\newpage
\subsection{Summary of Zero Disk Halo Fits}

In \S\ 5.2 and \S\ 5.3 we have presented updated zero disk halo fits 
to three galaxies from \citetalias{Kuzio} and six new galaxies 
observed with \Dpak.  Overall, we find the 
isothermal halo to be a better description of seven of these 
galaxies than the NFW halo.  Two of the new galaxies do not have data 
at large enough radii to put useful constraints on the halo models.  
When NFW fits could be made, the
concentrations were often beyond the range of values expected for
$\Lambda$CDM.    We find that the parameters of the halo fits to the 
three previously observed galaxies do not significantly change, but 
are better constrained by the additional \Dpak\ coverage.  The \Dpak\ 
rotation curves of the six new galaxies could all be fit by the 
isothermal halo, and an NFW fit could be made to only three.   Only one
of the three NFW fits had a concentration consistent with the range
expected for galaxies in a $\Lambda$CDM cosmology; the other two are
on the high end of expected values.  We also find that the quality of 
both the isothermal and NFW halo fits is greatly improved when the 
radial range of the data extends into the flat part of the rotation 
curve.  Of the six new galaxies, only NGC 959 has a \Dpak\ rotation 
curve which clearly turns over and flattens, and it is this galaxy 
that has the isothermal and NFW fits with the lowest $\chi^{2}_{r}$.  

\section{Mass Models}
Low surface brightness galaxies are dark matter-dominated down
to small radii.  Because of this, the velocity contribution from the
baryons is often disregarded when fitting dark matter halo models
to the galaxy rotation curves.  This type of fit which ignores the
contribution of the stars and gas, such as in \S\ 5,  is usually 
called the minimum-disk case.  While the dark matter is the dominant 
mass component of LSB galaxies at all radii, baryons are still 
important.  To accurately determine the distribution of the dark 
matter, it is necessary to properly account for the stars and gas 
in the galaxies and their contribution to the observed rotation.  
The velocity component coming from the stars is determined from the 
surface photometry scaled by the stellar mass-to-light ratio, \ml.  
This is a straightforward computation of the gravitational potential 
of the observed stars.  However,  the true value of  \ml\ is
difficult to determine, and as discussed below, can be assigned a
value following a number of techniques.  Similarly, the velocity
component from the gas is determined using \HI\ surface density
profiles.  The conversion from observed 21 cm luminosity to atomic 
gas mass is well understood from the physics of the spin-flip 
transition.    A scaling factor, the inverse of the hydrogen mass 
fraction,  is usually included to account for the
helium and metals also present in the galaxies.  

In \S\ 5 of this paper and in \citetalias{Kuzio}, we have so far ignored the 
baryons in the halo fits to LSB galaxies observed with \Dpak.  For
those galaxies with $R$-band photometry and \HI\ surface density
profiles we now include the contribution from the baryons and present
the dark matter halo fits for three determinations of \ml.

\begin{deluxetable}{lclcc}
\tabletypesize{\scriptsize}
\singlespace
\tablecaption{Galaxy Parameters}
\tablehead{
\colhead{Galaxy} &\colhead{$h$ (kpc)} &\colhead{($B$--$R$)} &\colhead{\ml\ (Pop)} &\colhead{References}
}
\startdata
UGC 4325 &1.6 &0.85 &1.14 &1,3\\
F563-V2 &2.1 &0.51$^{a}$ &0.88 &2,4\\
F563-1 &2.8 &0.96 &1.36 &2,5\\
DDO 64 &1.2 &0.9$^{b}$ &1.24 &1\\
F568-3 &4.0 &0.94 &1.32 &2,5\\
UGC 5750 &5.6 &0.9$^{b}$ &1.24 &2,2\\
NGC 4395 &2.3 &\nodata$^{c}$ &1.40 &1\\
F583-4 &2.7 &0.8 &1.06 &2,6\\
F583-1 &1.6 &0.9 &1.24 &2,6\\
\enddata
\tablecomments{Col.(2): Scale length (kpc). Col.(4): \ml\ for the popsynth case as determined from the colors.  Col.(5): References for $h$ and ($B$--$R$), respectively: (1) de Blok \& Bosma (2002) (2) de Blok, McGaugh, \& Rubin (2001) (3) van den Bosch \& Swaters (2001) (4) de Blok \& McGaugh (1997) (5) de Blok, van der Hulst, \& Bothun (1995) (6) de Blok, McGaugh, \& van der Hulst (1996).\\
$^{a}$ ($\bv$) color.\\
$^{b}$ Multicolor photometry unavailable; assuming ($B$--$R$)=0.9 for dwarf galaxies.\\
$^{c}$ Multicolor photomety unavailable; assuming  $\Upsilon_{*}$=1.4.}
\end{deluxetable}

\subsection{Dynamical Components}
There are three separate components in galaxy mass models that
contribute to the observed velocities:  the stars, the gas, and the
dark matter.  The stars, gas, and dark matter are added together in
quadrature to obtain the total velocity, $V_{total}^{2}$ =
\ml$v_{*}^{2}$ + $V_{gas}^{2}$ + $V_{DM}^{2}$.  In this section, we
describe each of these dynamical components.

\begin{figure*}
\plotone{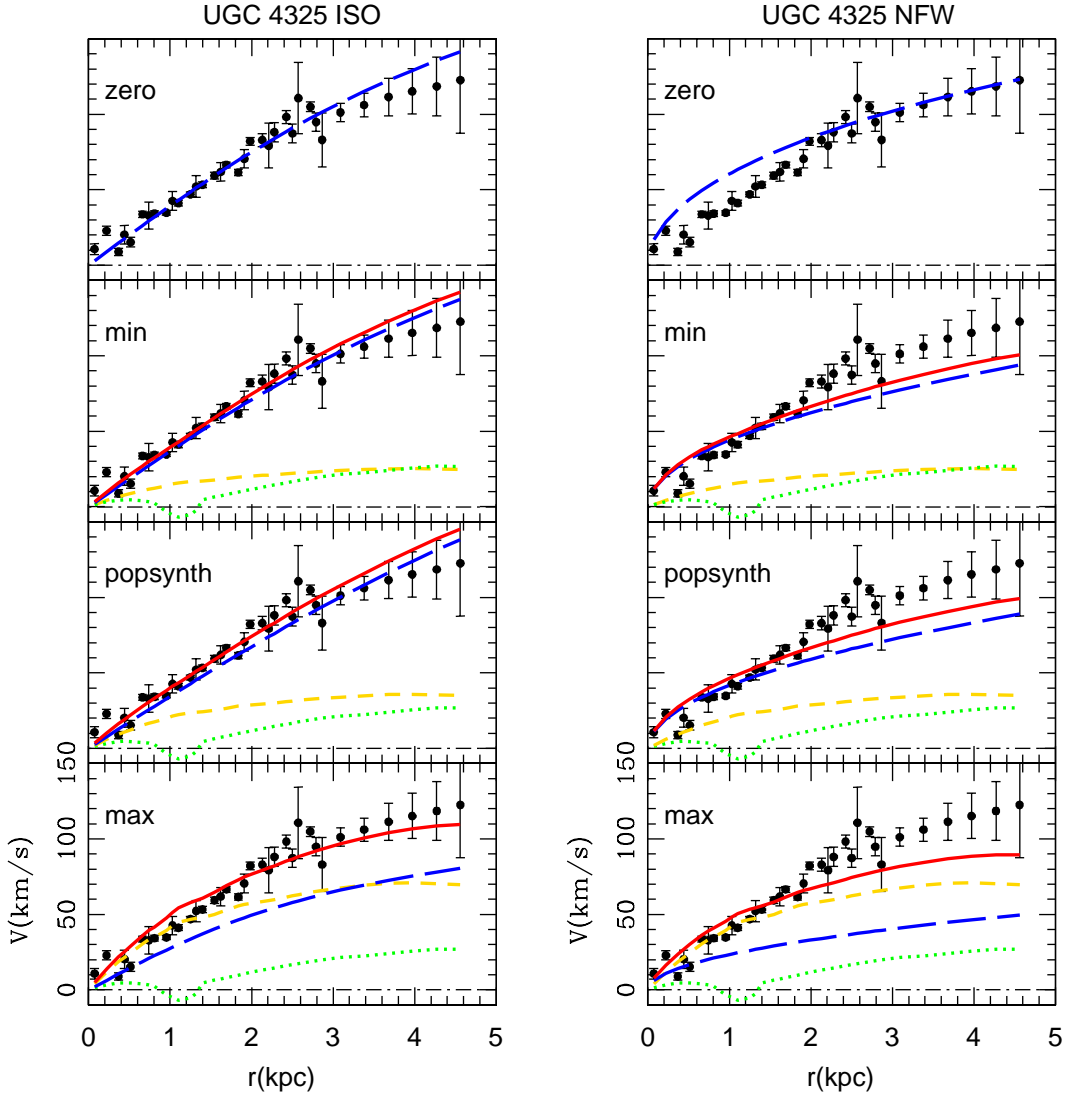}
\begin{quote}
\caption{Isothermal and NFW halo fits for
  UGC 4325.  The green
dotted line is the rotation curve of the gas disk, the gold
short-dashed line is the rotation curve of the stellar disk, the blue
long-dashed line is the rotation curve of the dark matter halo, and
the red solid line is the total model curve.  The velocity
contribution from both the stars and gas is ignored in the zero disk
case.  The minimal disk case considers the gas contribution and
assumes a lightweight IMF for the stars.  The popsynth case includes
the gas contribution and uses population synthesis models to determine
the stellar contribution.  The maximum disk case includes the gas
contribution and the stellar mass-to-light ratio is scaled up as far
as the data will allow.  [{\it See the
  electronic edition of the Journal for a color version of this figure.}]}
\end{quote}
\end{figure*}

\begin{figure*}
\plotone{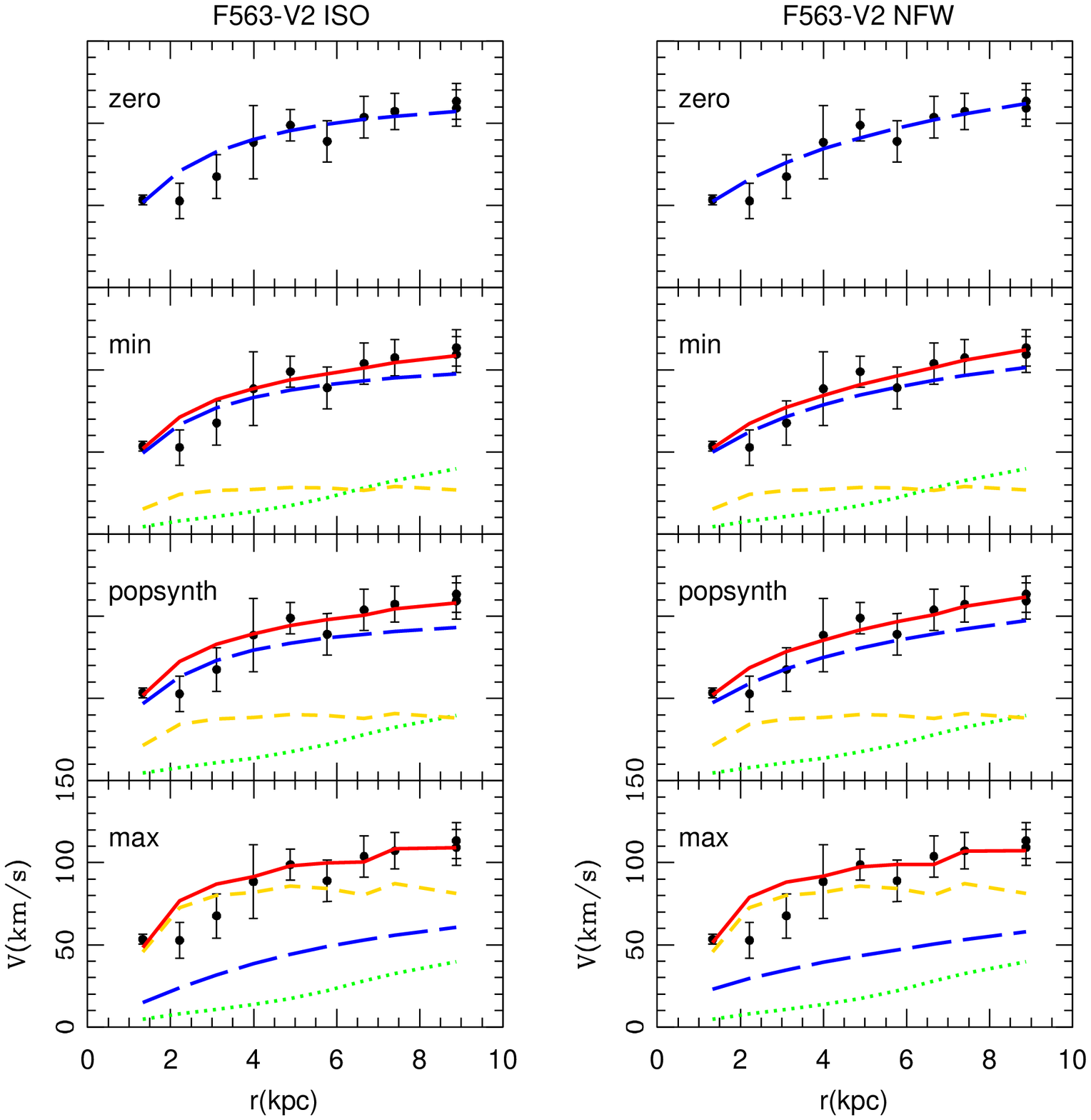}
\begin{quote}
\caption{Isothermal and NFW halo fits for F563-V2.  Line types are 
  described in Figure 5. [{\it See the
  electronic edition of the Journal for a color version of this figure.}]}
\end{quote}
\end{figure*}

\begin{figure*}
\plotone{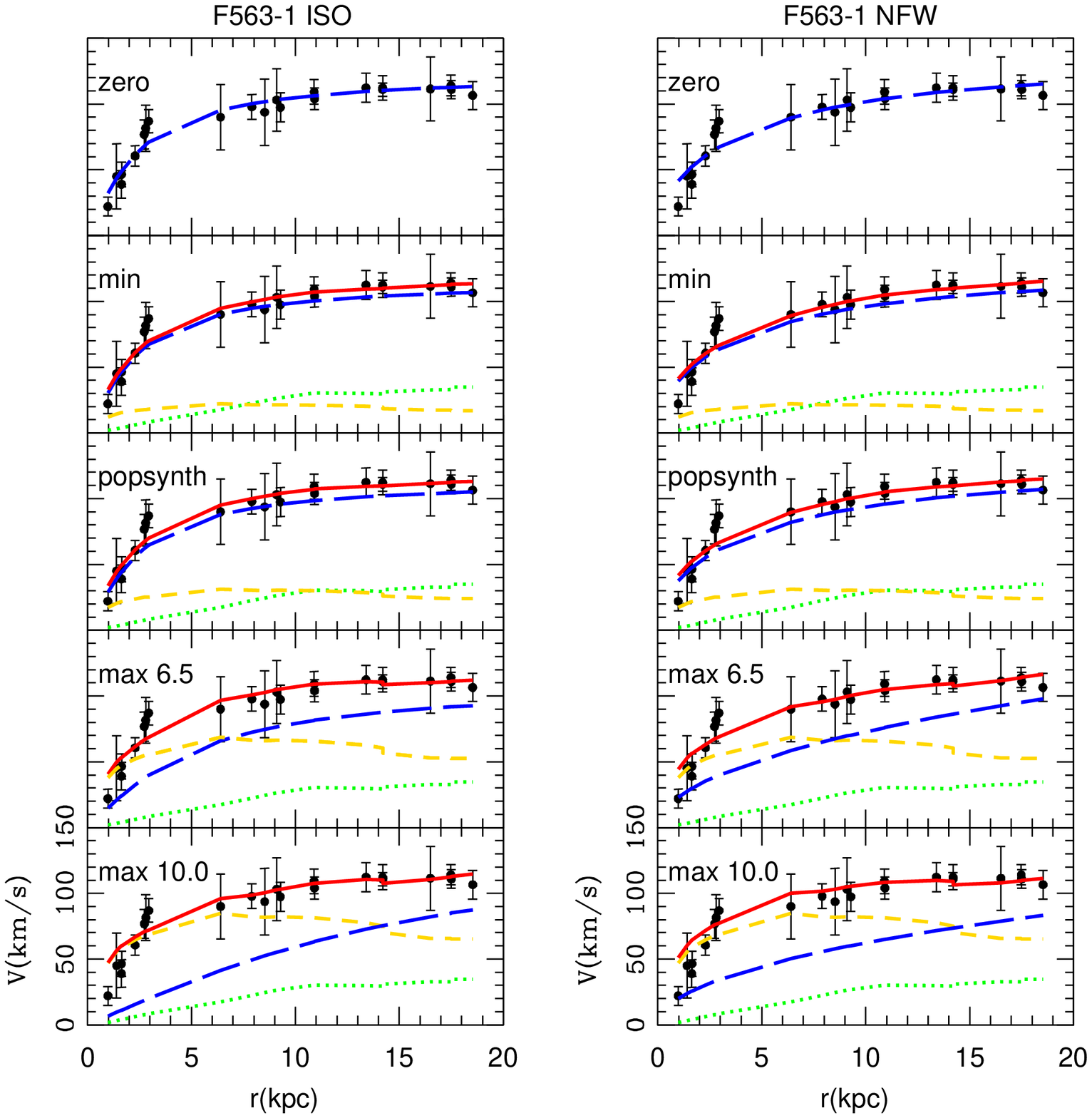}
\begin{quote}
\caption{Isothermal and NFW halo fits for
  F563-1.  Line types are described in Figure 5. [{\it See the
  electronic edition of the Journal for a color version of this figure.}]}
\end{quote}
\end{figure*}

\begin{figure*}
\plotone{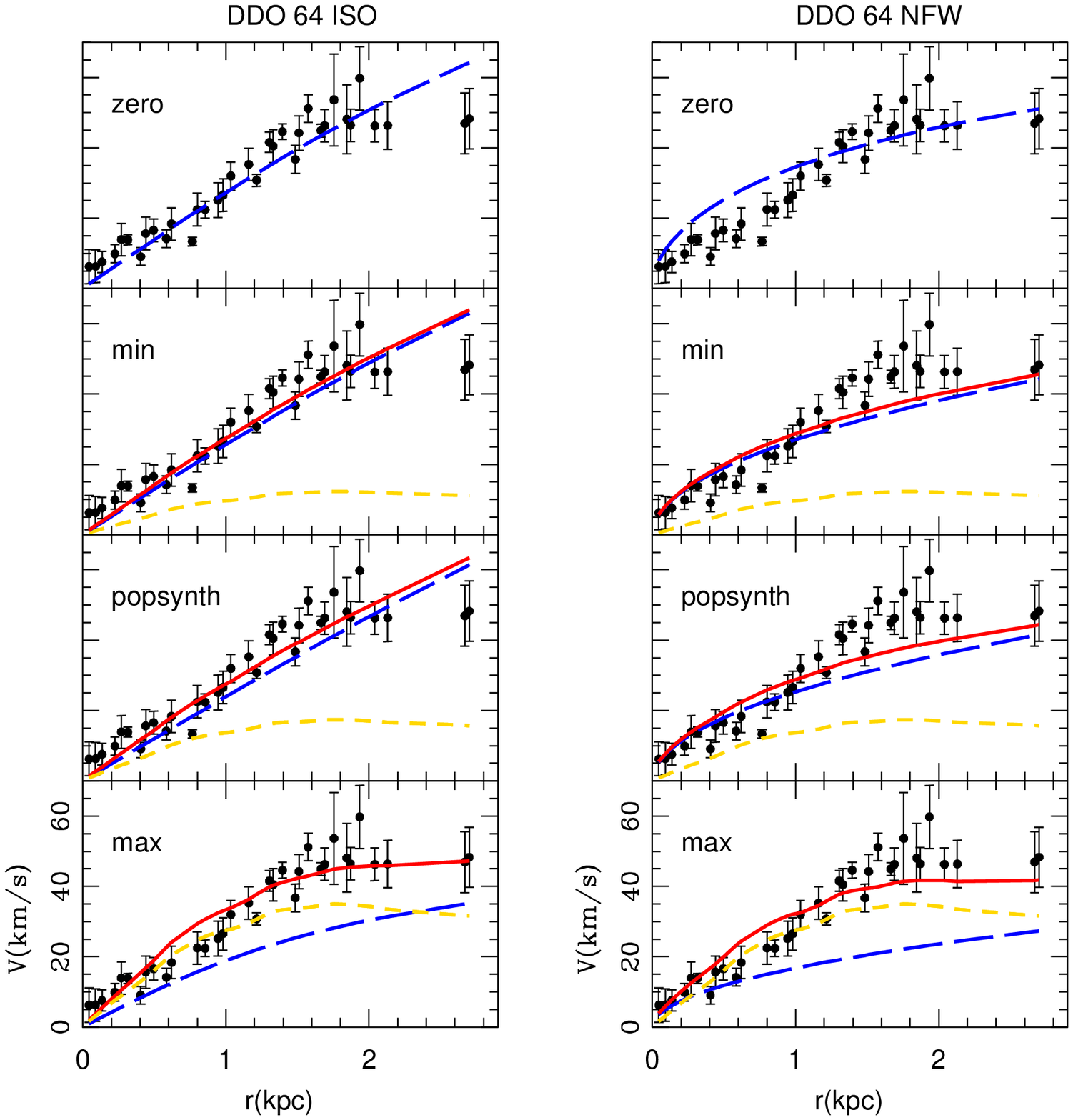}
\begin{quote}
\caption{Isothermal and NFW halo fits for DDO 64.  Line types are
  described in Figure 5. [{\it See the
  electronic edition of the Journal for a color version of this figure.}]}
\end{quote}
\end{figure*}

\begin{figure*}
\plotone{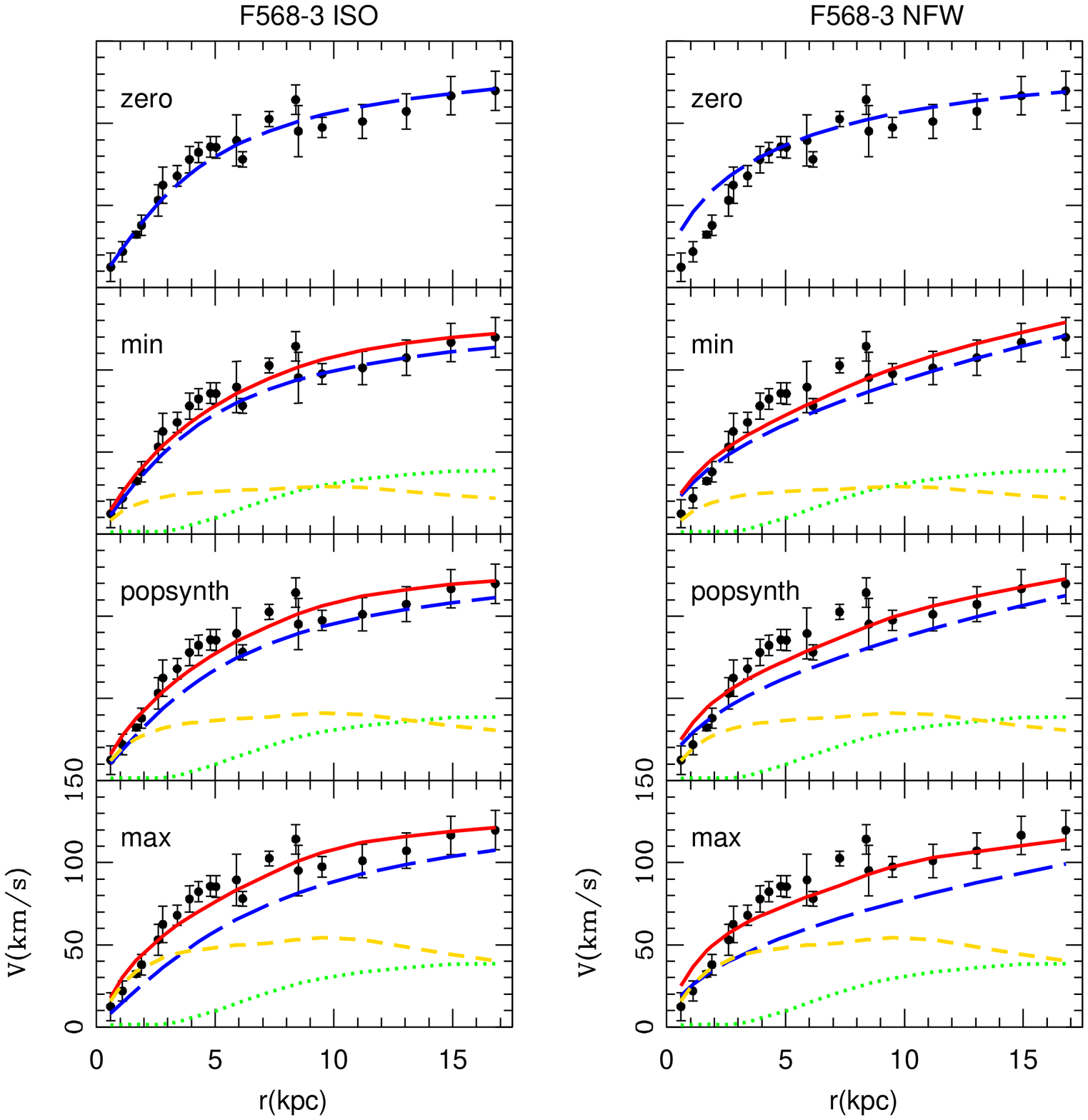}
\begin{quote}
\caption{Isothermal and NFW halo fits for F568-3.  Line types are
  described in Figure 5. [{\it See the
  electronic edition of the Journal for a color version of this figure.}]}
\end{quote}
\end{figure*}

\begin{figure*}
\plotone{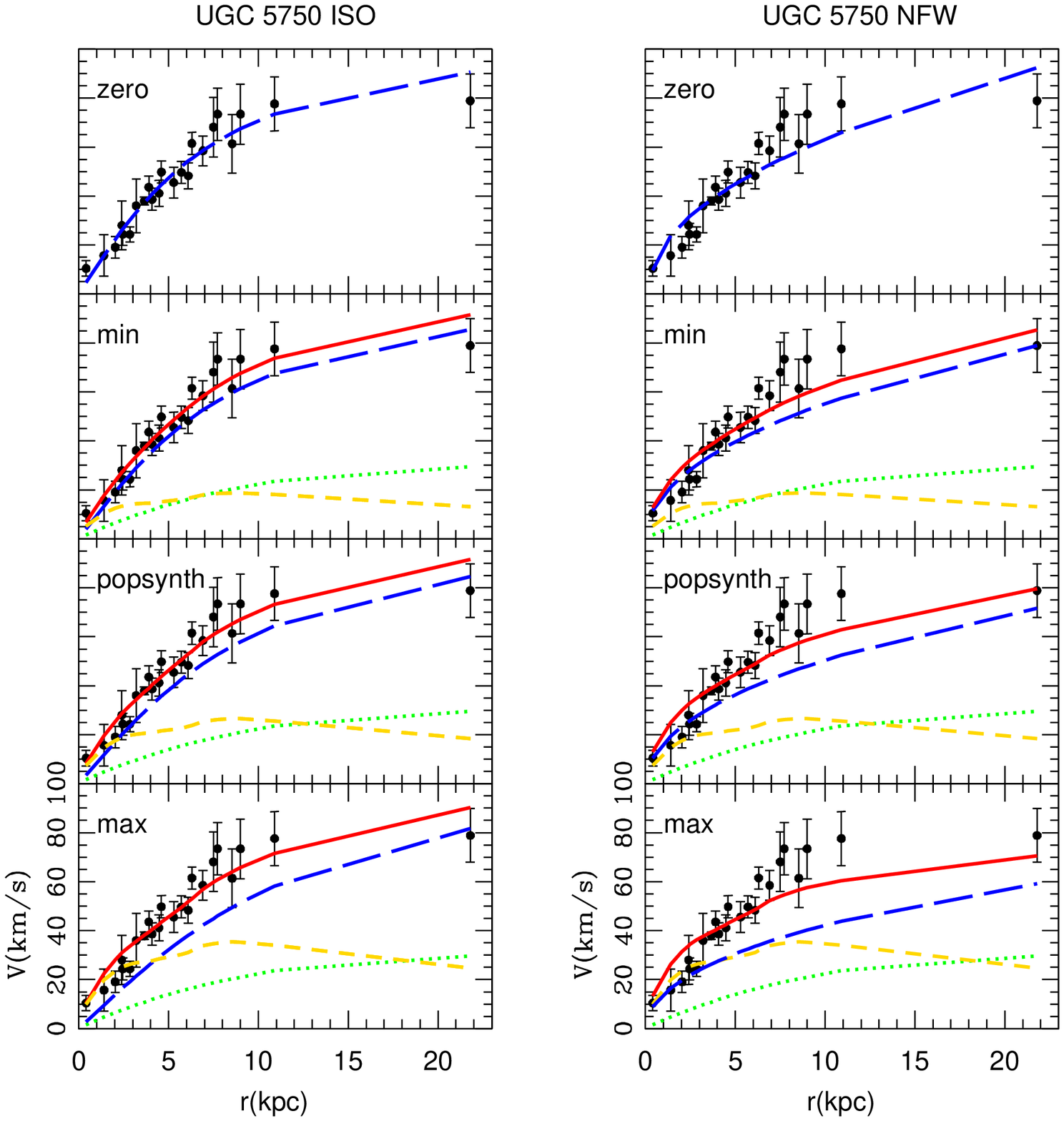}
\begin{quote}
\caption{Isothermal and NFW halo fits for
  UGC 5750.  Line types are described in Figure 5. [{\it See the
  electronic edition of the Journal for a color version of this figure.}]}
\end{quote}
\end{figure*}

\begin{figure*}
\plotone{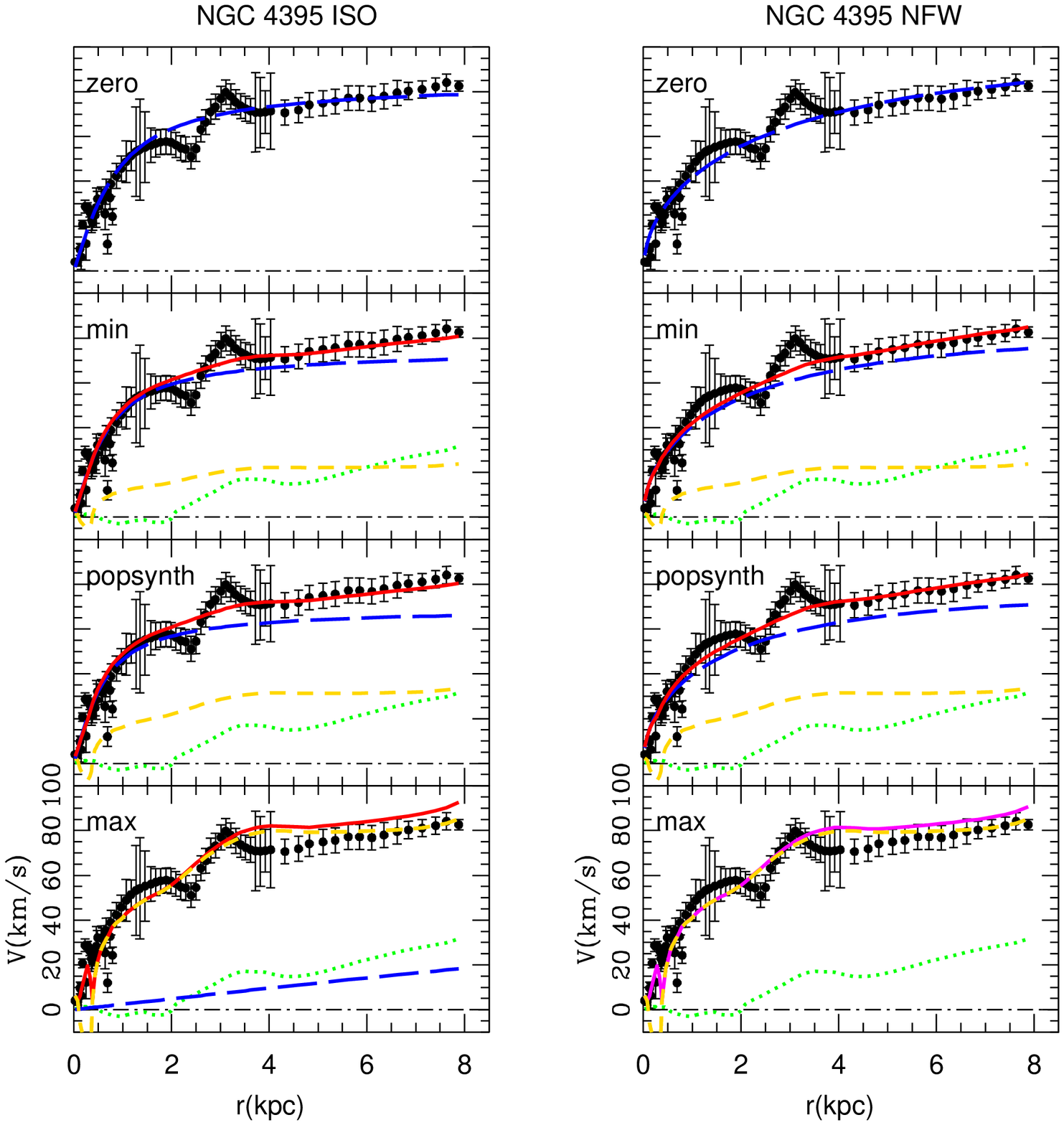}
\begin{quote}
\caption{Isothermal and NFW halo fits for
  NGC 4395.  Line types are described in Figure 5.  The solid magenta line
in the maximum disk NFW plot is the total baryonic rotation curve.  
See text for details. [{\it See the
  electronic edition of the Journal for a color version of this figure.}]}
\end{quote}
\end{figure*}

\begin{figure*}
\plotone{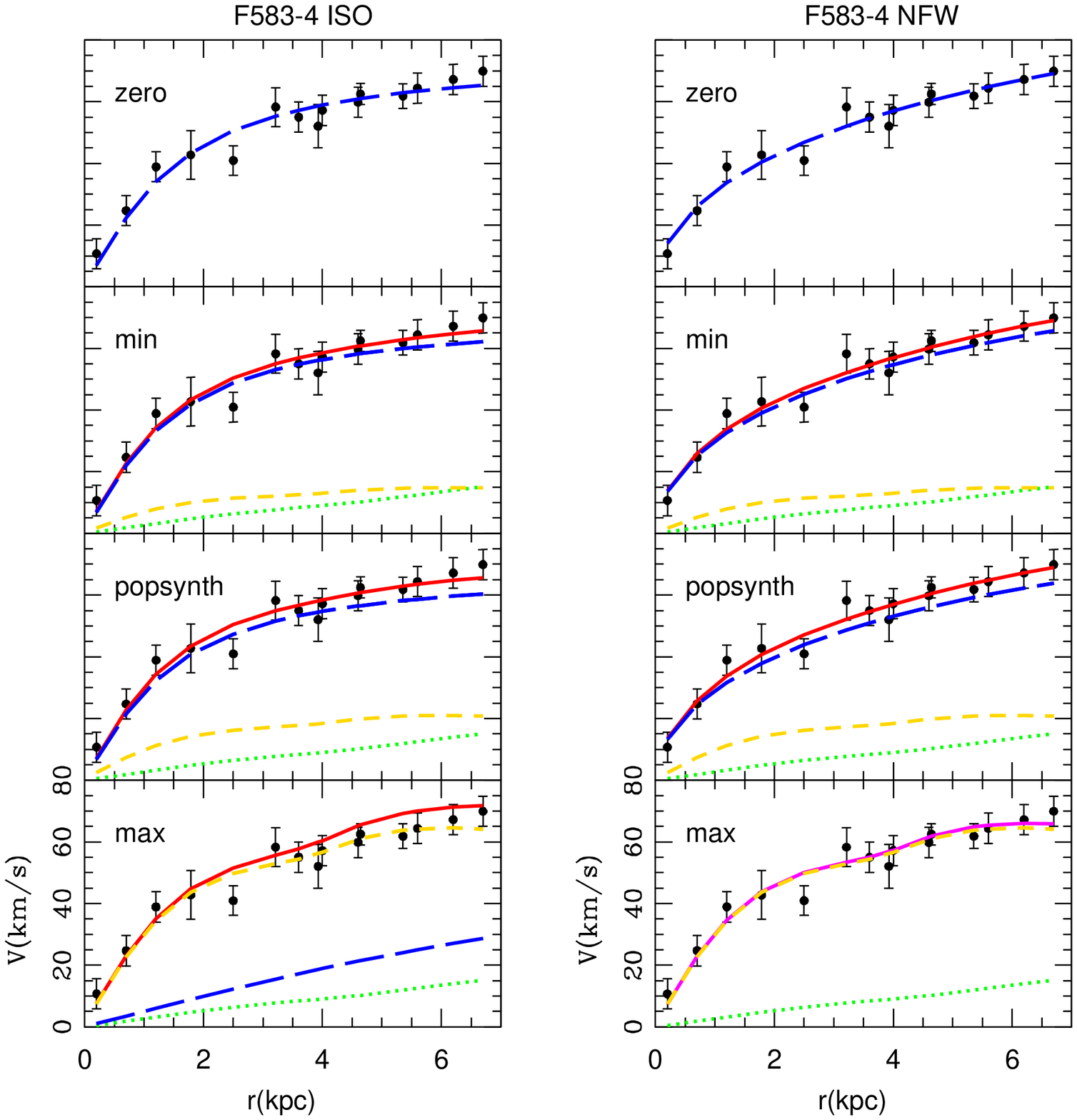}
\begin{quote}
\caption{Isothermal and NFW halo fits for
  F583-4.  Line types are described in Figure 5.  The solid magenta line
in the maximum disk NFW plot is the total baryonic rotation curve.  See
text for details.  [{\it See the
  electronic edition of the Journal for a color version of this figure.}]}
\end{quote}
\end{figure*}

\begin{figure*}
\plotone{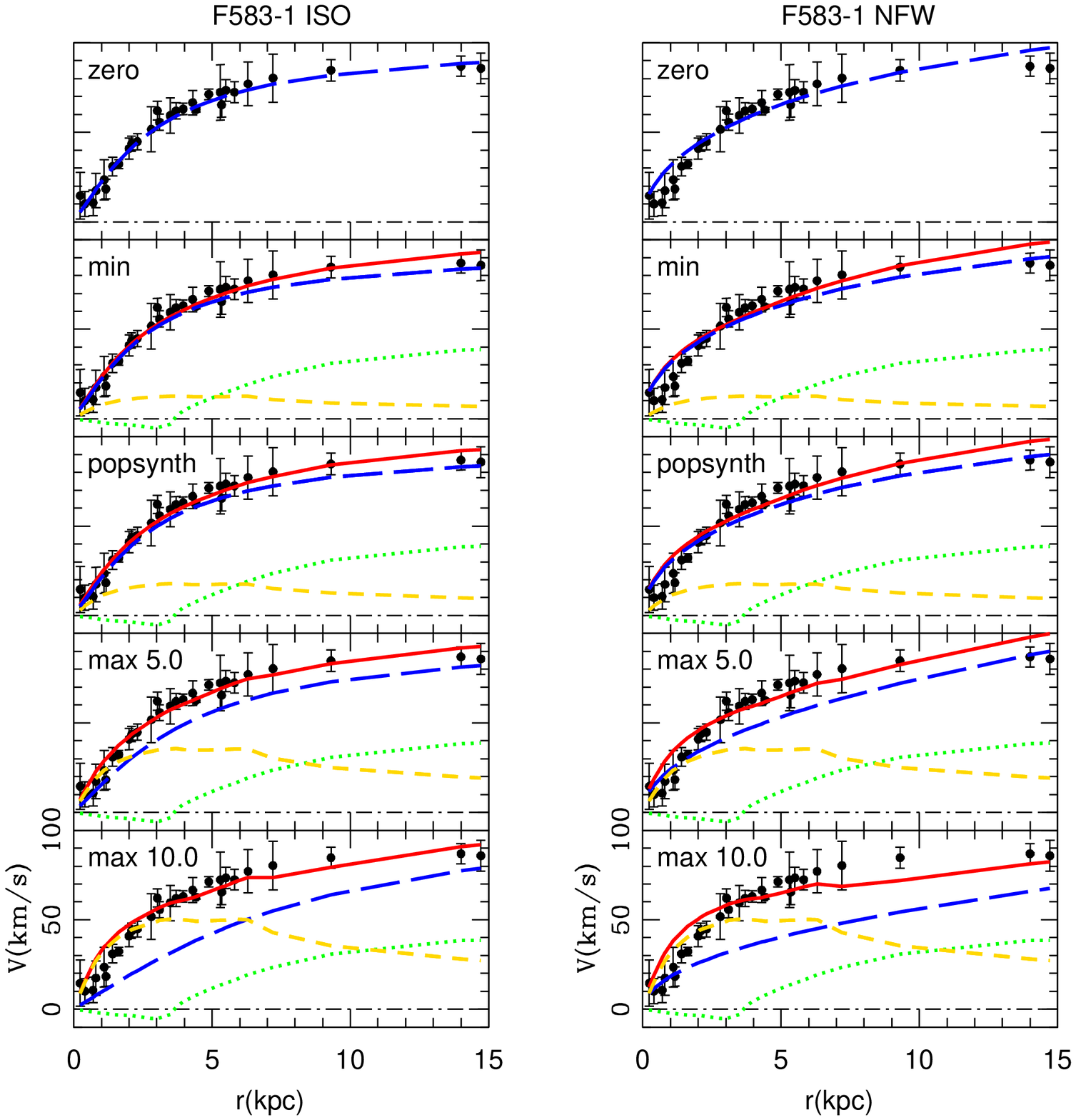}
\begin{quote}
\caption{Isothermal and NFW halo fits for F583-1.  Line types are
  described in Figure 5.  [{\it See the
  electronic edition of the Journal for a color version of this figure.}]}
\end{quote}
\end{figure*}

\begin{figure*}
\includegraphics[scale=0.35]{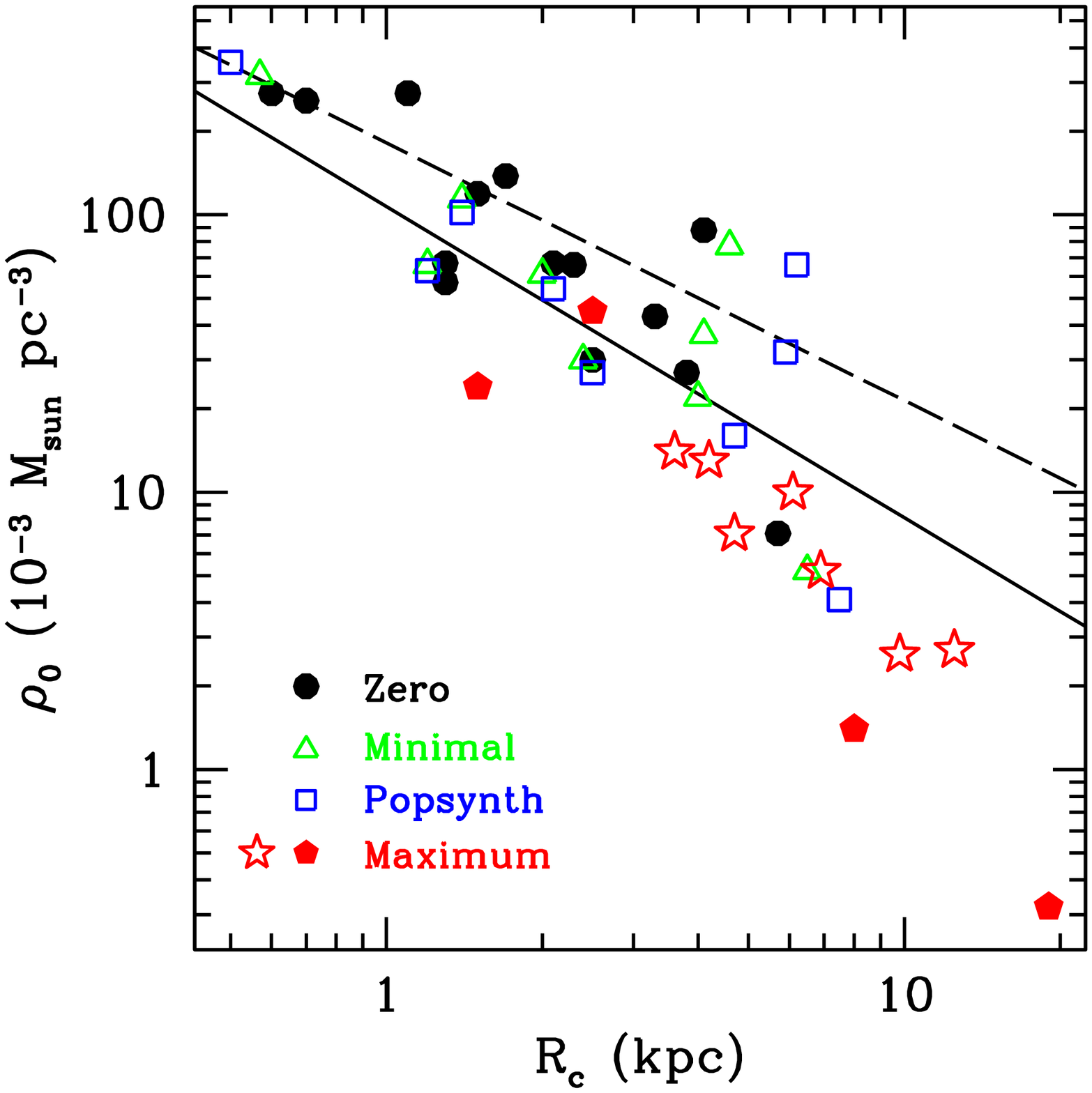}
\hfill
\includegraphics[scale=0.35]{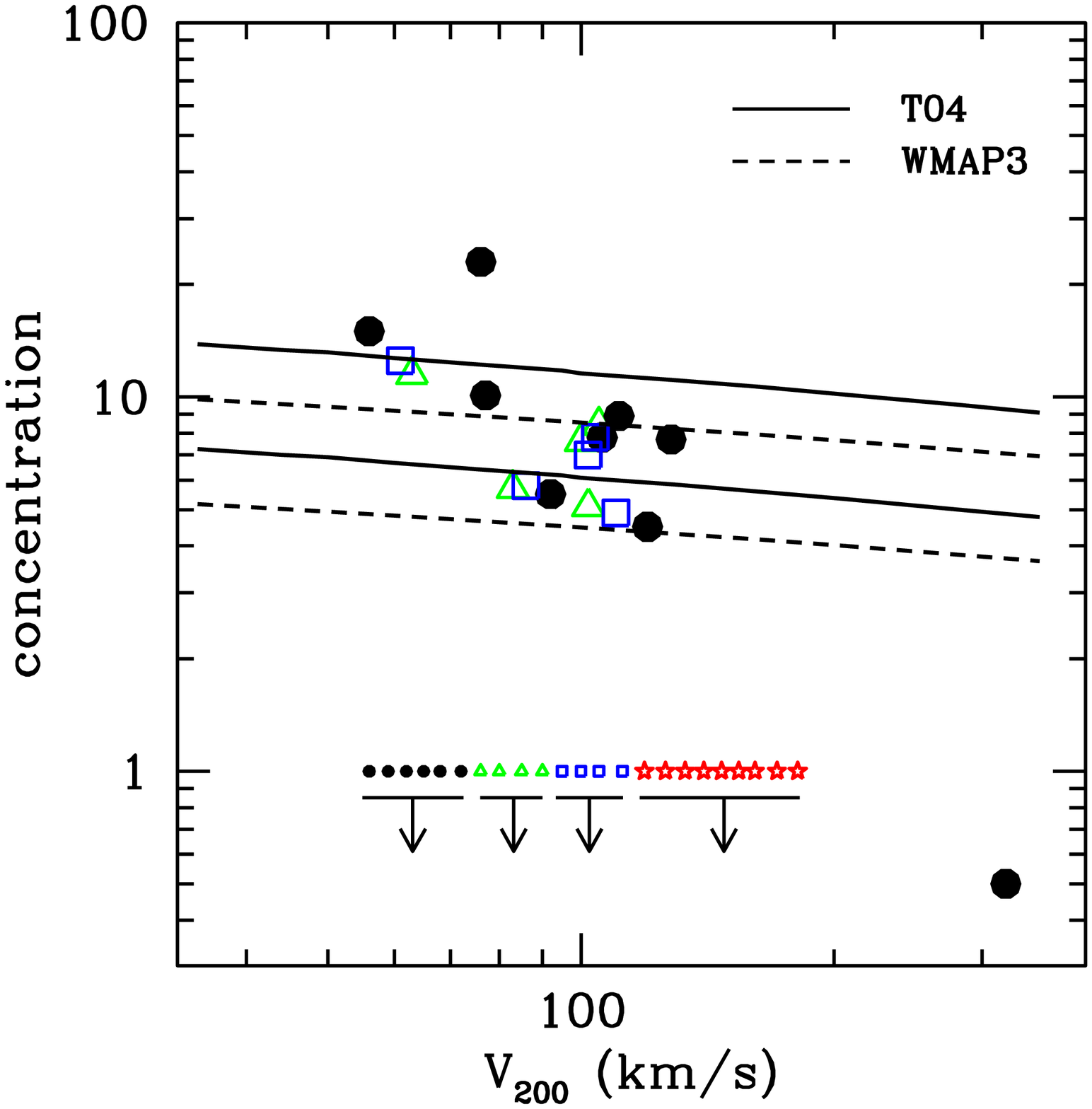}
\begin{quote}
\caption{$\textit{Left:}$ Isothermal halo parameters for different 
assumptions about the baryons.  The red pentagons are maximum disk 
fits where $R_{c}$ is a lower limit and $\rho_{0}$ is an upper limit.  
The data for NGC 959 are beyond the range of this plot and are not
shown.    The core radius is often $\gtrsim$ 1 kpc.  There is a trend towards
larger $R_{c}$ with increasing \ml.  For comparison, the solid line is
the $\rho_{0}$-$R_{c}$ scaling law of \citet{Kormendy} and the dashed line 
is the $\rho_{0}$-$R_{0}$ relation of \citet{Spano07}.
  $\textit{Right:}$ $NFW_{free}$ halo parameters for different 
assumptions about the baryons.  No NFW fits could be made in the 
maximum disk case.  The solid lines show the range of
$\textit{c-V$_{200}$}$ predicted by the cosmology of \citet{Tegmark}, 
whereas the dashed lines are the predictions for WMAP3
\citep{Spergel07}.  For both cosmologies, the width of the bands is 
 $\pm$1$\sigma$, assuming a scatter of $\sigma_{c}$ = 0.14 
\citep{Bullock01}.  The data show a steeper $\textit{c-V$_{200}$}$ 
relation, and favor the lower concentrations/lower $\sigma_{8}$ 
predicted by WMAP3.  [{\it See the
  electronic edition of the Journal for a color version of this figure.}]}
\end{quote}
\end{figure*}

\begin{figure*}
\includegraphics[scale=0.35]{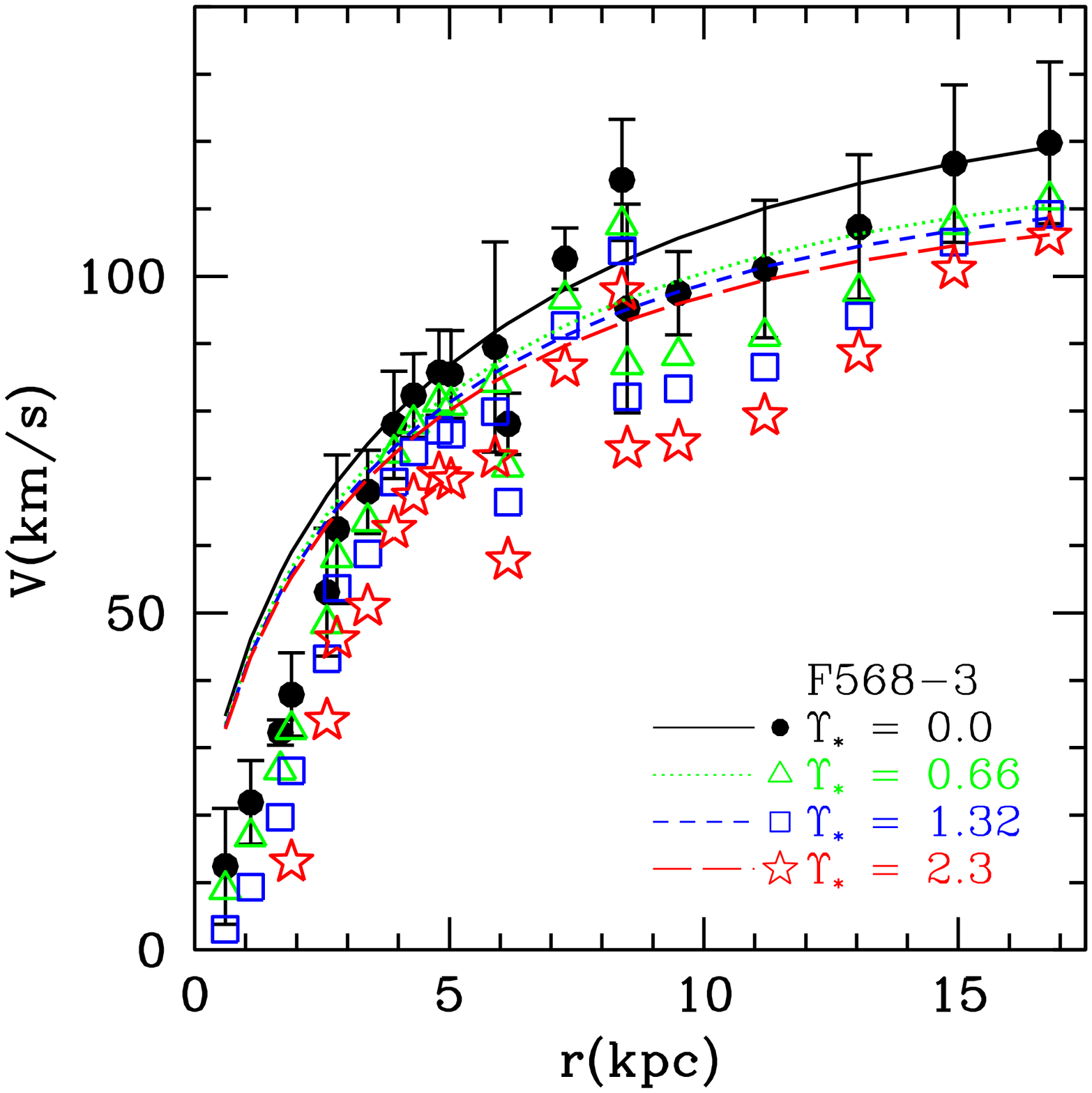}
\hfill
\includegraphics[scale=0.35]{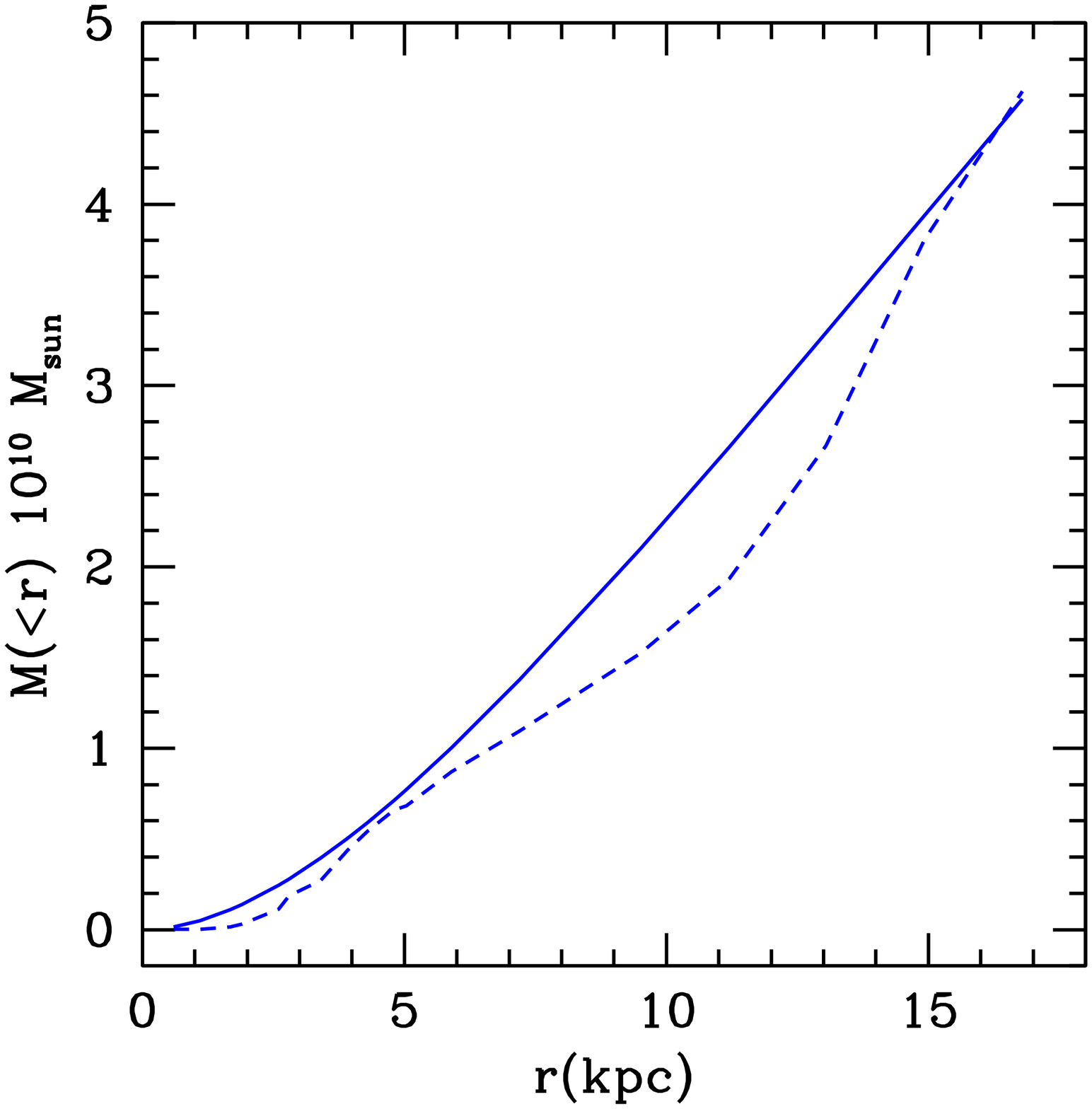}
\begin{quote}
\caption{$\textit{Left:}$ Comparison
  of the dark matter rotation curve for different values of \ml\ with
  the NFW rotation curves expected from cosmology for F568-3.  
  $\textit{Right:}$ Total dark
  matter mass as a function of radius for  \ml$_{POP}$.  The solid
  line is the dark matter mass predicted by CDM; the dashed line is
  the dark matter mass allowed by the data.  The two curves have been
  forced to meet at large radii; interior to this, CDM predicts more
  mass at all radii than is actually observed.  [{\it See the
  electronic edition of the Journal for a color version of this figure.}]}
\end{quote}
\end{figure*}

\begin{figure*}
\plotone{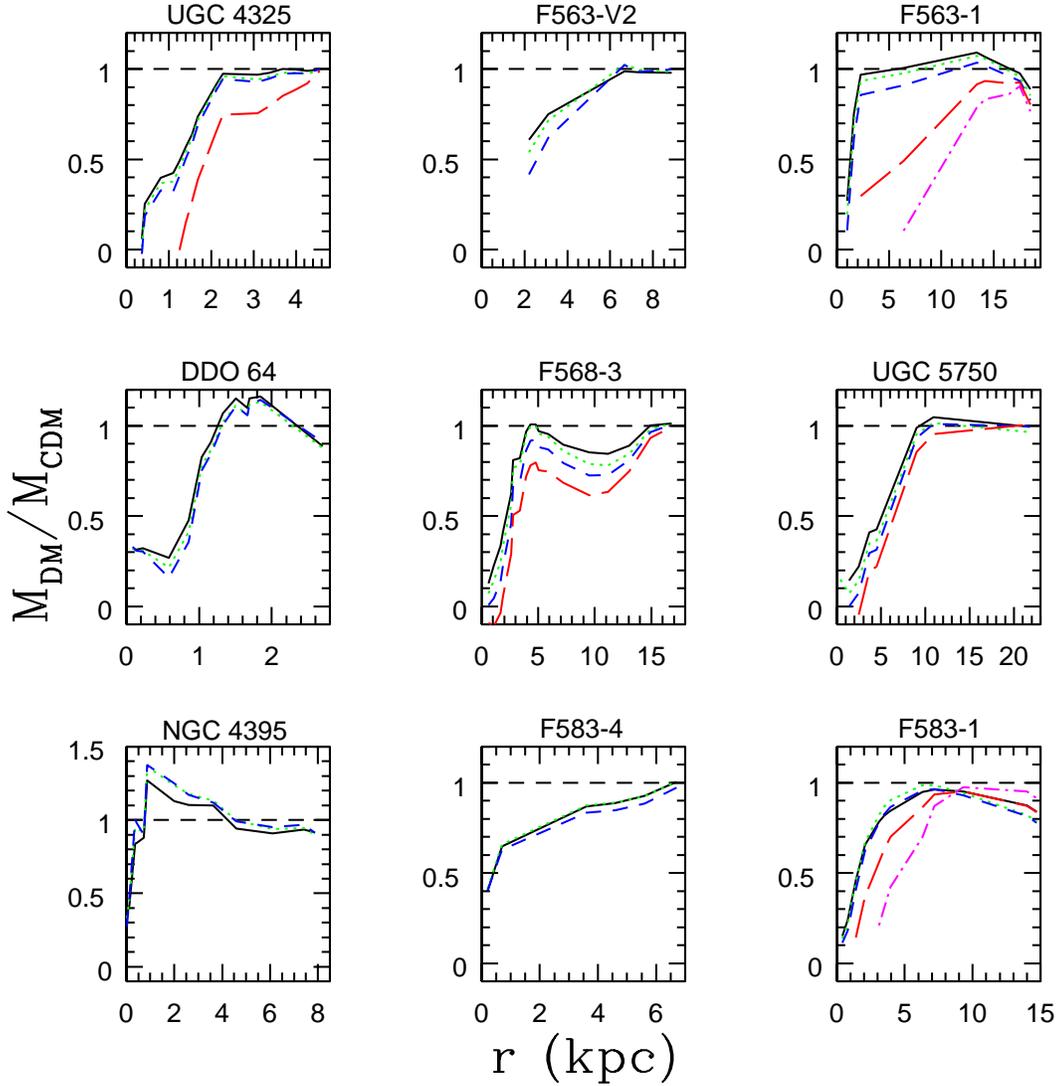}
\begin{quote}
\caption{Ratio of the observed dark matter mass
  to the constrained
cusp mass as a function of radius.  The black (solid), green (dotted),
blue (short dash), and red/magenta (long dash/dot-long dash) 
 lines are the zero, minimal, popsynth, and maximum disk
cases, respectively.  Near the centers of the galaxies there is a
substantial cusp mass excess with at least two times more mass
expected in the cuspy halo than is allowed by the data. [{\it See the
  electronic edition of the Journal for a color version of this figure.}]}
\end{quote}
\end{figure*}

\begin{figure*}
\plotone{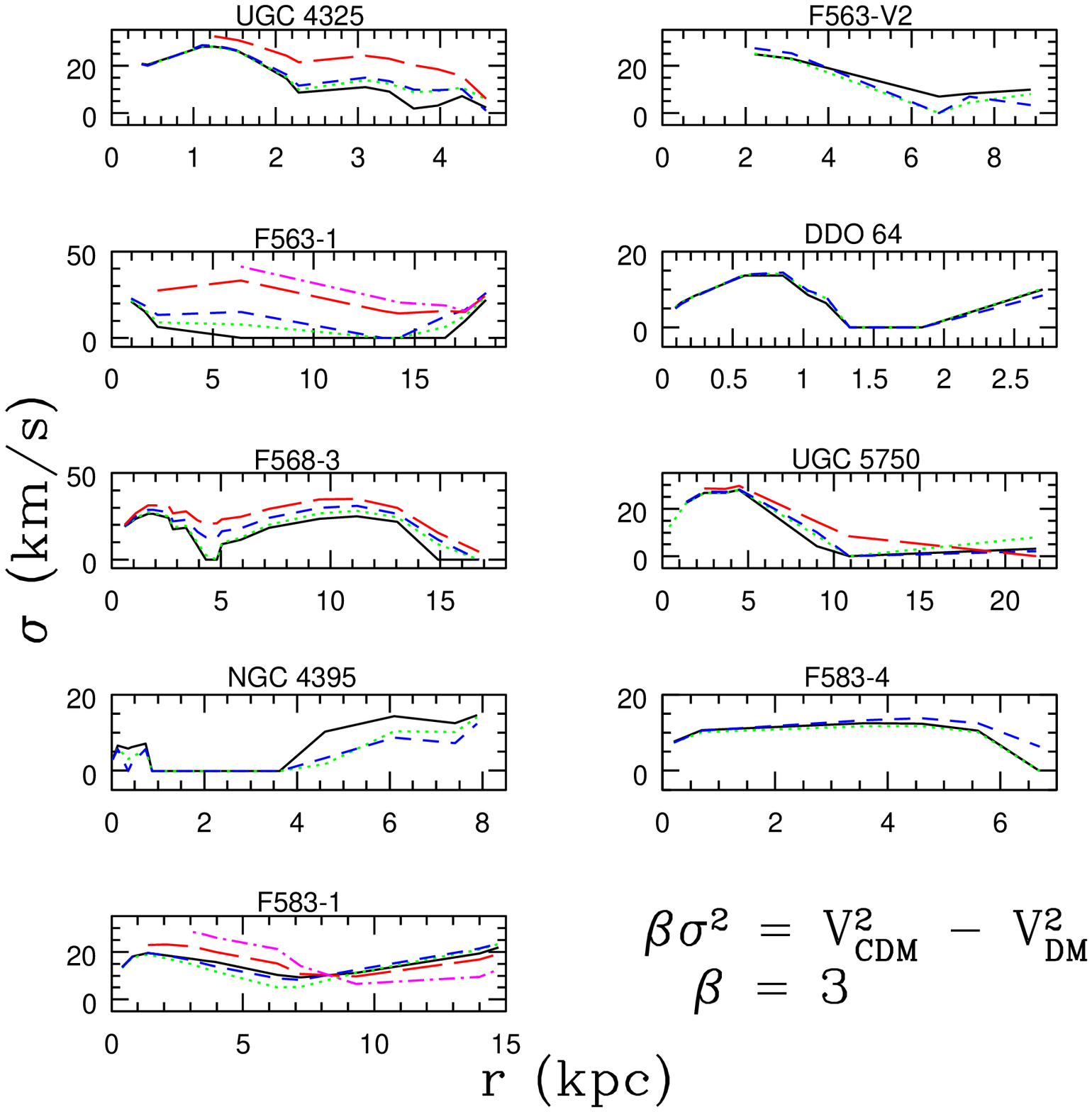}
\begin{quote}
\caption{The required noncircular motions, assuming an isotropic
  dispersion, to reconcile the difference between the $NFW_{constr}$
  velocity and the observed dark matter velocity.  The black (solid), 
  green (dotted), blue (short dash), and red/magenta (long dash/dot-long dash) 
 lines are the zero, minimal, popsynth, and maximum disk
cases, respectively. [{\it See the
  electronic edition of the Journal for a color version of this figure.}]}
\end{quote}
\end{figure*}

\subsubsection{Stars}
Galaxy photometry in combination with a stellar mass-to-light ratio,
\ml, is used to determine the stellar contribution to the observed
galaxy rotation.  It is difficult to determine the true value of  \ml\ 
of a galaxy because it depends on many factors ranging from the 
initial mass function to the extinction.  To address this uncertainty, 
we consider four different scenarios for the value of \ml.  We assume 
that \ml\ is constant with radius.

$\textbf{\textit{Zero disk}$-$}$  In this limiting case we ignore the 
contribution of the stars (\ml = 0) and gas and attribute all rotation
to dark matter.  This case is typically called `minimum disk' in the 
literature; we have chosen to refer to this case as zero disk because it aptly
describes the omission of the baryons.  The zero disk case puts an
upper limit on the slope and/or concentration of the halo density
profile.  The results for the zero disk case were presented in
\S\ 5 and also \citetalias{Kuzio} (where they were referred to as 
minimum-disk) and are reproduced in Tables 2 and 3 for completeness.   
Two galaxies from \citetalias{Kuzio}, UGC 1281 and UGC 477, are excluded 
from mass models beyond zero disk because of their high inclinations 
and possible associated line-of-sight integration effects.  Photometry 
is unavailable for the six new galaxies presented in \S\ 5; mass models 
beyond zero disk will not be presented.

$\textbf{\textit{Minimal disk}$-$}$  Though commonly employed, the 
zero disk case is unphysical.  We next consider a more realistic minimal 
contribution of the stars and the gas.  The \ml\ is determined from 
population synthesis models (see below) and then scaled by 0.5 
\citep[see][]{McGaugh05} to simulate a lightweight IMF.  

$\textbf{\textit{Popsynth}$-$}$  Here \ml\ is determined using the population 
synthesis models of \citet{Belletal03}.  Specifically, we use the 
relation between \ml\ and color as defined in their Table 7.  For a 
scaled Salpeter IMF, ($B$--$R$) 
and (\bv) colors are related to \ml.  This is our best estimate of the 
baryonic mass from the perspective of stellar populations.  The popsynth 
model also includes the gas contribution.  Colors and corresponding \ml\ 
are listed in Table 4.

$\textbf{\textit{Maximum disk}$-$}$ We also consider the case where the
\ml\ is scaled up as far as the data will allow.  In high surface
brightness galaxies, this approach usually fits the inner rotation
curve well.  In LSB galaxies, the shape of the stellar rotation curve
is often not well-matched to the observed rotation curve.  We have
chosen a \ml\ that allows the stellar rotation curve to match the
inner rotation curve points as well as possible, sometimes
overshooting the innermost point in order to hit the next few.  
These \ml\ are usually 
higher than those determined from the popsynth models, and in some cases 
(eg. F583-4) are substantially higher.  

To model the stellar disk we have used the $R$-band photometry
presented in \citet{dBetal95}, \citet{Stil}, 
\citet{Swatersthesis}, and \citet{Swatersetal02}.  The \texttt{GIPSY} task
\texttt{ROTMOD} was used to 
determine the rotation of the disk assuming a vertical sech$^{2}$ 
distribution with a scale height $z$$_{0}$=$h$/6
\citep{vanderKruit81}.  The stellar rotation curve computed from the 
photometry was resampled at
the same radii as the combined DensePak+long-slit+\HI\ rotation curves.  

Each estimator of \ml\ has its advantages and disadvantages.  The zero 
disk \ml\ is often adopted and is useful because it provides an upper 
limit on the dark matter.  Unfortunately, it is an unphysical 
assumption.  Population synthesis models best represent what we know 
about stars, however, sometimes the data will allow a larger \ml.  
And although \ml$_{MAX}$ often seems too large with respect to 
\ml$_{POP}$, disk features like bars and spiral arms require a large 
disk mass \citep{McGdB98, Fuchs} and support high \ml$_{MAX}$.

\subsubsection{Gas}
The gas present in galaxies also contributes to the observed galaxy
rotation.  \HI\ is the dominant gas component, but to include helium
and metals, the \HI\ data were scaled by a factor of 1.4.  
Substantial amounts of molecular gas are not obviously present in LSB 
galaxies \citep{dBvdH, Schombert}.  The total mass of H$_{2}$ is 
almost certainly much less than that of \HI\ \citep*{Mihos}.  
Moreover, in brighter galaxies H$_{2}$ is known to trace the stars 
\citep{Regan}, so at most represents a slight tweak to \ml.   The
contribution of the gas to the observed velocity is considered in the
minimal, popsynth, and maximum disk cases.  

The \HI\ surface density profiles presented in \citet{DMV}, 
\citet{vanderHulst}, and \citet{Swatersthesis} were used to model the
gas disk.  The \texttt{GIPSY} task \texttt{ROTMOD} was used to 
determine the rotation of the disk assuming a thin disk.  The gas
rotation curve was resampled at the same radii as the combined 
DensePak+long-slit+\HI\ rotation curves.  

\subsubsection{Dark Matter Halo}
In LSB galaxies there is usually a considerable amount of observed rotation 
unaccounted for after subtracting off the velocity of the stellar and gas 
disks, even in the maximum disk case.  The remaining rotation is 
usually attributed to dark matter.  While there have been a number of dark 
matter models proposed in the literature, we fit two of the most prominent 
competing profiles: the pseudo-isothermal halo and the NFW profile.  
Both halo profiles are described in \S\ 5.1.

\section{Mass Model Results for Individual Galaxies}
In this section, we present the isothermal and NFW halo fits for the 
four scenarios of \ml.  The halo parameters are listed in Tables 2
and 3.  Figures 5-13 plot the halo fits over the data. 

As \ml\ increases and the baryons become responsible for more of the
observed velocity, less room is available for dark matter and the NFW
halo becomes increasingly difficult to fit to the data.  The best-fit 
concentrations drop to very small, and sometimes negative, values.  In
these cases, we forced an NFW fit with $c$ = 1.0.  The halo parameters
of forced fits are italicised in the tables.

$\textbf{\textit{UGC 4325}$-$}$  UGC 4325 is clearly best described by
an isothermal halo; an NFW halo could not be fit to the data for any
\ml\ scenario.  With  \ml$_{MAX}$ = 4.5, the stellar rotation curve
is able to describe the data well out to a radius of $\sim$ 1.3 kpc, as
compared to $\lesssim$ 0.4 kpc for \ml$_{POP}$ = 1.14.  Because the
baryons are able to explain so much of the data in the maximum disk
case, there is little room left for a dark matter isothermal halo.
When this happens, the cored halo becomes very nearly hollow with
$\rho_{0}$ decreasing to a very small value and $R_{c}$ increasing to a
large number.  For the maximum disk case of UGC 4325, we forced an
isothermal halo fit by fixing $V_{h}$ to approximately the observed
velocities of the outer rotation curve and then varying $R_{c}$ such 
that $V_{TOT}$ follows the data.  In this fit, the resulting $R_{c}$ 
is a lower limit and $\rho_{0}$ is an upper limit.

$\textbf{\textit{F563-V2}$-$}$  As discussed in \citetalias{Kuzio} for the 
zero disk case, F563-V2 has too few data points to really distinguish 
between halo types.  This remains true for the other \ml\ scenarios as well.  
 The value of \ml\ can be turned up to 4.0 in the maximum disk
case, leaving essentially no room for an NFW halo; a fit was forced
with $c$ = 1.0.  NFW halos could be fit to the zero, minimal, and 
popsynth cases, however, and the values of the best-fitting 
concentrations, $c$ = 7.7, 8.4, and 7.8, respectively, are comparable 
to values expected from simulations.  Because of the $c$-$V_{200}$ 
degeneracy that allows halos of different ($c$, $V_{200}$) to look the 
same over a finite range of radius \citep{dBMR}, these three NFW fits 
are essentially indistinguishable.  This is an example of where the 
zero disk assumption is reasonable for LSB galaxies.  

$\textbf{\textit{F563-1}$-$}$  The isothermal halo fits the F563-1
data better than the NFW halo, though the values of the best-fitting
concentrations are consistent with values expected in $\Lambda$CDM.
The baryons can account for the majority of the observed rotation in
the maximum disk case and a forced NFW fit was made with the
concentration fixed at 1.0.  Stellar population models set \ml$_{POP}$
= 1.36 for F563-1.  With this value of the \ml, the stellar rotation
curve just grazes the innermost observed rotation curve point.  The
\ml$_{MAX}$ can be substantially turned up such that the stellar
rotation curve matches the data more closely.  If \ml$_{MAX}$ = 6.5,
$V_{*}$ overshoots the innermost rotation curve point, but matches the
cluster of points at $\sim$ 1.5 kpc and even crosses the lower
errorbars on points between 5 kpc and 10 kpc.  The shape of the
observed rotation curve is well-matched when \ml$_{MAX}$ = 10.0;
$V_{*}$ goes through the upper errorbars on points $\lesssim$ 2 kpc
and agrees well with the data between 5 kpc and 10 kpc.  Fitting the
inner observed rotation curve points or the overall rotation curve
shape are both equally plausible approaches to defining \ml$_{MAX}$ 
\citep{Palunas}, so we use both values in our halo fits.

F563-1 is a good example of how the choice of \ml\ affects the mass 
models, and the degeneracy between the luminous and dark components.  
While \ml$_{POP}$ is our best estimate for the stars, the data will 
clearly allow a higher \ml.  To constrain the degeneracy between the 
stars and dark matter in cases like this, we need more information.  
The stellar velocity dispersion perpendicular to the disk, for 
instance, would help to put limits on the disk mass \citep{Verheijen07}.

$\textbf{\textit{DDO 64}$-$}$  As \HI\ surface density profiles were 
unavailable for this galaxy, the baryons in the mass models are
represented by the stars only.  Multicolor photometry was also
unavailable, so a ($B$--$R$) = 0.9 color was assumed for this dwarf
galaxy \citep{dBetal95}.   DDO 64 is better described by the 
isothermal halo than 
the NFW halo; forced $c$ = 1 NFW fits are made for each value of \ml.  The
stellar rotation curve falls below the observed rotation curve with
the exception of two low points in the popsynth case, \ml$_{POP}$ =
1.24.  In the maximum disk case with \ml$_{MAX}$ = 5.0, however,
$V_{*}$ is able to follow the data very well out to just past 1 kpc,
and is even consistent with a data point at $\sim$ 1.5 kpc.  With
\ml$_{MAX}$ = 5.0, the stars are able to account for most of the
observed rotation, and the displayed isothermal halo fit is the upper
limit on $\rho_{0}$ and the lower limit on $R_{c}$.  

$\textbf{\textit{F568-3}$-$}$  F568-3 is fit well by isothermal halos;
only forced NFW fits could be made to the data.  The shape of the
stellar rotation curve is not well-matched to the observed rotation
curve and is only able to describe the inner 2 kpc of data, even in
the maximum disk case.

$\textbf{\textit{UGC 5750}$-$}$  Because multicolor photometry was
unavailable, a ($B$--$R$) = 0.9 color was assumed for this 
galaxy \citep{dBetal95}.  Excellent isothermal fits were made for 
UGC 5750; the NFW
halo provides very poor fits.  Like F568-3, there is not a substantial
difference between \ml$_{POP}$ and \ml$_{MAX}$.

$\textbf{\textit{NGC 4395}$-$}$  Multicolor photometry was unavailable
for this galaxy and \ml$_{POP}$ = 1.4 \citep{dBB} was assumed for 
the popsynth
model.  The NFW halo is a slightly better fit to NGC 4395 than the
isothermal halo, and the best-fitting concentrations are consistent
with expectations from  $\Lambda$CDM.  There is a substantial
difference between \ml$_{POP}$ and \ml$_{MAX}$.  For \ml$_{POP}$ =
1.4, $V_{*}$ is well below the observed rotation curve; however, for
\ml$_{MAX}$ = 9.0, $V_{*}$ is able to trace the data out to 8 kpc, the
entire length of the rotation curve.  Because the baryons can explain
the observed rotation so well in the maximum disk case, the data want
an isothermal halo with an almost hollow core.  We force a fit with an
upper limit on $\rho_{0}$ and a lower limit on $R_{c}$.  The baryons
do such a good job of explaining the observed rotation in the maximum
disk case that there is not even room for an NFW halo with $c$ = 1.0.
In the NFW maximum disk plot in Figure 11, the magenta line represents
the total baryonic rotation curve, $V_{disk}^{2}$ = $V_{*}^{2}$ + 
$V_{gas}^{2}$.  It should be noted that the nucleus of NGC 4395 is the 
least luminous known Seyfert 1 \citep{Filippenko} and may  
influence the inner rotation curve that we derive.  

$\textbf{\textit{F583-4}$-$}$  The isothermal halo is a good
description of the data.  The NFW halo fits are comparable, although
the values of the concentrations are on the low side of expected
values from simulations.  The stellar rotation curve is far below the
data at all radii for the popsynth \ml$_{POP}$ = 1.06.  The entire
observed rotation curve can be well described by the stellar rotation
curve when \ml$_{MAX}$ is turned up to 10.0.  In this case, there is
very little room left for a dark matter halo, and only a limiting
isothermal halo fit is made.  Like NGC 4395, there is not enough
velocity left for an NFW halo with $c$ = 1.0, and the magenta line in
the maximum disk NFW plot in Figure 12 represents the total baryonic
rotation curve.

$\textbf{\textit{F583-1}$-$}$  F583-1 is better fit by the isothermal
halo than the NFW halo.  Additionally, the best-fitting NFW
concentrations are on the low side of expected values from
$\Lambda$CDM.  Like F563-1, we consider two values of \ml$_{MAX}$.
With \ml$_{MAX}$ = 5.0, $V_{*}$ goes through the data within $\sim$
1.5 kpc.  $V_{*}$ goes through the upper errorbars on the inner
rotation curve, through the data at $\sim$ 2 kpc, and then through the
lower errorbars on the data out to $\sim$ 3.5 kpc when \ml$_{MAX}$ = 10.0.

\section{Discussion}

While one would like to know the true \ml\ for each galaxy, our data 
do not indicate that a particular estimator of \ml\ is any better than 
another.  In Figure 14 we show that the parameters of both the 
isothermal and NFW halo fits do not change much as \ml\ changes.  
This confirms that the details about what is assumed for the stars 
in LSB galaxies do not really matter.  Unfortunately, it also means 
that without additional information, we cannot constrain \ml\ in 
galaxies like F563-1 and F583-1 where a wide range of \ml\ are applicable.

Though the exact assumption about the stars may be unimportant, the
stars cannot be entirely ignored.  In reality, stars do not
have zero mass.  In fact, as the velocity contribution from the stars
becomes more important (ie. as \ml\ goes up), there is 
less room for dark matter at the centers of the galaxies.  The left 
panel of Figure 14 illustrates that the isothermal core radius is 
often larger than $\sim$ 1 kpc.  There is also a trend towards 
larger $R_{c}$ with increasing \ml.  This is important to recognize, 
as it shows that the cusp-core problem is not restricted to the 
innermost radii only, particularly when stars are allowed to have 
mass.  Also plotted in this figure is the $\rho_{0}$-$R_{c}$ scaling 
relation of \citet{Kormendy} (their equation 20) and the $\rho_{0}$-$R_{0}$ 
relation determined by \citet{Spano07} (their figure 2).  The \citet{Kormendy} 
relation follows the data more closely than the \citet{Spano07} relation, 
however the maximum disk case points tend to drift below the \citet{Kormendy} 
relation. 

The $\textit{c-V$_{200}$}$ plot in the right 
panel of Figure 14 
similarly shows the difficulty of fitting the centrally concentrated 
NFW halo to the data as the velocity contribution from the baryons 
becomes larger: as \ml\ increases, fewer galaxies can be fitted with 
an NFW halo.  In particular, the velocity contribution from the 
baryons is significant enough in the maximum disk case that maximum 
disk and the NFW halo are mutually exclusive.

For comparison, we have 
also included the predicted $\textit{c-V$_{200}$}$ lines for the  
high $\sigma_{8}$ cosmology of 
\citet[][hereafter T04]{Tegmark} and the low $\sigma_{8}$ WMAP3 cosmology \citep{Spergel07}.  
The width of the bands is $\pm$1$\sigma$, assuming a scatter of 
$\sigma_{c}$ = 0.14 \citep{Bullock01}.  The data appear to follow a steeper 
$\textit{c-V$_{200}$}$ relation than both predictions 
\citep[see also][]{McGaugh07}, and the concentrations of the \Dpak\ galaxies 
with NFW fits primarily cluster between the WMAP3 lines.  The median 
concentration of $\textit{all}$ the zero disk \Dpak\ galaxies is
$c$=4.5.  At the corresponding $V_{200}$, the median $c$ expected 
from T04 is $\sim$8.2, and the median $c$ expected from WMAP3 is 
$\sim$6.1.  An alternative method of measuring the halo central 
density is the $\Delta_{V/2}$ approach proposed by \citet*{Alam}.  
The median $\Delta_{V/2}$ of all the zero disk \Dpak\ galaxies is 
$\sim$2.9$\times$10$^{5}$.  At the corresponding $V_{max}$, the 
median $\Delta_{V/2}$ expected from T04 is $\sim$4.5$\times$10$^{5}$ 
and is $\sim$1.8$\times$10$^{5}$ from WMAP3.  The low 
concentrations/central densities observed in the data are more 
consistent with a power spectrum having a lower amplitude on small 
(galaxy) scales \citep[see also][]{McGaugh03}.

\subsection{Cusp Mass Excess}

In \S\ 5.2 we defined a constrained NFW halo, $NFW_{constr}$, 
which was constructed to have a cosmologically-consistent 
concentration by forcing the halo to match the velocities at the outer 
radii of each galaxy.  We can now find $NFW_{constr}$ halos for the 
dark matter rotation curves for each determination of \ml.  The 
rotation curves of the $NFW_{constr}$ halos over-predict the observed 
velocities interior to where they are forced to agree.  We can 
evaluate the difference between this expected CDM rotation curve and 
the observed dark matter rotation curve in terms of velocity 
difference, or alternatively, as a cusp mass excess.  We can ask what 
the difference is between the expected cuspy NFW halo mass and the 
dark matter mass that is allowed by the data.  In the left panel of 
Figure 15, we show the observed dark
matter rotation curves of F568-3 for the zero, minimal, popsynth, and
maximum disk models.  We also show the constrained NFW halo for each dark
matter rotation curve.  In the right panel of Figure 15, we plot the
same data in terms of mass using $M$ = $V^{2}$$R$/$G$.  To prevent a
cluttered plot, we show only the results for the popsynth model.  The
solid line is the enclosed mass as a function of radius for the
$NFW_{constr}$ halo.  The dashed line is the enclosed mass of the
observed dark matter.  We have not smoothed the data, so $V_{DM}$ is a
bit jittery between $\sim$ 6-8 kpc; $\textit{M(r)}$, however, is 
displayed using a smoother version of the data.  The $NFW_{constr}$ 
halo has been forced to match the data at large radii; interior to 
this, however, there is less mass observed at all radii than is expected.

In Figure 16 we plot the ratio of the observed dark matter mass to 
the constrained cusp mass as a function of radius for each galaxy.  
This ratio approaches 1 at large radii where the data have been forced 
to agree.  Near the centers of the galaxies there is a substantial 
cusp mass excess; there is at least two times more mass expected in 
the cuspy halo than is allowed by the data.  At all radii, the cusp 
mass excess becomes larger as the baryons become increasingly more 
important.

\subsection{Reconciling the Cusp Mass Excess with Noncircular Motions}

In our analysis so far, we have assumed that the observed velocity is 
the circular velocity ($V_{circ}$ = $V_{obsv}$).  If noncircular 
motions are present, the true circular velocity may be 
underestimated.  This is the argument commonly given to explain the 
discrepancy between the NFW halo and the observations at small radii 
\citep[e.g.][]{Rob, vandenB01}.  To determine the true circular 
velocity, the noncircular motions are added in quadrature to the 
observed velocities, $V^{2}_{circ}$ = $V^{2}_{rot}$ + 
$\beta$$\sigma^{2}$, usually assuming an isotropic dispersion 
($\beta$ = 3).  

We can invert Figure 16 and determine how large the noncircular 
motions must be in order to bring the observed dark matter velocity 
into agreement with the $NFW_{constr}$ velocity.  In Figure 17  we 
plot the required $\sigma$$(r)$, assuming an isotropic dispersion, for 
each galaxy.  In general, most of the galaxies need about 20 \kms\ 
noncircular motions at inner radii for the data to be consistent with 
the expectations of CDM.  This is roughly twice as large as the 
velocity dispersion in the \Dpak\ data \citepalias{Kuzio}.  It is also 
important to recognize the effect of stars not having zero mass.  
Instead of just having to convince ourselves that $\sim$20 \kms\ 
noncircular motions should be added to the data, we actually must be 
willing to consider noncircular motions an additional 5-10 \kms\ 
higher when \ml\ $\neq$ 0.  Because we have forced the halo velocity 
to match the observed velocity at large radius, our derived 
$\sigma$$(r)$ often show a steep drop to 0 \kms\ from $\sim$ 20 \kms.
Strictly speaking, this behavior is unphysical, and a more continuous 
model could be derived.  If $\sigma$ contributes at large radii, 
however, that implies that the halo has even more dark matter and 
$\sigma$ at small radii would have to be even higher.  In a 
forthcoming paper, we model the NFW halo and noncircular motions, 
and Figure 17 provides us with testable predictions.  For instance, 
we can use the results to test whether the observations are consistent 
with $NFW$+$\sigma(r)$.

\section{Conclusions}

We have presented  updated \Dpak\ velocity fields, rotation 
curves and zero disk halo fits for three galaxies from \citetalias{Kuzio}. 
We have also presented the velocity fields, derived
rotation curves and zero disk halo fits for six galaxies previously 
unobserved with \Dpak.   Overall, we find the isothermal halo to be 
a better description of both the shape of the rotation curves and the 
central densities of these nine galaxies in the limit of zero disk 
than the NFW halo.  When NFW fits could be made, the concentrations 
were often beyond the range of values expected for $\Lambda$CDM.  We 
also find that the quality of both the isothermal and NFW halo fits is 
greatly improved when the radial range of the data extends into the 
flat part of the rotation curve. 

For those \Dpak\ galaxies with photometry, we have also presented 
isothermal and NFW halo fits for four assumptions of the stellar 
mass-to-light ratio, \ml.  We have tested the zero, minimal, popsynth,
and maximum disk cases.  We have found that the NFW halo is a poor 
description of the data, and that the NFW halo and maximum disk are 
mutually exclusive. There 
is a substantial cusp mass excess near the centers of the galaxies 
with at least two times more mass expected in the cuspy CDM halo than 
is allowed by the data.  Most galaxies in the sample require $\sim$20 
\kms\ noncircular motions to reconcile the differences between 
observations and the NFW halo.  Even larger noncircular motions are 
required when stars are allowed to have mass.  We have also found 
the data to favor a low $\sigma_{8}$ cosmology.

\section{Acknowledgements}

We thank James Bullock for providing the theoretical predictions for 
$\Delta_{V/2}$.  The work of R.~K.~D. and S.~S.~M. was supported by NSF grant 
AST0505956. R.~K.~D. is also supported by an NSF Astronomy \&
Astrophysics Postdoctoral Fellowship under award AST0702496.  
This paper was part of R.~K.~D.~'s Ph.~D. dissertation at 
the University of Maryland.  The work of W.~J.~G.~d.~B. is supported 
by the South African Research Chairs Initiative of the Department of 
Science and Technology and National Research Foundation.      
This research has made use of the NASA/IPAC Extragalactic Database 
(NED) which is operated by the Jet Propulsion Laboratory, California 
Institute of Technology, under contract with the National Aeronautics 
and Space Administration.  Our velocity field plots were made using a 
modified version of the program found at 
\newline\url{http://www.astro.wisc.edu/$\sim$mab/research/densepak/}
\newline\url{DP/dpidl.html}


\begin{thebibliography}{}
\bibitem[Alam et al.(2002)Alam, Bullock, \& Weinberg]{Alam} Alam,
  S.~M.~K., Bullock, J.~S., \& Weinberg, D.~H. 2002, \apj, 572, 34
\bibitem[Avila-Reese et al.(1998)Avila-Reese, Firmani, \& Hernandez]
{AvilaReese} Avila-Reese, V., Firmani, C., \& Hernandez, X. 1998, 
\apj, 505, 37
\bibitem[Baggett et al.(1998)Baggett, Baggett, \& Anderson]{Baggett}Baggett, W.~E., 
Baggett, S.~M., \& Anderson, K.~S.~J. 1998, \aj, 116, 1626
\bibitem[Begeman(1989)]{Begeman} Begeman, K.~G. 1989, \aap, 223, 47
\bibitem[Bell et al.(2003)]{Belletal03} Bell, E.~F., McIntosh, D.~H., Katz, 
N., \& Weinberg, M.~D. 2003, \apjs, 149, 289
\bibitem[Borriello \& Salucci(2001)]{Borriello} Borriello, A., \& Salucci, 
P. 2001, \mnras, 323, 285
\bibitem[Bullock et al.(2001)]{Bullock01} Bullock, J.~S., Kolatt,
  T.~S., Sigad, Y., Somerville, R.~S., Kravtsov, A.~V., Klypin, A.~A.,
  Primack, J.~R., \& Dekel, A. 2001, \mnras, 321, 559
\bibitem[Chemin et al.(2004)]{Chemin04} Chemin, L., et al. 2004, IAUS, 220, 333
\bibitem[Cole \& Lacey(1996)]{ColeLacey} Cole, S., \& Lacey, C. 1996,
  \mnras, 281, 716
\bibitem[de Blok \& Bosma(2002)]{dBB} de Blok, W.~J.~G., \& Bosma, A. 
2002, \aap, 385, 816 
\bibitem[de Blok et al.(2003)de Blok, Bosma, \& McGaugh]{dBBM} de 
Blok, W.~J.~G., Bosma, A., \& McGaugh, S.~S. 2003, \mnras, 340, 657
\bibitem[de Blok \& McGaugh(1996)]{dBM96} de Blok, W.~J.~G., \& McGaugh, 
S.~S. 1996, \apj, 469, L89
\bibitem[de Blok \& McGaugh(1997)]{dBM97} ---------- . 1997, \mnras, 290, 533
\bibitem[de Blok et al.(2001)de Blok, McGaugh, \& Rubin] {dBMR} de Blok, W.~J.~G., 
McGaugh, S.~S., \& Rubin, V.~C. 2001, \aj, 122, 2396 
\bibitem[de Blok et al.(1996)de Blok, McGaugh, \& van der Hulst]{DMV} de Blok, 
W.~J.~G., McGaugh, S.~S., \& van der Hulst, J.~M. 1996, \mnras, 283, 18
\bibitem[de Blok \& van der Hulst(1998)]{dBvdH} de Blok, W.~J.~G., \& 
van der Hulst, J.~M. 1998, \aap, 336, 49
\bibitem[de Blok et al.(1995)de Blok, van der Hulst, \& Bothun]{dBetal95} de Blok, W.~J.~G., 
van der Hulst, J.~M., \& Bothun, G.~D. 1995, \mnras, 274, 235
\bibitem[de Jong(1996)]{deJong} de Jong, R.~S. 1996, \aaps, 118, 557
\bibitem[Diemand et al.(2005)]{Diemand} Diemand, J., Zemp, M., Moore, B.,
Stadel, J., \& Carollo, M. 2005, \mnras, 364, 665
\bibitem[Filippenko \& Sargent(1989)]{Filippenko} Filippenko, A.~V., 
\& Sargent, W.~L.~W. 1989, \apjl, 342, L11
\bibitem[Flores \& Primack(1994)]{Flores} Flores, R.~A., \& Primack, J.~R.
1994, \apjl, 427, L1
\bibitem[Fuchs(2003)]{Fuchs} Fuchs, B. 2003, \apss, 284, 719
\bibitem[Gentile et al.(2005)]{Gentile05} Gentile, G., Burkert, A.,
  Salucci, P., Klein, U., \& Walter, F. 2005, \apj, 634, L145
\bibitem[Gentile et al.(2007)]{Gentile07} Gentile, G., Salucci, P.,
  Klein, U., \& Granato, G.~L. 2007, \mnras, 375, 199
\bibitem[Gentile et al.(2004)]{Gentile} Gentile, G., Salucci, P., 
Klein, U., Vergani, D., \& Kalberla, P. 2004, \mnras, 351, 903
\bibitem[Heraudeau \& Simien(1996)]{Heraudeau} Heraudeau, P.,  \&
  Simien, F. 1996, \aaps, 118, 111
\bibitem[James et al.(2004)]{James} James, P.~A., et al. 2004, \aap,
  414, 23
\bibitem[Klypin et al.(2001)]{Klypin} Klypin, A., Kravtsov, A.~V.,
  Bullock, J.~S., \& Primack, J.~R. 2001, \apj, 554, 903
\bibitem[Kormendy \& Freeman(2004)]{Kormendy} Kormendy, J., \&
  Freeman, K.~C. 2004, IAUS, 220, 377
\bibitem[Kuzio de Naray et al.(2006)]{Kuzio} Kuzio de Naray, R.,
  McGaugh, S.~S., de Blok, W.~J.~G., \& Bosma, A. 2006, \apjs, 165,
  461 (K06)
\bibitem[Marchesini et al.(2002)]{Marchesini} Marchesini, D., 
D'Onghia, E., Chincarini, G., Firmani, C., Conconi, P., Molinari, E., 
\& Zacchei, A. 2002, \apj, 575, 801
\bibitem[McGaugh(2005)]{McGaugh05} McGaugh, S.~S. 2005, \apj, 632, 859
\bibitem[McGaugh et al.(2003)McGaugh, Barker, \& de Blok]{McGaugh03} McGaugh, S.~S.,
  Barker, M.~K., \& de Blok, W.~J.~G. 2003, \apj, 584, 566
\bibitem[McGaugh \& de Blok(1998)]{McGdB98} McGaugh, S.~S., \& de Blok, 
W.~J.~G. 1998, \apj, 499, 41
\bibitem[McGaugh et al.(2007)]{McGaugh07} McGaugh, S.~S., de Blok,
  W.~J.~G., Schombert, J.~M., Kuzio de Naray, R., \& Kim, J.~H. 2007, ApJ,
  659, 149
\bibitem[McGaugh et al.(2001)McGaugh, Rubin, \& de Blok]{MRdB} McGaugh, S.~S., Rubin, 
V.~C., \& de Blok, W.~J.~G. 2001, \aj, 122, 2381 
\bibitem[Mihos et al.(1999)Mihos, Spaans, \& McGaugh]{Mihos} Mihos, J.~C., Spaans, 
M., \& McGaugh, S.~S. 1999, \apj, 515, 89
\bibitem[Moore et al.(1999)]{Moore} Moore, B., Quinn, T., Governato, 
F., Stadel, J., Lake, G. 1999, \mnras, 310, 1147
\bibitem[Navarro et al.(1996)Navarro, Frenk, \& White]{NFW96} Navarro, J.~F., Frenk, 
C.~S., \& White, S.~D.~M. 1996, \apj, 462, 563 
\bibitem[Navarro et al.(1997)Navarro, Frenk, \& White]{NFW97} ---------- . 1997,
  \apj, 490, 493 
\bibitem[Navarro et al.(2004)]{Navarro2004} Navarro, J.F., et
  al. 2004, \mnras, 349, 1039
\bibitem[Osterbrock et al.(1996)]{Osterbrock} Osterbrock, D.~E., 
Fulbright, J.~P., Martel, A.~R., Keane, M.~J., Trager, S.~C., \& Basri, 
G. 1996, \pasp, 108, 277
\bibitem[Palunas \& Williams(2000)]{Palunas} Palunas, P., \& Williams, 
T.~B. 2000, \aj, 120, 2884
\bibitem[Reed et al.(2003)]{Reed} Reed, D., Gardner, J., Quinn, T., 
Stadel, J., Fardal, M., Lake, G., \& Governato, F.  2003, \mnras, 346, 565
\bibitem[Regan et al.(2001)]{Regan} Regan, M.~W., Thornley, M.~D., 
Helfer, T.~T., Sheth, K., Wong, T., Vogel, S.~N., Blitz, L., \& Bock, 
D.~C.~J. 2001, \apj, 561 218
\bibitem[Rhee et al.(2004)]{Rhee04} Rhee, G., Valenzuela, O., Klypin, A.,
Holtzman, J., \& Moorthy, B. 2004, \apj, 617, 1059
\bibitem[Rosenberg \& Schneider(2003)]{Rosenberg} Rosenberg, J.~L., \&
  Schneider S.~E. 2003, \apj, 585, 256
\bibitem[Schombert et al.(1990)]{Schombert} Schombert, J.~M., Bothun, 
G.~D., Impey, C.~D., \& Mundy, L.~G. 1990, \aj, 100, 1523
\bibitem[Simon et al.(2003)]{Simon03} Simon, J.~D., Bolatto, A.~D., 
Leroy, A., \& Blitz, L. 2003, \apj, 596, 957
\bibitem[Simon et al.(2005)]{Simon05} Simon, J.~D., Bolatto, A.~D., 
Leroy, A., Blitz, L., \& Gates, E.~L. 2005, \apj, 621, 757
\bibitem[Spano et al.(2007)]{Spano07} Spano, M., Marcelin, M., Amram, P., 
Carignan, C., Epinat, B., \& Hernandez, O. 2007, \mnras, in press 
(arXiv:0710.1345)
\bibitem[Spergel et al.(2007)]{Spergel07} Spergel, D.~N., et al. 2007,
  ApJS, 170, 377
\bibitem[Stil(1999)]{Stil} Stil, J. 1999, Ph.D. Thesis, University of 
Leiden
\bibitem[Swaters(1999)]{Swatersthesis} Swaters, R.~A. 1999,
Ph.D. Thesis, University of Groningen
\bibitem[Swaters \& Balcells(2002)]{SwatersBalcells} Swaters, R.~A., \&
  Balcells, M. 2002, \aap, 390, 863
\bibitem[Swaters et al.(2003a)]{Rob} Swaters, R.~A., Madore, B.~F., 
van den Bosch, F.~C., \& Balcells, M. 2003a, \apj, 583, 732
\bibitem[Swaters et al.(2002)]{Swatersetal02} Swaters, R.~A., van
  Albada, T.~S., van der Hulst, J.~M., \& Sancisi, R. 2002, \aap,
  390, 829
\bibitem[Swaters et al.(2003b)]{Swaters03} Swaters, R.~A., Verheijen, 
M.~A.~W., Bershady, M.~A., \& Andersen, D.~R. 2003b, \apj, 587, L19
\bibitem[Tegmark et al.(2004)]{Tegmark} Tegmark, M., et al. 2004, 
Phys.\ Rev.\ D, 69, 103501
\bibitem[Tully(1988)]{Tully} Tully, R.~B. 1988, Nearby Galaxies 
Catalogue (Cambridge: Cambridge Univ. Press) 
\bibitem[van den Bosch et al.(2000)]{vandenB00} van den Bosch, F.~C., 
Robertson, B.~E., Dalcanton, J.~J., \& de Blok, W.~J.~G. 2000, \aj, 119,
1579
\bibitem[van den Bosch \& Swaters(2001)]{vandenB01} van den Bosch, F.~C., 
\& Swaters, R.~A. 2001, \mnras, 325, 1017
\bibitem[van der Hulst et al.(1993)]{vanderHulst} van der Hulst, J.~M.,
  Skillman, E.~D., Smith, T.~R., Bothun, G.~D., McGaugh, S.~S., \& de
  Blok, W.~J.~G. 1993, \aj, 106, 548 
\bibitem[van der Kruit \& Searle(1981)]{vanderKruit81} van der Kruit, 
P.~C., \& Searle, L. 1981, \aap, 95, 105
\bibitem[van Zee \& Haynes(2006)]{vanZeeHaynes} van Zee, L., \&
  Haynes, M.~P. 2006, \apj, 636, 214
\bibitem[van Zee et al.(1997)]{vanzee97} van Zee, L., Haynes, M.~P.,
  Salzer, J.~J., \& Broeils, A.~H. 1997, \aj, 113, 1618
\bibitem[Verheijen \& de Blok(1999)]{VerheijendB} Verheijen, M., \& de
  Blok, W.~J.~G. 1999, \apss, 269, 673
\bibitem[Verheijen et al.(2007)]{Verheijen07} Verheijen, M.~A.~W., 
Bershady, M.~A., Swaters, R.~A., Andersen, D.~R., \& Westfall, K.~B. 
2007, ISLAND UNIVERSES, Astrophysics and Space Science
Proceedings.~ISBN 978-1-4020-5572-0.~Springer, 2007, p.~95

\end{thebibliography}
\end{document}